\documentclass[manuscript]{aastex}

\shorttitle{}
\shortauthors{K. Lakhchaura et al.}
\usepackage{psfig}
\usepackage{epsfig}
\usepackage{graphicx}
\usepackage{rotate}
\usepackage{subfigure}
\usepackage{natbib}
\usepackage{threeparttable}
\usepackage[abs]{overpic}
\usepackage{amsmath}
\usepackage{amssymb}

\begin{document}
\title{Intracluster medium of the merging cluster Abell 3395}

\author{Kiran Lakhchaura\altaffilmark{1} and K. P. Singh}
\affil{Department of Astronomy and Astrophysics, Tata Institute of Fundamental Research, \\1 Homi Bhabha Road, Mumbai 400005, India}

\author{D. J. Saikia}
\affil{National Centre for Radio Astrophysics, Tata Institute of Fundamental Research, Pune University Campus, Pune 411007, India}

\and

\author{R. W. Hunstead}
\affil{Sydney Institute for Astronomy, School of Physics, University of Sydney, NSW 2006, Australia}

%
\altaffiltext{1}{e-mail : kiran\_astro@tifr.res.in}

\begin{abstract}
We present a detailed imaging and spectral analysis of the merging environment of the bimodal cluster A3395 using X-ray and radio 
observations. X-ray images of the cluster show five main constituents of diffuse emission : A3395 NE, A3395 SW, A3395 NW, A3395 W, 
and a filament connecting NE to W. X-ray surface-brightness profiles of the cluster did not show any shock fronts in the cluster. 
Temperature and entropy maps show high temperature and high entropy regions in the W, the NW, the filament and between the NE and SW 
subclusters. The NE, SW and W components have X-ray bolometric luminosities similar to those of rich clusters of galaxies but have 
relatively higher temperatures. Similarly, the NW component has X-ray bolometric luminosity similar to that of isolated groups but 
with much higher temperature. It is, therefore, possible that all the components of the cluster have been heated by the ongoing 
mergers. The NE subcluster is the most massive and luminous constituent and other subclusters are found to be gravitationally bound 
to it. The W component is most probably either a clump of gas stripped off the SW due to ram pressure or a separate subcluster that 
has merged or is merging with the SW. No X-ray cavities are seen associated with the Wide Angle Tailed (WAT) radio source near the 
centre of the SW subcluster. Minimum energy pressure in the radio emission-peaks of the WAT galaxy is comparable with the external 
thermal pressure. The radio spectrum of the WAT suggests a spectral age of $\sim$∼10Myr.

\end{abstract}


\section{Introduction}
\label{sec:Intro}
  Clusters of galaxies, the largest gravitationally-bound structures, are believed to form hierarchically in a sequence of cosmic 
structure formation where smaller groups of galaxies merge to form larger and richer systems (Geller \& Beers 1982; Dressler \& 
Shectman 1988; Girardi et al. 1997; Kriessler \& Beers 1997; Jones \& Forman 1999; Schuecker et al. 2001; Burgett et al. 2004). The 
intra-group or intra-cluster medium (ICM) of these clusters contains a hot (T$\sim10^{7}-10^{8}$K) and tenuous (n$\sim10^{-3}$ 
cm$^{-3}$) plasma which emits in X-rays mainly through bremsstrahlung and line emission (Kellogg et al. 1972; Mitchell et al. 1979; 
also see reviews by Sarazin 1988; McNamara \& Nulsen 2007). Although many rich clusters show smooth X-ray surface-brightness 
distributions with a central peak, a large fraction (at least, $\sim$30\% (Forman \& Jones 1990) or even up to $\sim$40\%-60\% (Jones 
\& Forman 1992, 1999; Schuecker et al. 2001)) shows double or multiple peaks in their X-ray surface brightness maps. As X-ray 
surface-brightness is more sensitive to density than temperature these observations effectively show peaks in the density distribution
 of the ICM itself, which suggests the presence of subclusters or substructures which are possibly in the process of merging to form 
richer groups and clusters of galaxies (e.g. Nakamura et al. 1995). Presence of large inhomogeneities in the X-ray temperature of the 
ICM (greater than a few keV) on scales less than 0.5 Mpc provides additional observational evidence of mergers (Roettiger et al. 
1996). As dynamical evolution rapidly removes substructures, their existence proves that such clusters are dynamically young systems 
(Flin 2003). 

  For studies of cosmology and the formation and evolution of large-scale structures, it is important to study the physical properties
 of clusters of galaxies containing sub-groups which are young and in the process of merging. Mergers of subgroups can produce shocks,
 bulk gas flows and turbulence in the ICM, and can disrupt cooling flows as has been seen in many clusters (see Fabian 1994; 
Markevitch \& Vikhlinin 2007; Owers et al. 2011; Maurogordato et al. 2011). Mergers may also cause metal enrichment of the ICM through
 enhanced ram-pressure stripping during mergers (Domainko et al. 2005), thereby affecting the evolution of the galaxies and the ICM. 
Subclustering information has also been used for estimating cosmological parameters such as the density parameter $\Omega_{0}$ 
(Richstone et al. 1992; Kauffmann \& White 1993, Angrick \& Bartelmann 2011).

  At radio wavelengths, besides the diffuse emission which may be in the form of halos and relics, the tailed radio sources associated
 with individual galaxies can provide further insights towards understanding these systems. For example, the Wide Angle Tail (WAT) 
galaxies are usually associated with the dominant galaxy in a group or a cluster, with the jets flaring into tails of emission which 
can extend to over a Mpc in size (e.g. Roettiger et al. 1996; Douglass et al. 2008; Mao et al. 2010). Their relatively high 
luminosity, close to the upper range of the Fanaroff-Riley I sources, and large sizes make these good tracers of galaxy clusters at 
moderate and high redshifts (e.g. Blanton et al. 2003; Douglass et al. 2008). Since the dominant galaxies with which WATs are 
associated are likely to have at most small peculiar velocities, their jets are unlikely to be bent by ram pressure caused by the 
motion of the galaxy (Burns 1981; Eilek et al. 1984; O'Donoghue et al. 1990). Cluster mergers where relative velocities between the 
systems can exceed $\sim$1000 km s$^{-1}$, causing significant bulk motion of the ICM, have been invoked as the primary scenario for 
understanding the WAT structures (Eilek et al. 1984; Burns et al. 1994; Roettiger et al. 1996).

  X-ray observatories, particularly Chandra and XMM-Newton, have brought to light new information leading to a more detailed 
understanding and new insights into the physical processes occurring in galaxy clusters (Weratschnig 2010). While Chandra with its 
unprecedented spatial resolution of 0.5$^{\prime\prime}$ can study small-scale phenomena such as cold fronts, merger shocks and AGN 
cavities,  XMM-Newton with its larger collecting area and sensitivity can study clusters to a much farther extent and can detect 
fainter X-ray signals. Because of its large FOV of $\sim30^{\prime}$ and moderate resolution (6$^{\prime\prime}$ FWHM), XMM-Newton can 
also study the nearby clusters very well. 

  We have chosen A3395 for a detailed study of its merging environment since it is a well-known bimodal merging cluster (Henriksen \& 
Jones 1996; Markevitch et al. 1998; Donnelly et al. 2001; Flin \& Krywult 2006) and it has been observed with XMM-Newton, Chandra and 
Australia Telescope Compact Array (ATCA).

  The paper is organized as follows. Detailed information about A3395 and a summary of its earlier X-ray observations with the 
Roentgen Satellite (ROSAT) and the Advanced Satellite for Cosmology and Astrophysics (ASCA) and the previous radio observations are 
given in \S\ref{sec:A3395}. The details of the recent X-ray and unpublished radio observations and their data analyses, are presented 
in \S\ref{sec:Obs_n_data_red}. The results including the X-ray and radio morphology, X-ray surface brightness profiles, X-ray 
temperature, metallicity, density, entropy and pressure maps, mass, luminosity and cooling time estimates are provided in 
\S\ref{sec:results}. A discussion of the results and the merging environment of A3395 is given in \S\ref{sec:discussion}. A lambda 
cold dark matter cosmology with $H_{0}$ = 70 km s$^{-1}$ Mpc$^{-1}$ and $\Omega_{M}$ = 0.3 ($\Omega_{\Lambda}$ = 0.7) has been 
assumed, so that 1 arcsec corresponds to 0.96 kpc at z=0.0498, the redshift of the cluster.

\section{A3395}
\label{sec:A3395}
  Located nearby, at a distance of  211 Mpc and redshift of 0.0498, the positional coordinates of A3395  are : R.A.(J2000) $= 06^{\rm 
h} 27^{\rm m} 31.1^{\rm s} $, Dec.(J2000)$ = -54^{\rm d} 23^{\prime} 58^{\prime\prime}$ (l=262.9589$^{\circ}$, b=-25.007$^{\circ}$) (from Set of 
Identifications, Measurements, and Bibliography for  Astronomical Database (SIMBAD)). A3395 is a regular cluster with 54 members 
(cluster members between m$_{3}$ (magnitude of $3^{\rm rd}$ brightest galaxy) and m$_{3}$+2) and richness class 1 (Abell, Corwin, \& 
Olowin, 1989). It has a large extent with $r_{180}$ (radius within which mean density of the cluster equals 180 times the critical 
density at the redshift of the cluster) $=$ 34.6$^{\prime}$ (Markevitch et al. 1998) and hence can be studied with XMM-Newton very 
well.

  A3395 has been observed previously in X-rays with the ROSAT Position Sensitive Proportional Counter (PSPC) by Henriksen \& Jones 
(1996). They reported temperature of 2 to 4 keV for the cluster and calculated the cooling time for the cluster as $\sim10$ Gyr, which
 is much larger than the time predicted for the cluster to fully merge. Hence, they suggested that either there are no cooling flows 
at the centre of the cluster or the cooling flows are disrupted by the merger. Markevitch et al. (1998) reanalyzed the ROSAT PSPC data
 and combined it with the analysis of ASCA observations. They found some evidence for the presence of shock-heated gas in the 
cluster and suggested that a collision between the two subclusters may be responsible for heating the gas. They calculated the 
emission-weighted gas temperature $T_{X}$ of A3395 (excluding cooling flow and other contaminating components) to be $4.8\pm0.4$ keV. 
Donnelly et al. (2001) reanalyzed both the ROSAT PSPC and the ASCA data and reinforced the evidence for a merger in A3395, as they 
detected a rise in the gas temperature ($\sim30\%$ above the average temperature) in the region between the two main subcluster 
components. By using Newtonian energy considerations Donnelly et al. proved that the system of the two subclusters is in a bound 
state and determined the virial masses of the two main subclusters (NE and SW) as $4.5^{+1.1}_{-0.9}\times10^{14} \rm M_{\odot}$ and 
$3.1^{1.8}_{-1.4}\times10^{14} \rm M_{\odot}$ respectively. Tittley \& Henriksen (2001) report the presence of a group of lower
 redshift galaxies (z=0.048), at $\sim12^{\prime}$ northwest from the centre of the galaxy distribution, which is elongated in the NW-SE 
direction. They found an increase in the X-ray emission from this group and classified this as the third component of X-ray emission 
viz., A3395 NW (centred at R.A.(J2000) $= 06^{\rm h} 26^{\rm m} 35^{\rm s} $, Dec.(J2000)$ = -54^{\rm d} 20^{\prime}$). This is in addition
 to the two main X-ray peaks A3395 NE and A3395 SW, that coincide with subclusters, at slightly higher redshifts of 0.051 and 0.052
 respectively. Using ASCA and wide field of view ROSAT observations, they also detected a filamentary structure that connects the 
clusters A3395 and A3391, shown by the presence of excess X-ray emission in a region between the two clusters and aligned with the 
distribution of galaxies.

  A3395 has also been observed in radio with the Molonglo Cross Telescope at 408 MHz (Large et al. 1981) and with the Molonglo 
Observatory Synthesis Telescope (MOST) at 843 MHz (Jones \& McAdam 1992; Burgess \& Hunstead 2006). The observations showed a WAT 
inside the cluster. The cluster was also observed by ATCA at a frequency of 4.79 GHz (Gregorini et al. 1994). Reid (2000) used ATCA 
observations of this cluster at 1348 MHz and 2374 MHz, and detected eight radio sources of which five had optical counterparts. 
There are 2 prominent radio sources in the SW subcluster, a WAT at the centre (close to the X-ray peak) and a 
Head-Tail (HT) galaxy at the periphery. We have used the ATCA data to understand the interaction of the WAT radio emission with the 
ICM.

\section{Observations and Data Reduction}
\label{sec:Obs_n_data_red}
  A journal of the XMM-Newton and Chandra observations of A3395 is given in Table~\ref{tab:observation_table}.

\subsection{XMM-Newton}
\begin{figure}
\begin{center}$
 \begin{array}{cc}
  \includegraphics[width=3.5in,height=3.5in]{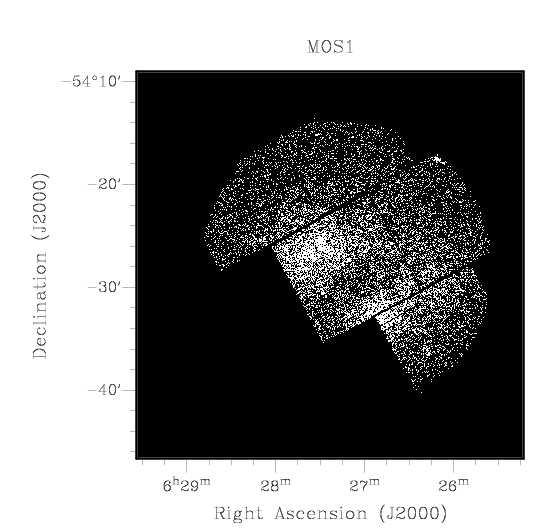}
  \includegraphics[width=3.5in,height=3.5in]{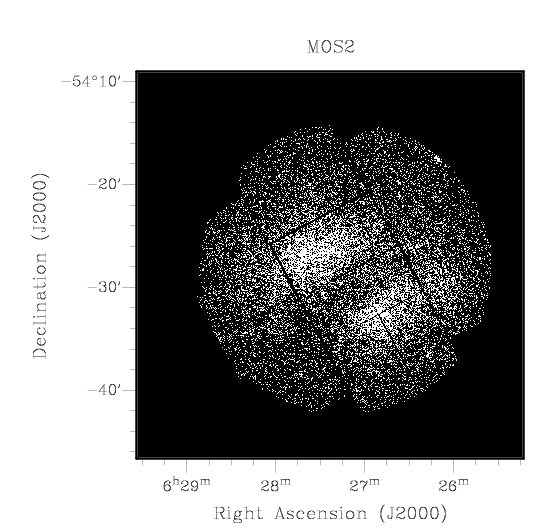}\\
  \includegraphics[width=3.5in,height=3.5in]{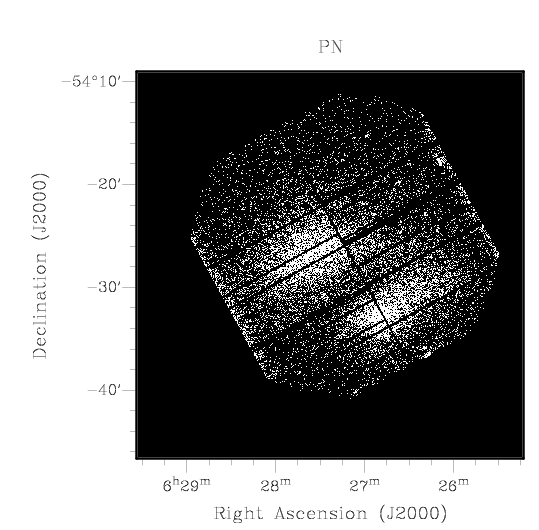}
  \includegraphics[width=3.3in,height=3.3in]{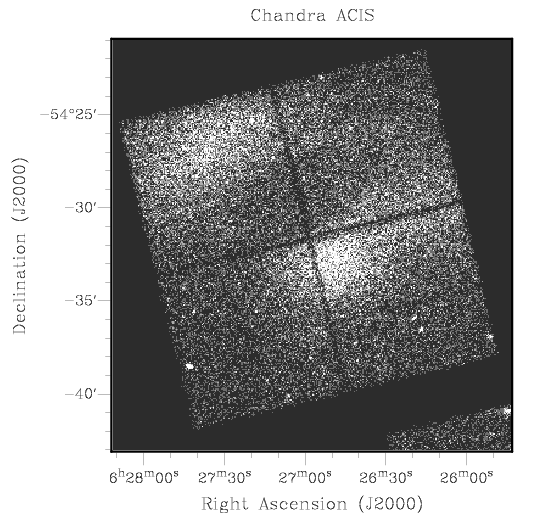}
 \end{array}$
\end{center}
\caption{Raw images from XMM-Newton MOS1 (top left), MOS2 (top right) and PN (bottom left) detectors and from Chandra ACIS detector
(bottom right). Note that, the MOS1 image is missing two CCDs viz. CCD\# 5 and CCD\# 6 due to enhanced background at E $<$ 1keV and 
meteorite hit respectively.}
\label{fig:Raw_MOS1_MOS2_PN_Chandra_raw_images}
\end{figure}

 The cluster was observed with XMM-Newton on 2007 January 24 (Table~\ref{tab:observation_table}). The three EPIC cameras MOS1, MOS2 
(Turner et al. 2001) and PN (Str\"{u}der et al. 2001) were operated in full frame mode with the thin1 filter. The data have been 
obtained from the HEASARC archives. The raw MOS1, MOS2 and PN images are shown in 
Figure~\ref{fig:Raw_MOS1_MOS2_PN_Chandra_raw_images}. Note that the observation has been performed after the loss of MOS1 CCD\# 6 due
 to a meteorite hit in 2006 June. Also, during the analysis, MOS1 CCD\# 5 was found to be in anomalous mode (because of a strong 
enhancement of the background at $E < 1$ keV) and hence, has not been used anywhere in the imaging and spectral analysis. 

  All data analysis has been done using the standard procedures from the Science Analysis System (SAS) software version 9.0. 
Calibrated photon event files were produced from the raw data using the SAS tasks \textbf{epchain and emchain} and the latest 
calibration files. These files were then filtered for the good time intervals using the SAS tasks \textbf{mos-filter} and 
\textbf{pn-filter}. Good time interval found for MOS1, MOS2 and PN CCDs are 27.16 ks, 27.28 ks and 22.44 ks respectively (see 
Table~\ref{tab:observation_table}).

\subsubsection{Background Treatment}
  The histograms of lightcurves from all three detectors had very well defined Gaussian shapes showing that the data are not 
much affected by the soft-proton contamination. Also the temporal filtering using the tasks \textbf{mos-filter} and \textbf{pn-filter} 
removes the soft proton contamination sufficiently. The residual soft proton contamination left after this step was 
removed from the data by adding powerlaws to the models (Snowden et al. 2008) used in the spectral analyses done in 
\S\ref{sec:Total-Spec-Analy}, \S\ref{sec:Azimuth_spec_analys} and \S\ref{sec:box_thermodynamic_maps}. The details have been provided 
in \S\ref{sec:Total-Spec-Analy}. For removing the quiescent particle induced background and the cosmic background component we have 
used the blanksky observations, which consist of a superposition of pointed observations that have been processed with SAS version 
7.1.0 (Carter \& Read, 2007). Blank-sky event files were obtained for all three detectors by submitting an XMM-Newton EPIC Background 
Blank Sky Products Request Form with the request of a Galactic column in the range 3.5 to $8.5\times10^{20}$ cm$^{-2}$. The 
blanksky event files from the MOS(PN) detectors were then filtered for flares using an upper threshold of 0.35(0.40)
 counts second$^{-1}$ and then again using the selection criteria of $\rm PATTERN\leq 12$ ($\rm PATTERN\leq 4$) and $\rm FLAG=0$. 
The files were then recast to have the same sky coordinates as A3395. The resulting event files were used for generating all the 
background images and background spectra for this paper. The high energy (E$\sim$10-12 keV) count-rates for the MOS1, MOS2 and PN 
detectors from the filtered blanksky images were found to be very close to those from the source images showing that the observations 
were not affected by flares. We could not do the local background subtraction as the source fills almost the entire field of view of 
the detectors and it was impossible to find emission-free regions and therefore, using local background could have led to 
oversubtraction.

\subsubsection{Point source subtraction}
\begin{figure}
\centering
\includegraphics[width=7.0in]{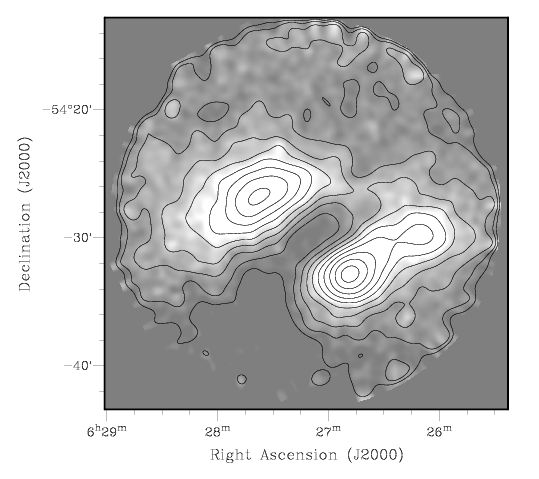}
\caption{The exposure corrected combined image from MOS1 and MOS2 detectors after removing point sources and smoothing with a 
Gaussian kernel of width 17.5$^{\prime\prime}$. Overlaid contours are logarithmically distributed between $1.72\times10^{-6}-8.78\times10^{-6}$ 
counts s$^{-1}$arcsec$^{-2}$.}
\label{fig:Combined_MOS1_MOS2_smoothedimage}
\end{figure}

 The MOS1 and MOS2 images were combined together using the SAS tasks \textbf{merge} and \textbf{evselect} to increase the 
signal-to-noise ratio of the sources and reach fainter flux levels. Only those events that were spread over less than 4 pixels (i.e., 
pattern =0-12) were selected for producing the combined image. Point source extraction was done simultaneously on the MOS1, MOS2 
combined image and the PN detector image using the SAS task \textbf{edetect\_chain} which is a combination of several SAS tasks viz. 
\textbf{eboxdetect} (in local mode), \textbf{esplinemap}, \textbf{eboxdetect} (in map mode), and \textbf{emldetect}. When run in local 
mode \textbf{eboxdetect} produced source lists by collecting source counts in the cells of $5\times5$ pixels and using a value of 8 for 
the minimum detection likelihood (\textbf{eboxl\_likemin}) (i.e. a value of e$^{-8}$ for the minimum value of the probability of the 
Poissonian random fluctuation of the counts in the detection cell which would result in the observed number of source counts). Then 
\textbf{esplinemap} was used to generate background maps by using the source list generated by \textbf{eboxdetect} (in local mode) to 
remove point sources from the original merged image. Then \textbf{eboxdetect} (in map mode) uses these background maps to generate a 
new point source list (with fainter sources included). A total of 258 point sources were thus detected. Each of these detected sources
 was checked for each of the detectors and spurious sources (sources that did not look like actual sources in the individual detector 
images) were removed. Finally 77 sources for MOS1, 67 for MOS2, and 44 sources for PN were detected. Of these, 22 sources were found 
to be common in MOS1 and MOS2 and 13 were common in all three detectors. These detected sources were then removed from individual 
MOS1, MOS2 and PN images to create the cheesed images. The cheesed MOS1 and MOS2 images were combined using the procedure described 
in Snowden et al. (2008). A contour 
map of diffuse X-ray emission from A3395 using the combined MOS1 and MOS2 image after the removal of point sources and after applying 
the smoothing function is shown in Figure~\ref{fig:Combined_MOS1_MOS2_smoothedimage}. Contours of the smoothed X-ray emission are also
 overlaid on the optical image of A3395 from the SuperCOSMOS survey in the $B_J$ band and shown in 
Figure~\ref{fig:A3395_SuperCOSMOS_optical_image}. The optical image shows both the bright central galaxies (BCGs) for the NE and SW 
regions. The position of the BCG for the NE region coincides with the X-ray emission peak for the NE region whereas the BCG for the 
SW region is offset from the  X-ray emission peak for the SW region by about $\sim16.5^{\prime\prime}$ which is equivalent to $\sim16$ kpc at the
 redshift of the cluster.

\subsection{Chandra X-ray Observatory}
  A3395 was observed with Chandra on 2004 July 11 (ObsID 4944) with ACIS-I detector for 22.2 ks (Table~\ref{tab:observation_table}). 
The raw Chandra ACIS image is shown in Fig.~\ref{fig:Raw_MOS1_MOS2_PN_Chandra_raw_images}. The data were
 analyzed with the CIAO version 4.3 and CALDB version 4.4.0. No reprocessing of data and time-dependent corrections were required as 
the ASCDSVER (keyword that stores the processing version information) is DS 7.6.7.2 and time-dependent corrections have become a part 
of standard data processing after DS 7.3.0. The standard Charge Transfer Inefficiency (CTI) corrections have been applied. Point 
sources were detected using the CIAO task \textbf{wavdetect} with the detection threshold fixed to the default value $10^{-6}$. A 
total of 47 bright sources were detected; 16 of these sources were common with the sources detected in the MOS 2 detector image, 
discussed in the previous section. The smoothed, point-source-removed and exposure-corrected Chandra image of the cluster in the 
0.3-7.0 keV band is shown in Figure \ref{fig:Chandra_smoothedimage}. 

\subsection{Australia Telescope Compact Array}
  A3395 was observed by A.D. Reid and R.W. Hunstead with ATCA on 1995 January 09. The pointing centre was 06$^{\rm h}$ 26$^{\rm m}$ 
58.0$^{\rm s}$, $-$54$^{d}$ 31$^{\prime}$ 12.0$^{\prime\prime}$ (J2000) and centre frequencies were 1348 and 2374 MHz with 
bandwidths of 128 MHz. The array was in the 6A configuration, with baselines ranging from 337 to 5939 metres. The total integration 
time was 5.6 hours, consisting of multiple cuts spread over a wide range of hour angles. Primary flux density calibrator was 
B1934$-$638 and the phase calibrator was B0647$-$475.

  Data reduction was carried out in {\tt MIRIAD} (see Sault et al. 1995) using standard techniques. The restored beams were 
11.0$\times$8.9 arcsec$^2$ along PA=$-$30.5$^\circ$  at 1348 MHz, and 6.2$\times$5.0 arcsec$^2$ PA $-$32.5$^\circ$ at 2374 MHz. The 
rms noise is 1.4 mJy beam$^{-1}$ in the 1348 MHz image and 1.3 mJy beam$^{-1}$ in the 2374 MHz image. A contour map of the 1348 MHz 
ATCA radio emission has been overlaid on the image from the SuperCOSMOS survey in the $B_J$ band and is shown in 
Fig.~\ref{fig:A3395_SuperCOSMOS_optical_image}.

\section{Analysis and Results}
\label{sec:results}

\subsection{X-ray Morphology}
\label{X-ray_morphology}
\begin{figure}
\centering
\includegraphics[width=7.0in]{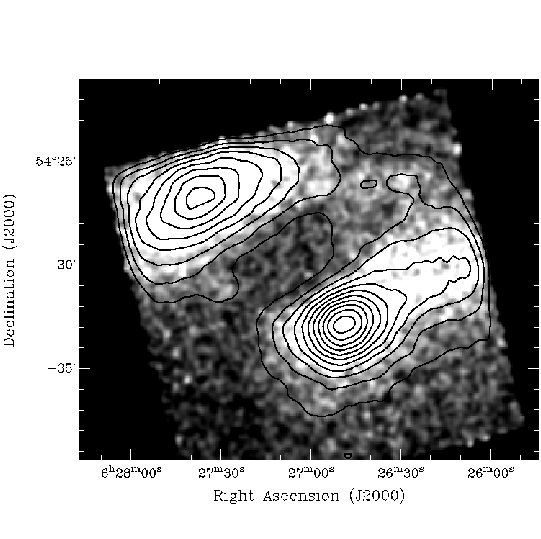}
\caption{Exposure-corrected Chandra ACIS image of A3395 in the 0.3-7.0 keV band (after point source removal and smoothing with a 
Gaussian kernel of width 8$^{\prime\prime}$). The overlaid X-ray emission contours are at 2, 4, 6, 8, 10, 12 , 14, 16, 18, 20, 22, and 24 times 
$\sigma=1.16\times10^{-8}$ counts cm$^{-2}$s$^{-1}$pixel$^{-1}$ level above the mean background level (=$4.67\times10^{-8}$ counts 
cm$^{-2}$s$^{-1}$pixel$^{-1}$).}
\label{fig:Chandra_smoothedimage}
\end{figure}

\begin{figure}
\centering
\includegraphics[width=6.0in]{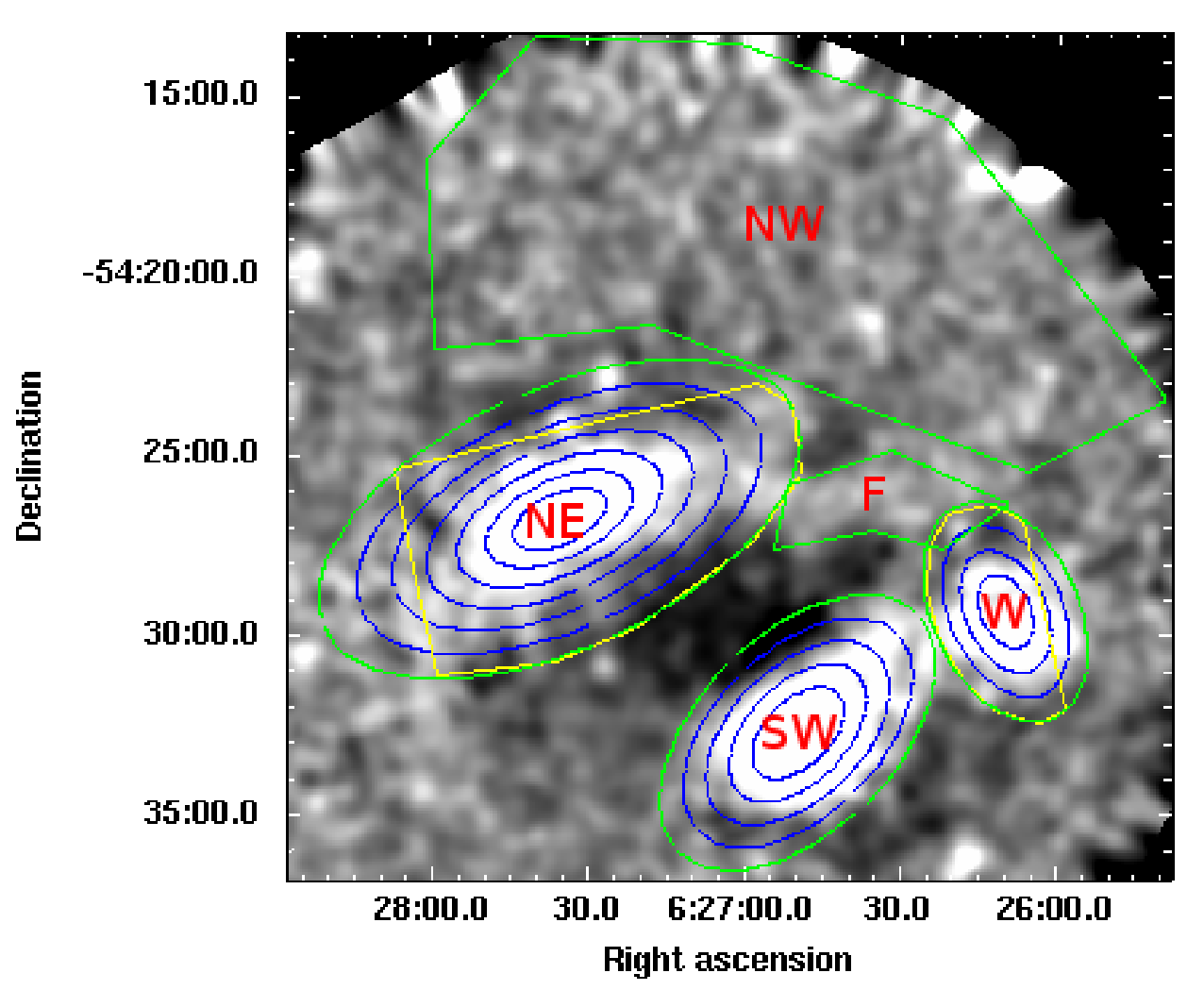}
\caption{The unsharp-masked image of A3395 from the combined MOS1 and MOS2 detectors (produced by subtracting a large scale (100$^{\prime\prime}$) 
smoothed image from a small scale (15$^{\prime\prime}$) smoothed image) showing the four principal regions of emission : NE, SW, W, the filament  
connecting NE to W region (marked with an 'F') and the NW region (reported by \citep{2001ApJ...563..673T}) used for extraction of 
spectra using XMM-Newton MOS1, MOS2 and PN detectors in section  \S\ref{sec:Total-Spec-Analy} with green. For extraction 
of spectra using Chandra data polygon shaped approximations (yellow) were made for the NE and W regions as Chandra did not cover these
 parts sufficiently, while regions used for the SW and filament regions were kept same. All the annuli regions, shown with green plus 
blue ellipses in all three subclusters were used in the azimuthally averaged spectral analysis using XMM-Newton and Chandra 
data in \S\ref{sec:projection_analysis} and \S\ref{sec:deprojection_analysis}. Only the innermost 4 annuli in the NE and innermost 2 
annuli in the W region were used for Chandra data.}
\label{fig:A3395_unsharp_mask_comb_MOS1_MOS2_and_tot_spec_regions}
\end{figure}

  Figure \ref{fig:A3395_unsharp_mask_comb_MOS1_MOS2_and_tot_spec_regions} shows the unsharp-masked image produced from the combined 
MOS1, MOS2 detector image, by subtracting a large scale (100$^{\prime\prime}$) smoothed image from a small scale (15$^{\prime\prime}$) smoothed image. Five 
distinct regions can be seen: the A3395 NE, A3395 SW, A3395 W, A3395 NW and a filament joining the NE region to the W region.
 The W region is a small and relatively faint subclump of diffuse emission visible towards the west of the SW region. Only two main 
X-ray peaks NE and SW, could be seen in the earlier X-ray images and are believed to be subclusters. 
Figs.~\ref{fig:Combined_MOS1_MOS2_smoothedimage} and \ref{fig:Chandra_smoothedimage} show a strong gradient in the surface-brightness 
in the southeast part of the SW component evident from the compression of X-ray contours in this part, indicating that the SW 
component is moving along the southeast direction. It appears that the small clump 
in the W region may also be a subcluster participating in the merger process. The Chandra image of the cluster (Fig. 
\ref{fig:Chandra_smoothedimage}) does not fully cover the NE and W regions. None of the three regions in A3395 cluster show a circular
 symmetry. Therefore, for the analyses that follow, we have made elliptical approximations for them. The NE, SW, and W regions have 
been approximated as ellipses with semi-major-axis lengths 432.0$^{\prime\prime}$, 280.8$^{\prime\prime}$, and 194.4$^{\prime\prime}$, ellipticities 0.87, 0.81, and 0.78, and 
major-axis position angles $115^{\circ}$, $135^{\circ}$, and $25^{\circ}$ respectively, measured from the North in an anti-clockwise 
direction (shown with the green ellipses in Fig.~\ref{fig:A3395_unsharp_mask_comb_MOS1_MOS2_and_tot_spec_regions}). A3395 NW 
can be seen as a region of weak X-ray emission towards the north and northwest of the cluster in 
Fig.\ref{fig:A3395_unsharp_mask_comb_MOS1_MOS2_and_tot_spec_regions}. This excess emission was first reported by Tittley \& Henriksen 
(2001). The optical image of A3395 NW shows a clustering of galaxies, however, its X-ray emission is very diffuse and does not seem to
 clump. The mean velocity of the group of galaxies associated with the NW, from the velocity data given in Teague, Carter, \& Gray 
(1990) is $14540\pm70$ km s$^{-1}$, while the same calculated from the velocity data given in Donnelly et al. (2001) is $15110\pm130$ 
km s$^{-1}$. The data by Donnelly et al. put this group at almost the redshift of the NE and SW subclusters. Considering that 
A3395 NW lies along the filament joining the clusters A3395 and A3391 (Tittley \& Henriksen 2001), it appears to be a part of the 
supercluster network.

\subsection{X-ray Surface Brightness Profiles}
\label{X-ray_sb_profile}

\begin{figure}
\centering
\subfigure[]
{
\includegraphics[height=1.65in]{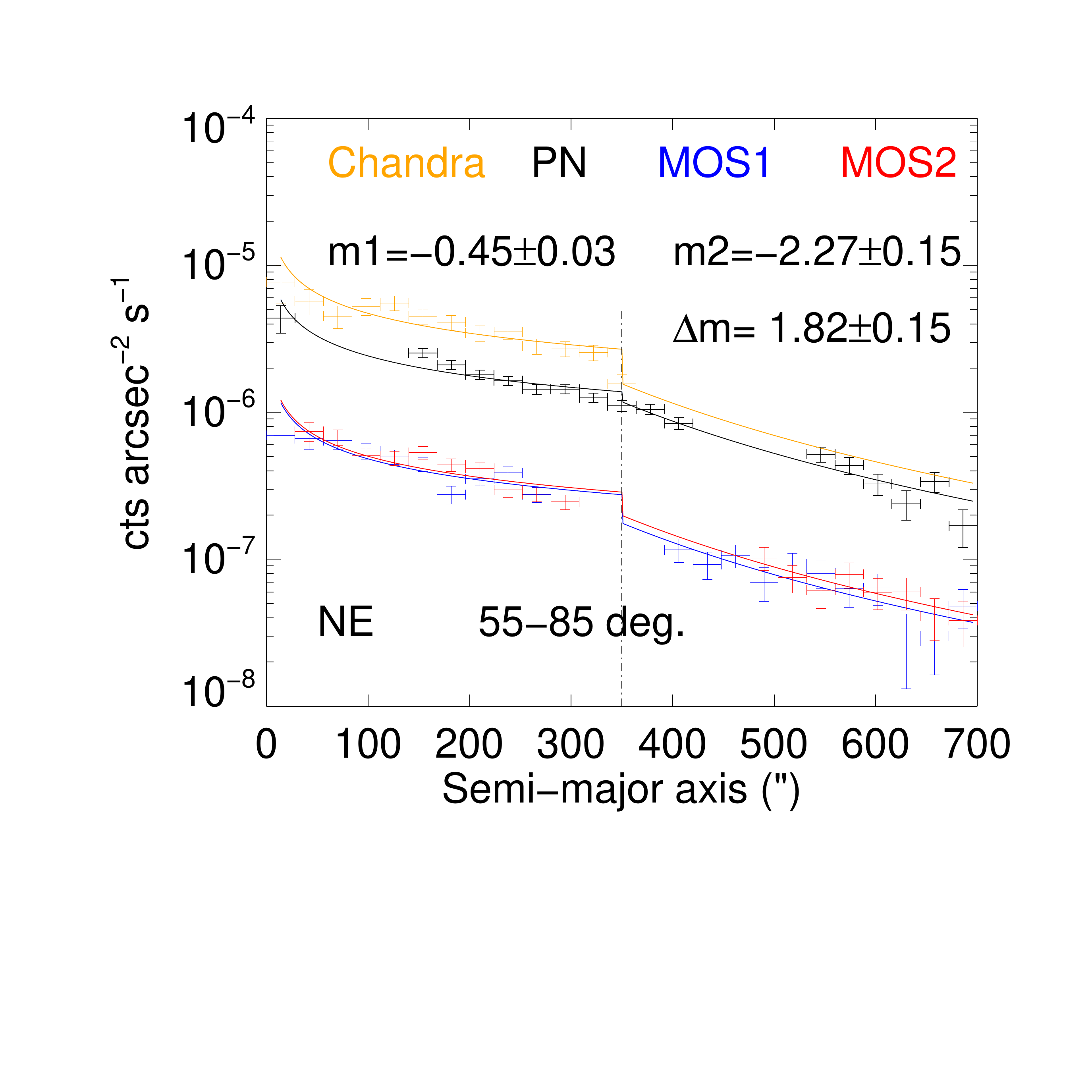}
\label{fig:NE_surface_brightness_profiles_sect5}
}
\subfigure[]
{
\includegraphics[height=1.65in]{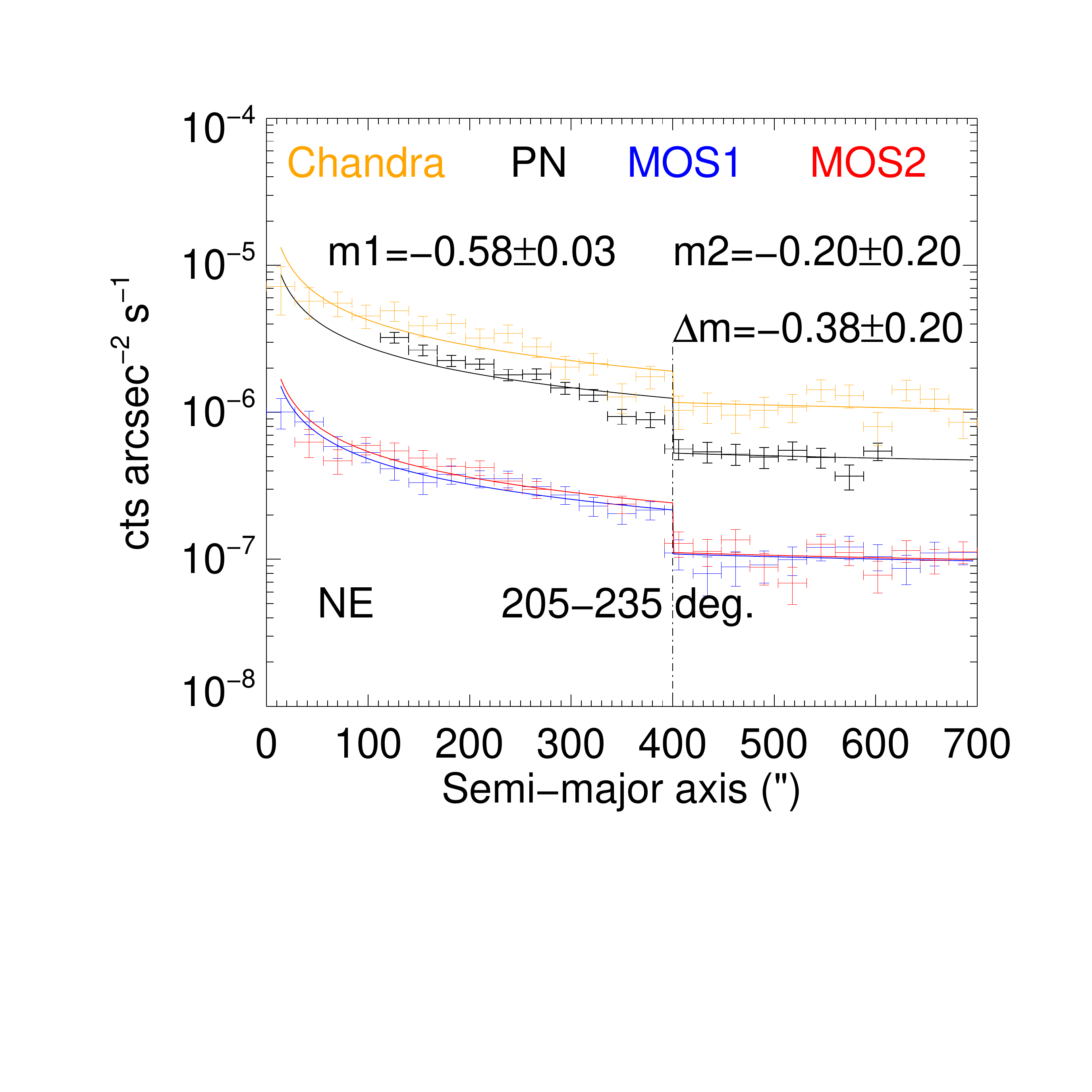}
\label{fig:NE_surface_brightness_profiles_sect10}
}
\subfigure[]
{
\includegraphics[height=1.65in]{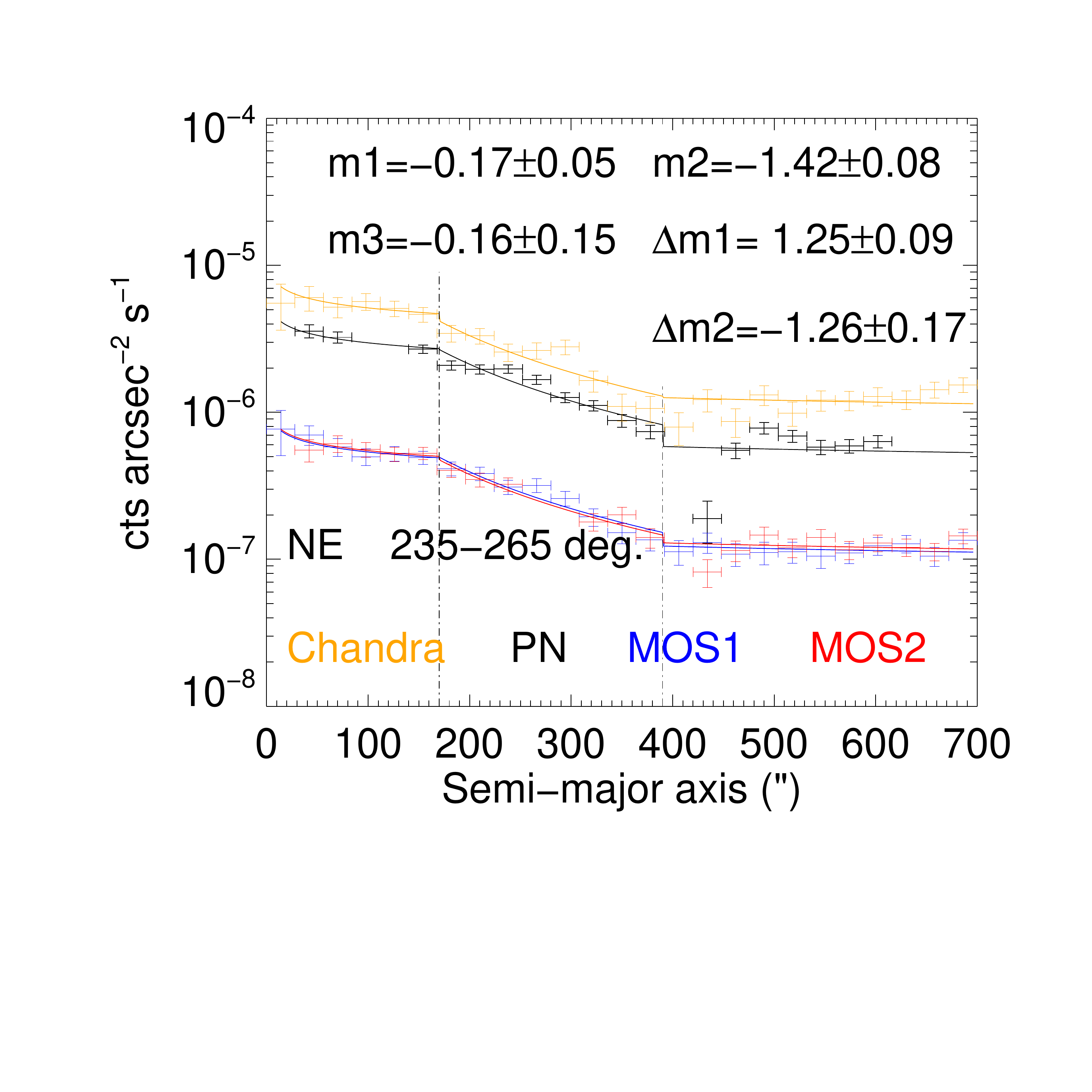}
\label{fig:NE_surface_brightness_profiles_sect11}
}
\subfigure[]
{
\includegraphics[height=1.65in]{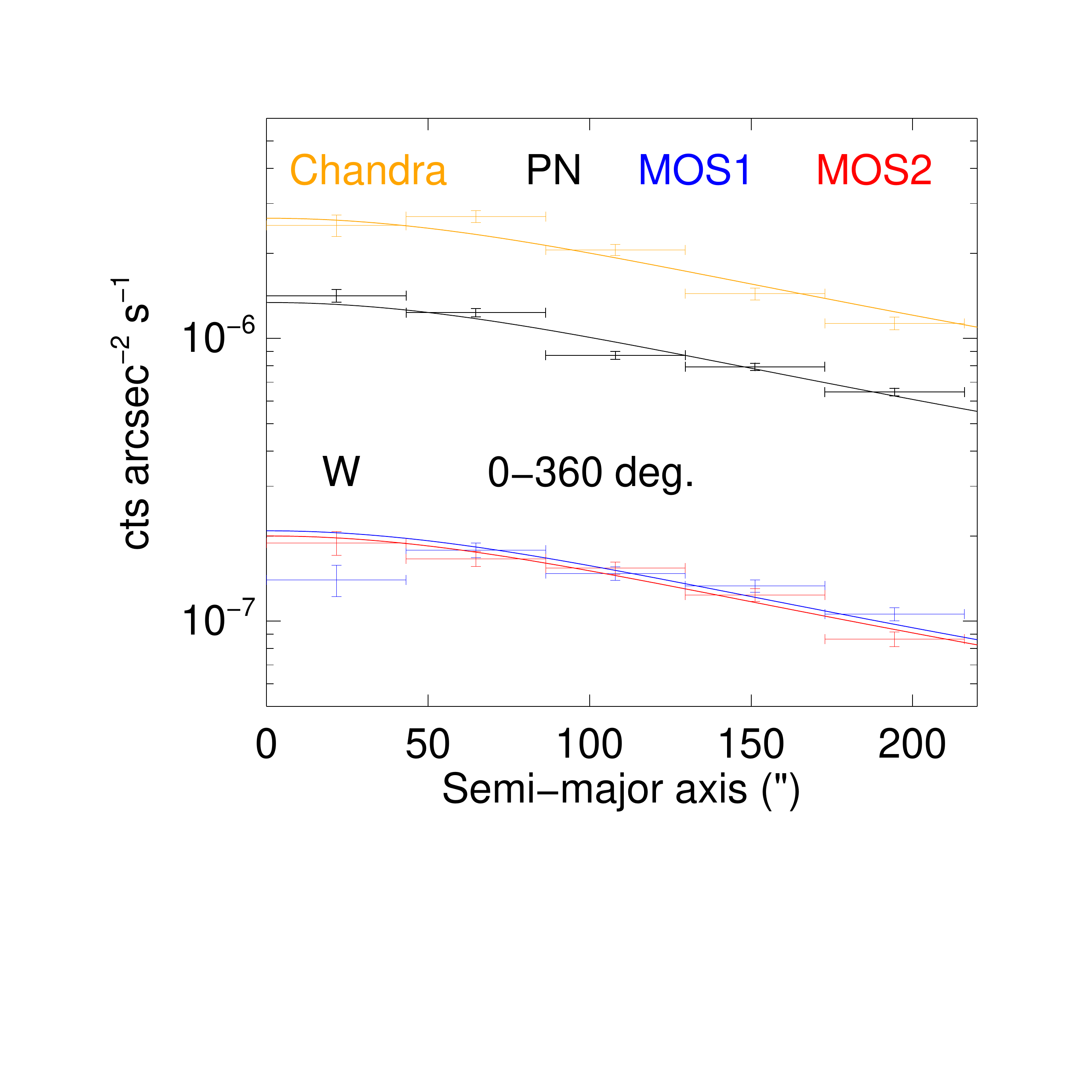}
\label{fig:W_surface_brightness_profiles_full}
}
\subfigure[]
{
\includegraphics[height=1.65in]{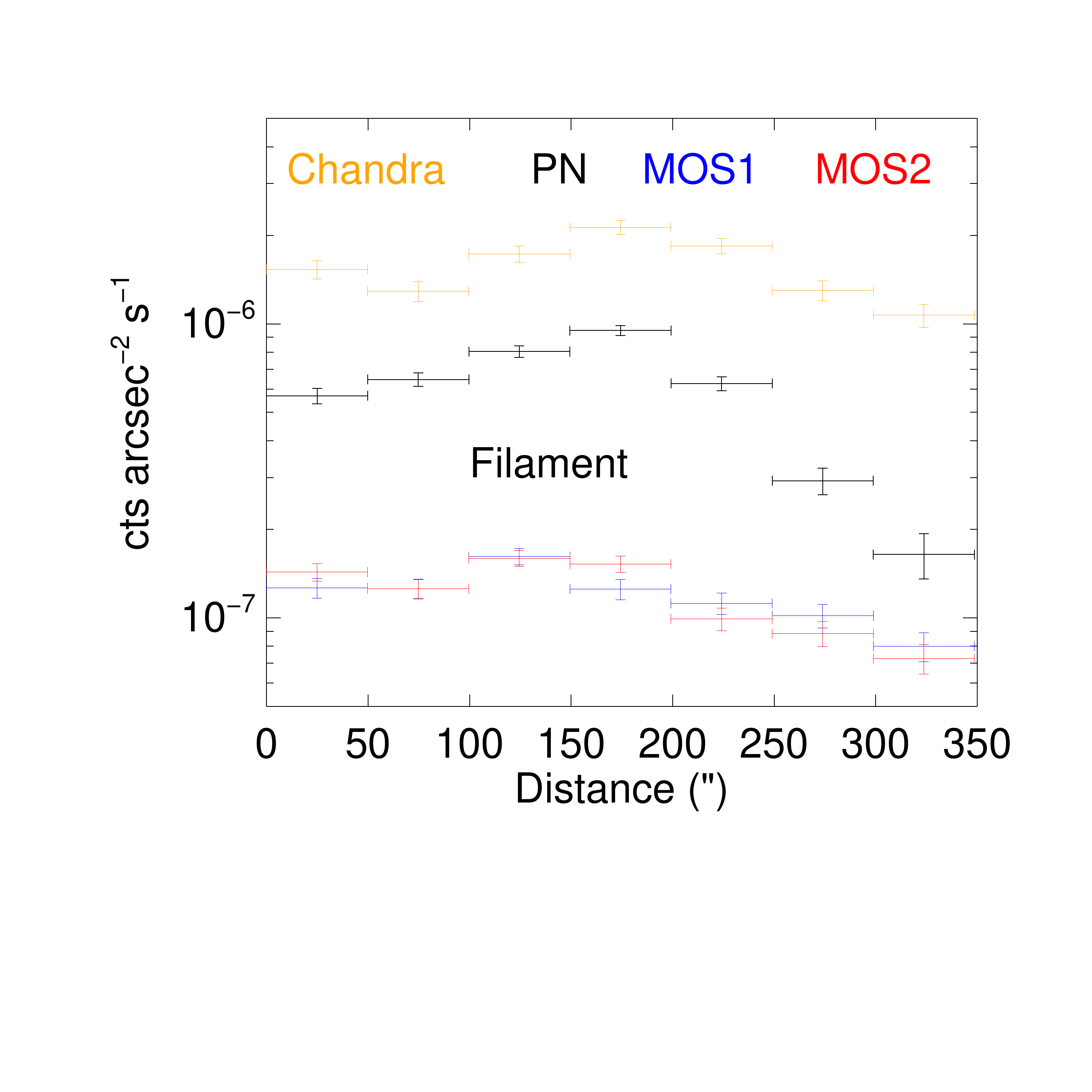}
\label{fig:bridge_surface_brightness_profiles_full}
}
\caption{Surface brightness profiles from three sectors in the NE subcluster (Figures~\ref{fig:NE_surface_brightness_profiles_sect5},
 \ref{fig:NE_surface_brightness_profiles_sect10} and \ref{fig:NE_surface_brightness_profiles_sect11}) made by using MOS1 (Blue), MOS2 
(Red), PN (Black) and Chandra ACIS (Orange) raw images after removing point sources, obtained from 25 elliptical annuli, each 
divided into 12 sectors. The profiles have been fitted with single or multiple powerlaws which are shown using solid lines and 
the positions of discontinuities are shown with dotted lines. The SW 
subcluster shows uniformly decreasing surface brightness profiles in all sectors (not shown here). The surface brightness profiles of 
the W region from the full $0^{\circ}-360^{\circ}$ range using 5 elliptical annuli and the filament from 8 
rectangular regions approximately parallel to the length of the filament have been shown in 
Figures~\ref{fig:W_surface_brightness_profiles_full} and \ref{fig:bridge_surface_brightness_profiles_full} respectively. Points 
affected by point source removal, individual CCD overlaps in a detector, and those lying within MOS1 CCD5 and MOS1 CCD6 have been 
removed. All angles are measured from the North and in the anti-clockwise direction.}
\label{fig:A3395_surface_brightness_profiles}
\end{figure}

  Merging clusters like A3395 can host both shock and cold fronts which can be confirmed 
by surface-brightness and temperature discontinuities. The discontinuities are seen by a change in the slope of the profile. In order 
to search for these features, surface-brightness profiles for the NE, SW, and W regions were made. We have used the raw MOS1, 
MOS2, PN detector images (without any smoothing applied but with the point sources removed by applying the cheese masks generated by 
the $\textquoteleft$cheese$\textquoteright$ task) and the point-source-removed raw Chandra ACIS detector image to derive the 
surface-brightness profiles. We used 25 annuli in the NE, 10 annuli in the
 SW, and 9 annuli in the W regions centred at the intensity peaks in each of the NE, SW, and W regions, which were further divided 
into 12 sectors of $30\,^{\circ}$ each. The ellipses are such that the $n^{\rm th}$ ellipse for the NE region has a semi-major axis 
of $n\times 28.0^{\prime\prime}$, the SW region has a semi-major axis of $n\times 28.1^{\prime\prime}$, and the W region has a semi-major axis of $n\times 
21.6^{\prime\prime}$. The major-axis position angles and ellipticities for the ellipses in NE, SW, and W regions were the same as given in section 
\S\ref{X-ray_morphology} and the $30^{\circ}$ sectors have been arranged symmetrically about the major-axes. To analyze the 
surface-brightness in the region between NE and SW subclusters, (which is not covered by the green ellipses in 
Fig.~\ref{fig:A3395_unsharp_mask_comb_MOS1_MOS2_and_tot_spec_regions}) and also to avoid overlaps between the three subclusters, the 
area spanned by the ellipses in the NE subcluster is larger, whereas that for the ellipses in the SW and W subclusters is almost same 
as compared to the green ellipses in Fig.~\ref{fig:A3395_unsharp_mask_comb_MOS1_MOS2_and_tot_spec_regions}. In the SW region, certain 
sectors had holes (due to point source removal) at the centre. Also, in the NE region, a few outermost annuli in the sectors 
with position angles from $55^{\circ}$ to $145^{\circ}$ had no MOS1 data (because of the missing CCD6 in MOS1), and with position 
angles from $295^{\circ}$ to $205^{\circ}$ had no Chandra data. In addition to this, the raw images had spurious discontinuities at 
the positions where the individual CCDs in a detector overlap. Points from all affected annular sectors were removed from the plots, 
but the resulting gaps made it more difficult to identify discontinuities. Surface-brightness profiles for all four detectors for 
each and every sector were obtained and were fitted by using single or multiple (for profiles showing more than one discontinuity) 
power laws. In order to make confident detections of discontinuities, we focused only on those discontinuities which are seen in all 
four detectors. All sectors in the NE and SW regions, in general, show a fairly uniform decreasing profile, except for a few sectors 
that show a change in the slope. The surface-brightness profiles of all sectors in the W region show a nearly constant profile. Only those surface-brightness 
profiles that show evidence for a significant change in slope are shown in Figure~\ref{fig:A3395_surface_brightness_profiles}.\\

  The $55^{\circ}$-$85^{\circ}$ sector (Figure~\ref{fig:NE_surface_brightness_profiles_sect5}) in the NE region shows a 
discontinuity at semi-major axis of $\sim350^{\prime\prime}$. The sectors $205^{\circ}$-$235^{\circ}$ 
(Figure~\ref{fig:NE_surface_brightness_profiles_sect10}), and $235^{\circ}$-$265^{\circ}$ 
(Figure~\ref{fig:NE_surface_brightness_profiles_sect11}), in the NE region, show a flattening of the slope in all four detectors at 
a semi-major axis of $\sim400^{\prime\prime}$. Starting from the centre of the NE component, this discontinuity is found at $\sim3/4$ 
times the distance between the centres of the NE and SW regions ($d_{NE-SW}\sim540^{\prime\prime}$). The observed flatness or increase
 in the surface-brightness in a region between the NE and SW subclusters is found to be due to the co-addition of the emission from 
the two subclusters. This result was confirmed by analyzing the sum of the $\beta$-models fitted to the surface brightness profiles of
 the $25^{\circ}-55^{\circ}$ sector in the NE (away from SW) and the $265^{\circ}-295^{\circ}$ sector in SW (away from NE), which 
showed a similar flatness in the same region between the NW and SW subclusters. A weak discontinuity in the sector 
$235^{\circ}$-$265^{\circ}$ is also seen at a semi-major axis of $\sim170^{\prime\prime}$. All the sectors in the SW subcluster show a
 uniformly decreasing profiles. The surface-brightness profiles in all the sectors of A3395 W show a nearly constant profile. We also 
obtained the total surface brightness profiles of A3395 W over the full $0^{\circ}-360^{\circ}$ range by using 5 annuli centred at the
 emission peak of A3395 W and the filament by using 8 rectangular regions parallel to the length of the filament and placed 
almost symmetrically about it. The surface brightness profile of the W subcluster shows a decreasing profile from the innermost to 
the outermost annulus with a peak at the centre (Figure~\ref{fig:W_surface_brightness_profiles_full}) while that of the filament shows
 a significant hump above the background at the centre (Figure~\ref{fig:bridge_surface_brightness_profiles_full}).

\subsection{Global X-ray Spectra}
\label{sec:Total-Spec-Analy}
  Average spectra of the various components of A3395 were extracted from both the XMM-Newton and Chandra data. For XMM-Newton,
 these spectra were extracted from elliptical regions for the NE, SW and W components, and from polygon shaped regions for the NW 
component and the filament. The ellipses were centred on the peaks of the surface-brightness. The ellipse for the NE region had a 
semi-major axis of 432$^{\prime\prime}$ and an ellipticity of 0.87, the SW region had a semi-major axis of 281$^{\prime\prime}$ and an ellipticity of 0.81, and the 
W region had a semi-major axis of 194$^{\prime\prime}$ and an ellipticity of 0.78.  For the spectral extraction using Chandra data, the regions used 
for the SW and the filament were the same as those used for XMM-Newton, while for the NE and W, maximum possible areas were picked by 
using a polygon shaped region, since Chandra field does not cover these regions fully. As the NW region was not covered by Chandra, 
its spectral analysis was done using only XMM-Newton detectors. All these regions are shown in 
Figure~\ref{fig:A3395_unsharp_mask_comb_MOS1_MOS2_and_tot_spec_regions}. All spectra were extracted in the energy band 0.5-9.0 keV. 
For MOS data, events with $\rm PATTERN\leq12$ were used, whereas for PN data, events with $\rm PATTERN\leq4$ were selected. For 
XMM-Newton detectors, the response matrices and effective areas were generated using the tasks \textbf{rmfgen} and \textbf{arfgen}. 
The neutral hydrogen column density along the line of sight to the cluster (i.e., along $\alpha = 06^{\rm h} 27^{\rm m} 31.1^{\rm s}$,
 $\delta = -54^{\rm d} 23^{\prime} 58^{\prime\prime}$) was taken to be $6.3\times 10^{20}$ cm$^{-2}$ based on Leiden/Argentine/Bonn 
(LAB) Galactic HI survey (Kalberla et al. 2005) and redshift was frozen to the value of 0.0498 (SIMBAD astronomical database).

\begin{figure}
\centering
\subfigure[]
{
\includegraphics[width=2.0in]{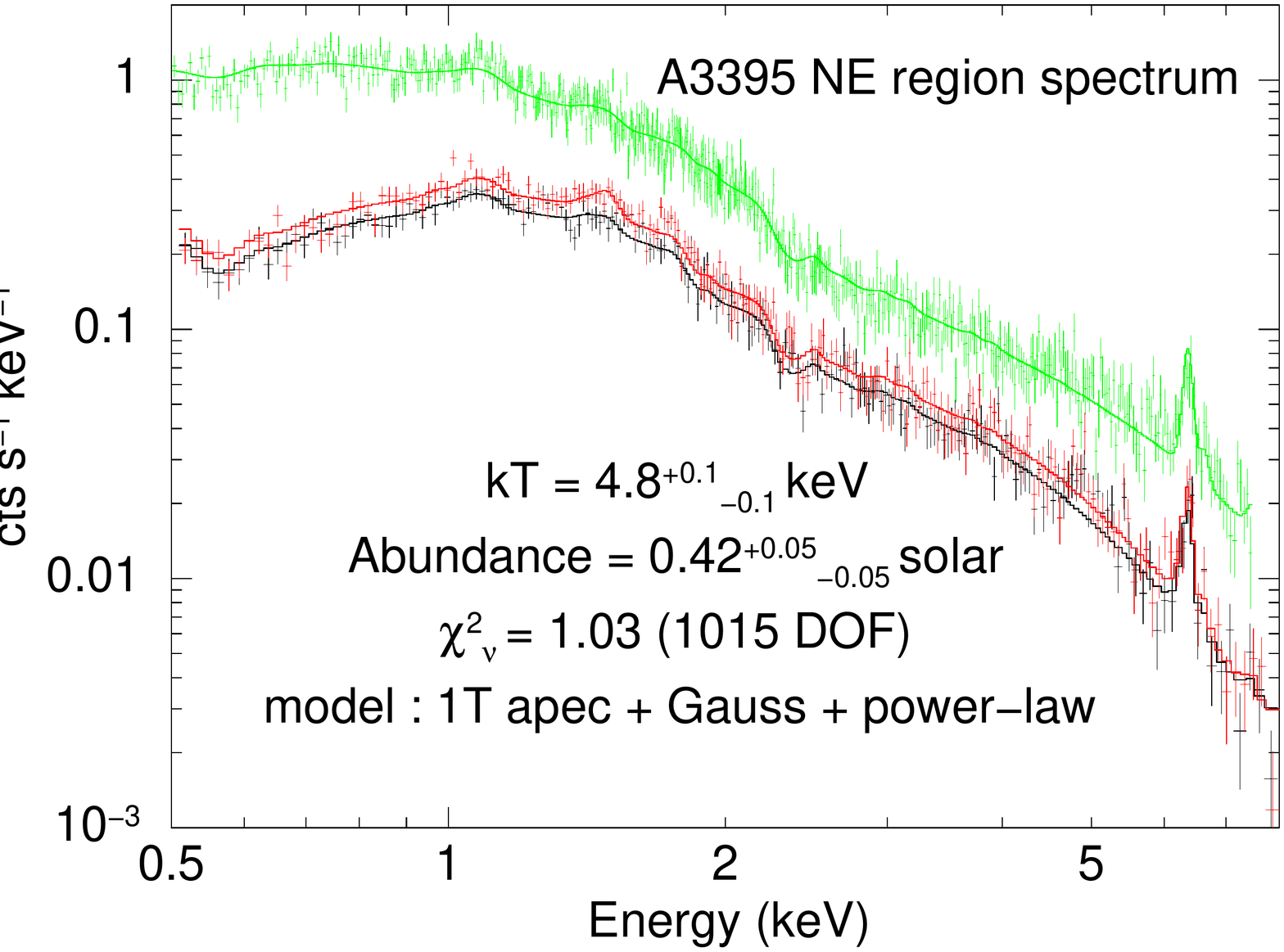}
\label{fig:NE_XMM_Newton_spectra}
}
\subfigure[]
{
\includegraphics[width=2.0in]{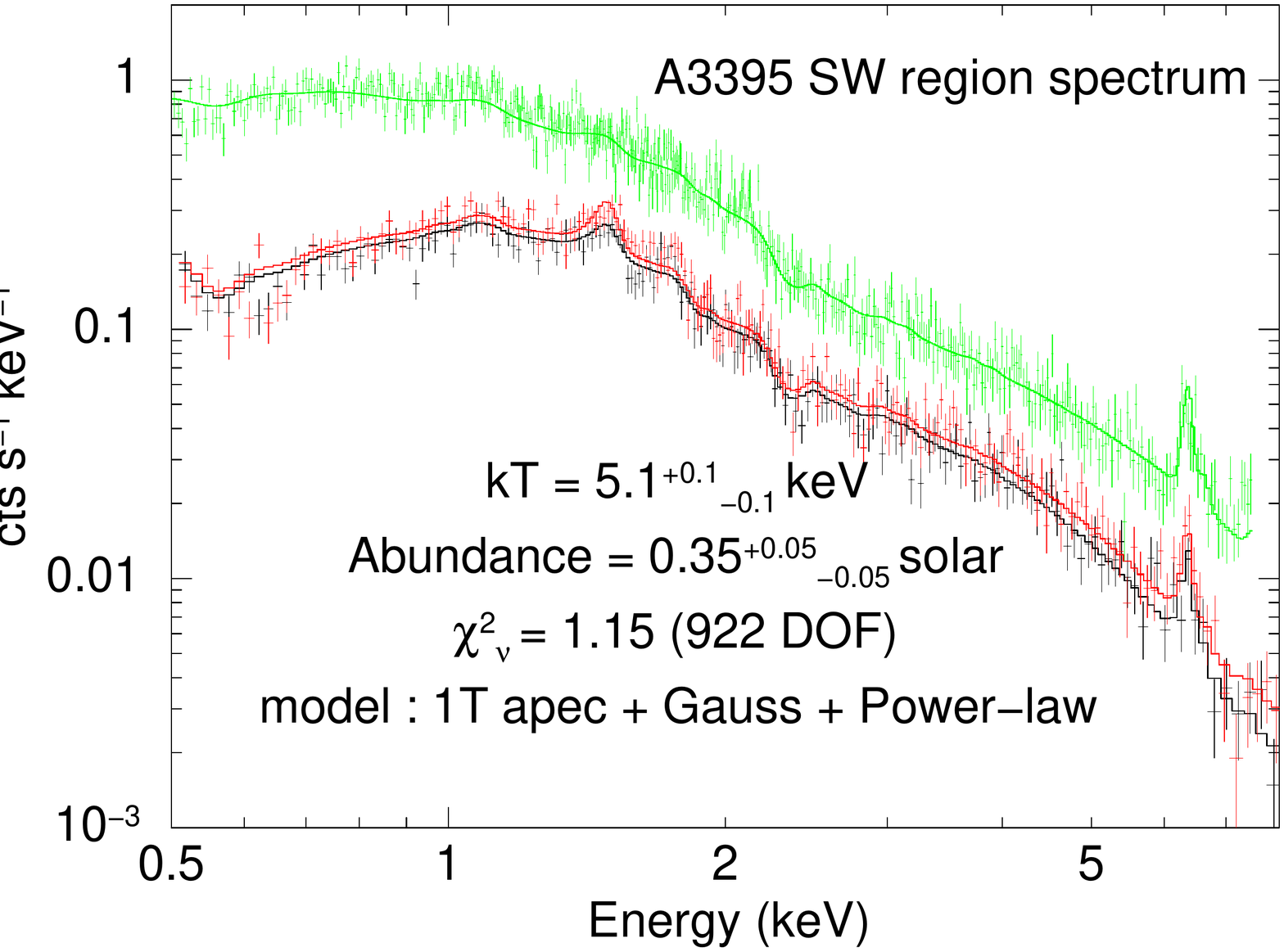}
\label{fig:SW_XMM_Newton_spectra}
}
\subfigure[]
{
\includegraphics[width=2.0in]{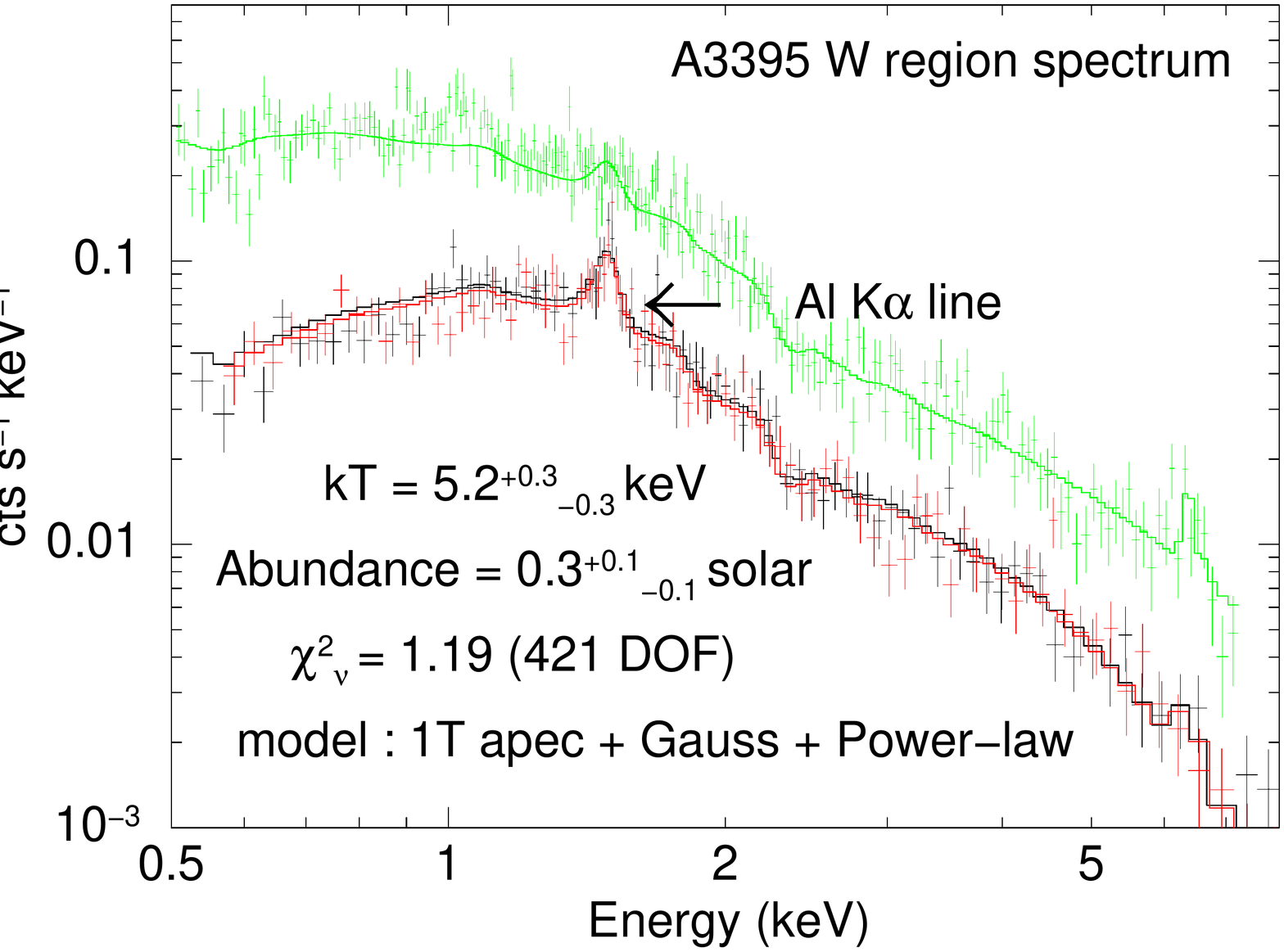}
\label{fig:W_XMM_Newton_spectra}
}
\subfigure[]
{
\includegraphics[width=2.0in]{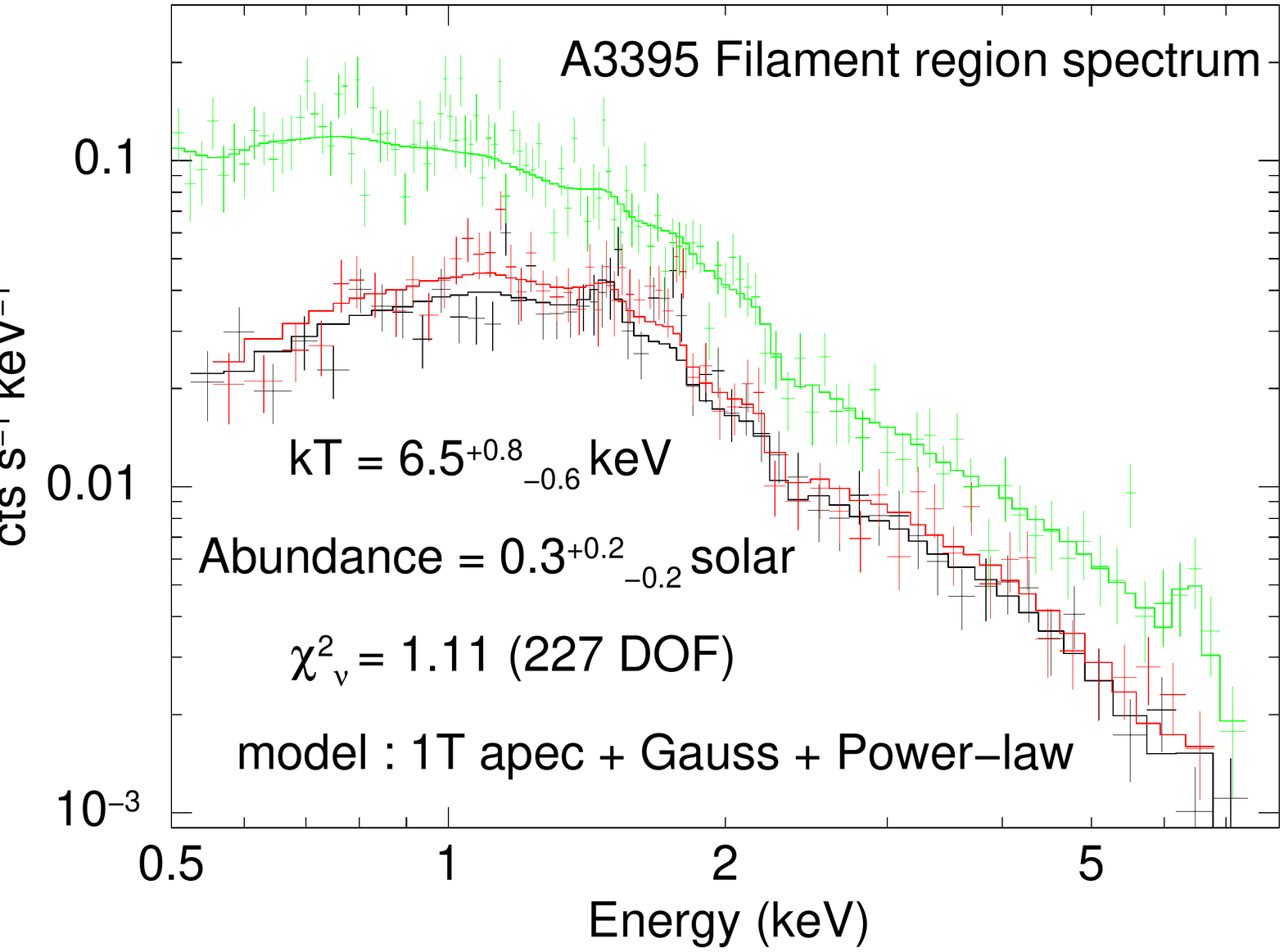}
\label{fig:Filament_XMM_Newton_spectra}
}
\subfigure[]
{
\includegraphics[width=2.0in]{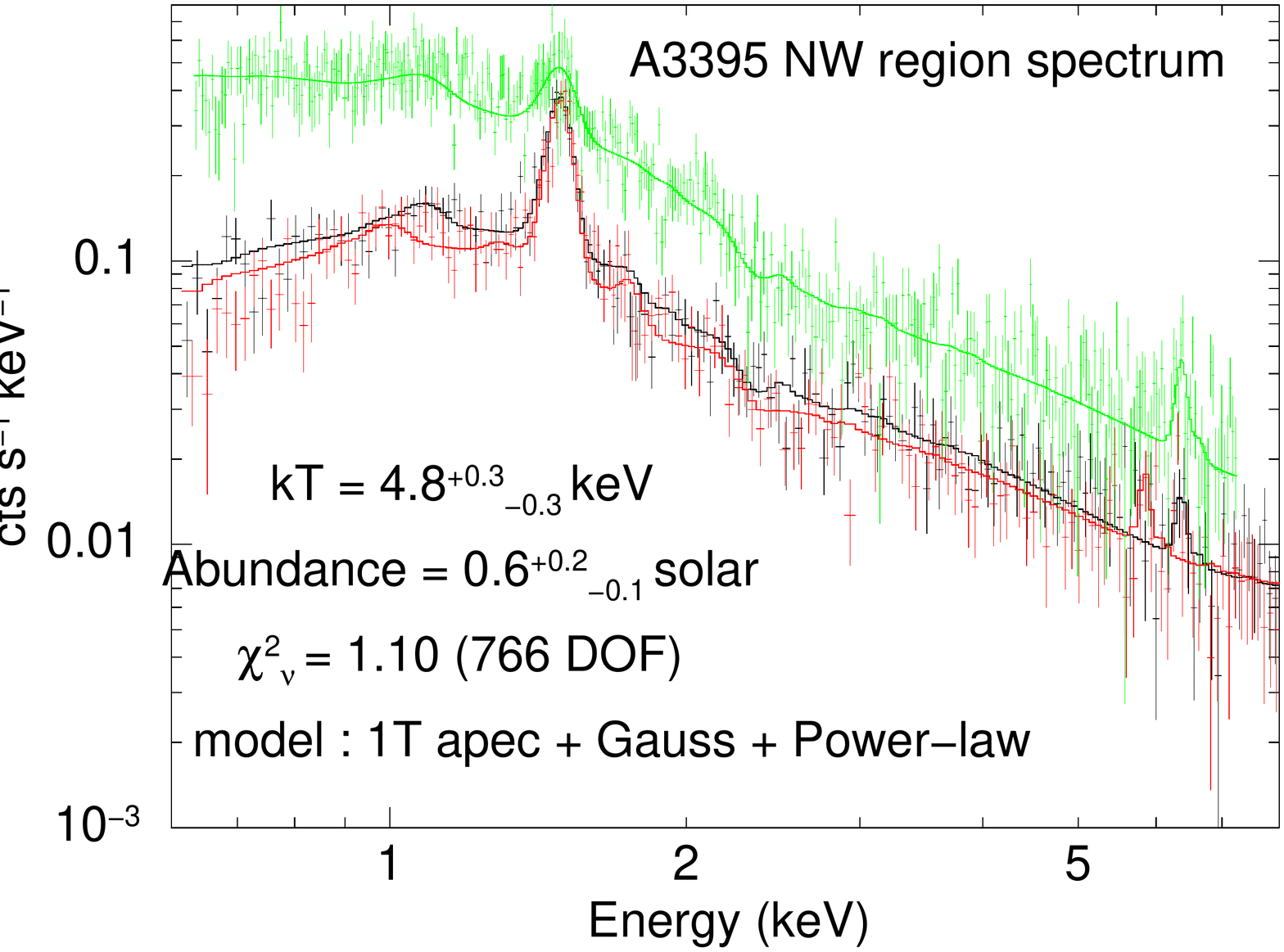}
\label{fig:NW_XMM_Newton_spectra}
}
\subfigure[]
{
\includegraphics[width=2.0in]{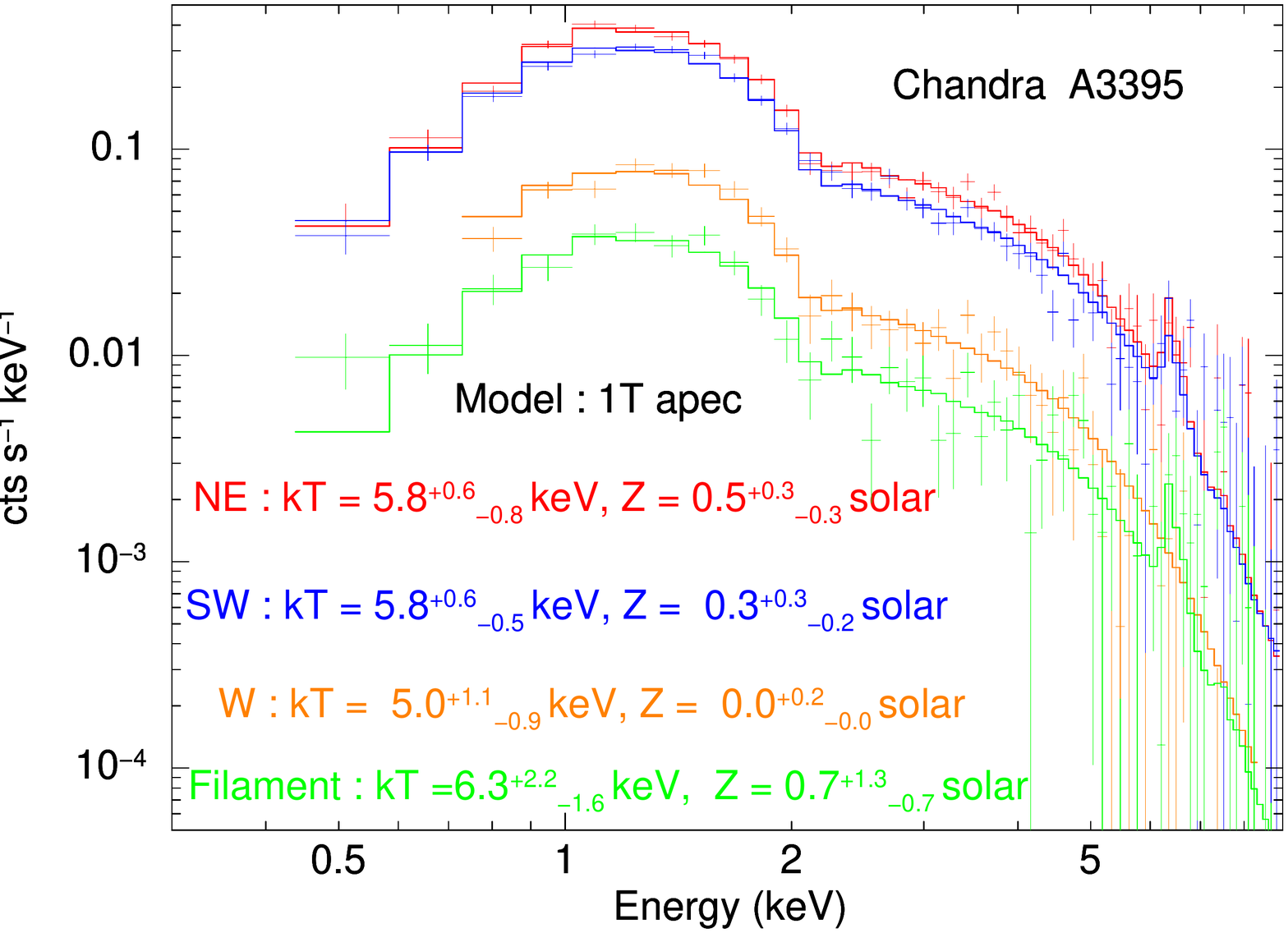}
\label{fig:NE_SW_W_filament_Chandra_Newton_spectra}
}
\caption{Figure~\ref{fig:NE_XMM_Newton_spectra} to \ref{fig:NW_XMM_Newton_spectra}:  Average spectra of the A3395 NE, SW, W 
subclusters, the filament connecting NE to W, and the NW region (reported by \citep{2001ApJ...563..673T}), produced by using the 
XMM-Newton PN (Green), MOS1 (Black), and MOS2 (Red). Figure \ref{fig:NE_SW_W_filament_Chandra_Newton_spectra} : Average spectra of NE,
 SW, W subclusters and the filament produced using Chandra data. All the spectra have been fitted with \textbf{wabs*apec} model shown 
as a histogram. The powerlaw and Gaussian components in the XMM-Newton spectra have been added to model the residual soft proton 
contamination and the instrumental Al line at 1.49 keV. Details of spectral analysis are given in \S\ref{sec:Total-Spec-Analy}, and 
the best fit parameters are shown here as insets.}
\label{fig:XMM_Chandra_NE_SW_W_bridge_average_spectra}
\end{figure}

\begin{figure}
\begin{center}
\subfigure[]
{
\includegraphics[width=3.0in]{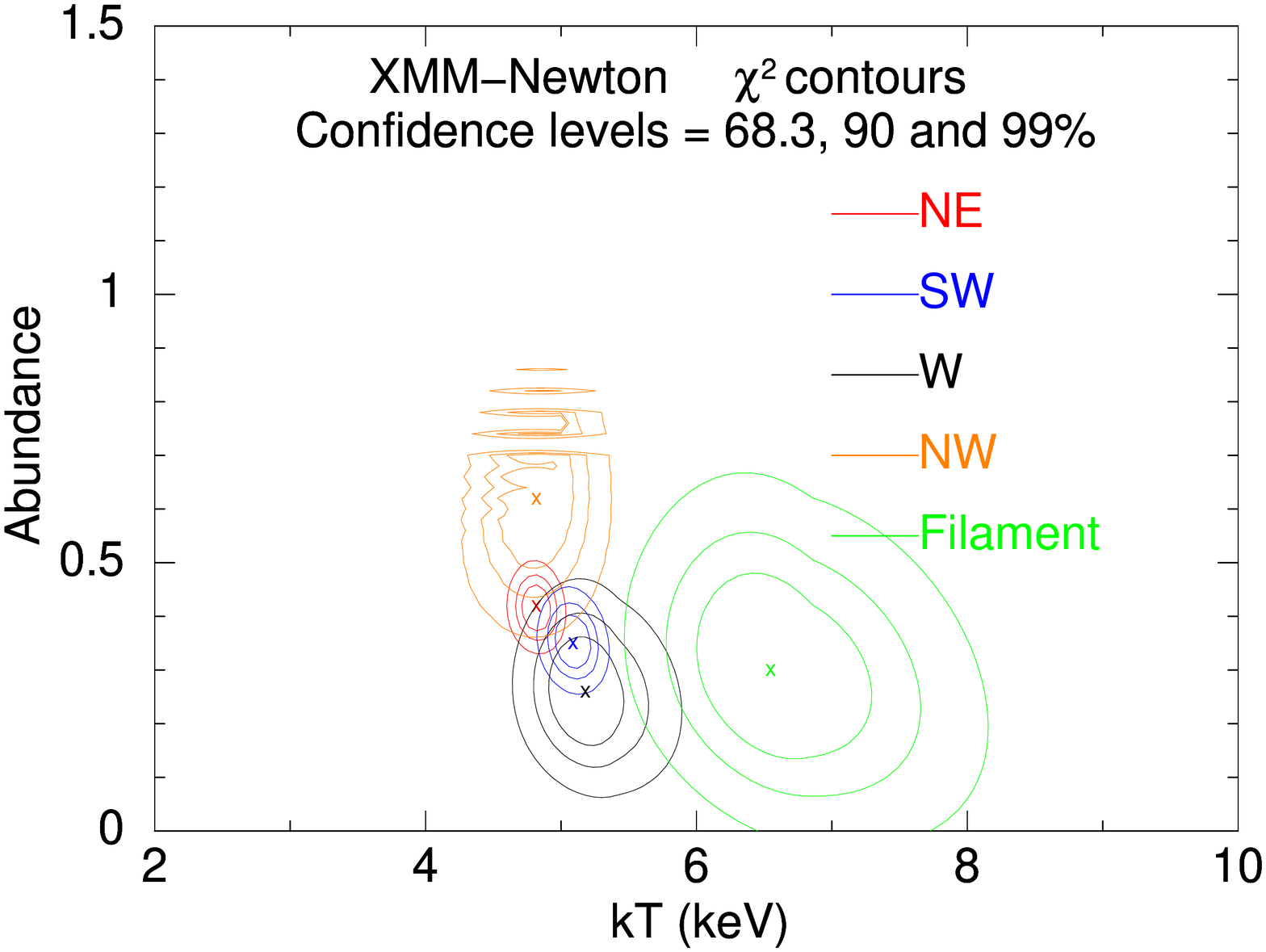}
\label{fig:XMM_NE_SW_S_W_chisq_cont}
}
\subfigure[]
{
\includegraphics[width=3.0in]{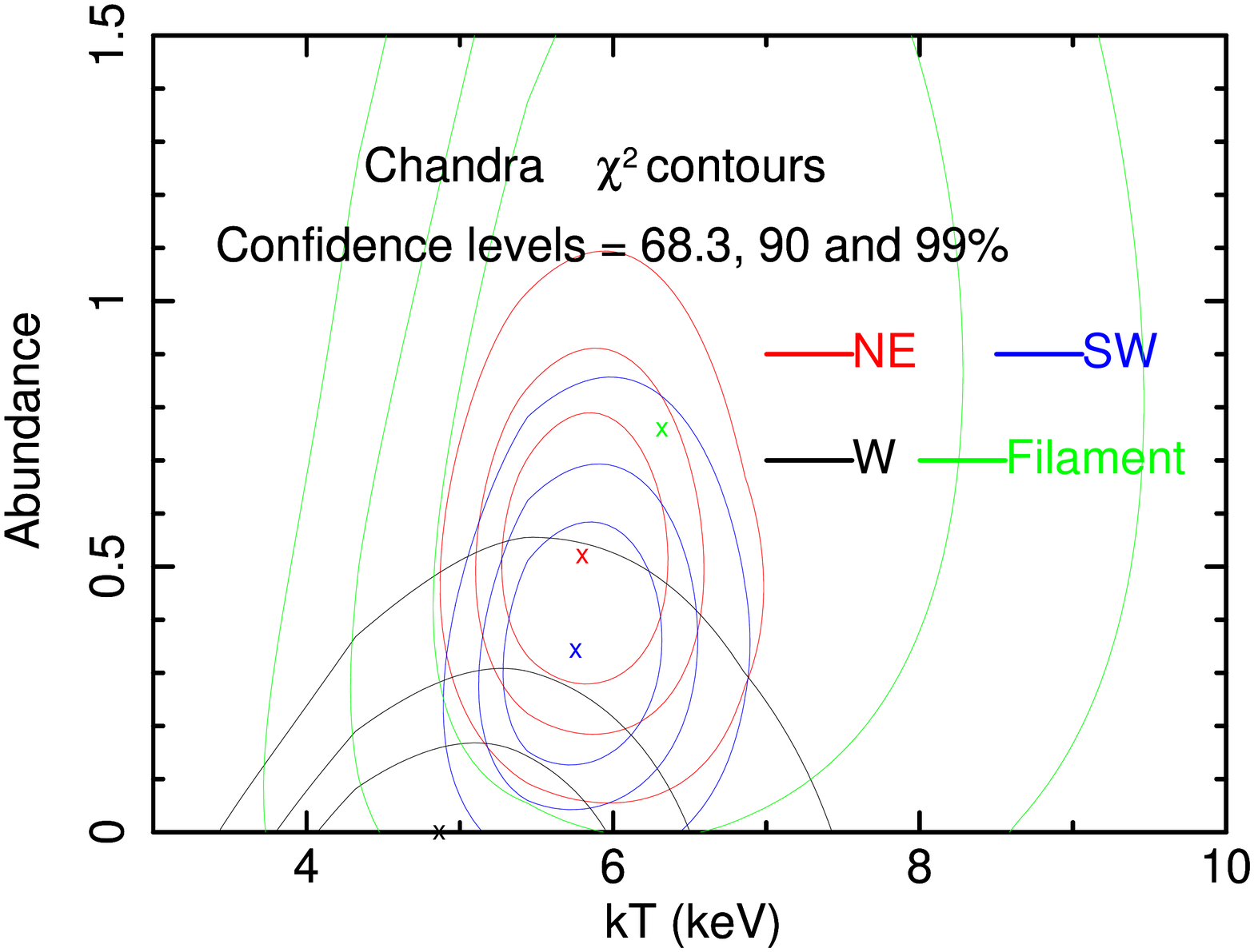}
\label{fig:Chandra_NE_SW_S_W_chisq_cont}
}
\end{center}
\caption{\ref{fig:XMM_NE_SW_S_W_chisq_cont} : $\chi^{2}$ contours of the temperature and abundance measurements for the NE (red), SW 
(blue), W (black) and NW (orange) regions of A3395 and the filament connecting NE to W (green) from the spectral analysis done using 
XMM-Newton data. \ref{fig:Chandra_NE_SW_S_W_chisq_cont} : $\chi^{2}$ contours of the temperature and abundance measurements for the 
NE (red), SW (blue) and W (black) regions and the filament (green) from the spectral analysis done using Chandra data. Details of the 
spectral analyses are given in \S\ref{sec:Total-Spec-Analy}. The confidence levels for the innermost, middle, and outermost contours 
for each of the 9 sets of contours are at 68.3\%, 90\% and 99\% respectively.}
\label{fig:XMM_Chandra_NE_SW_S_W_chisq_cont}
\end{figure}

  Spectral analyses have been performed using the X-ray spectral fitting package \textbf{Xspec} (version 12.5.1). Spectra extracted 
from all detectors were fitted using the \textbf{wabs} photoelectric absorption model (Morrison \& McCammon 1983) and \textbf{apec} 
plasma emission model (Smith et al. 2001). The relative elemental abundances used in \textbf{wabs} are as given by Anders \& Ebihara 
(1982). The MOS1, MOS2 and PN spectra for each of the regions were fitted 
simultaneously using three separate \textbf{wabs*apec} models. The values of abundance, temperature and \textbf{apec} normalizations 
for the models were linked together but were not frozen. The MOS1, MOS2 and PN spectra of the weak emission regions viz. the NW,
 the W and the filament showed presence of residual soft proton contamination (shown by prominent residues in the high energy end of 
the spectra) and the residual instrumental Al K-$\alpha$ line at 1.49 keV. The residual soft proton contamination was modeled by using
 separate powerlaw models with diagonal RMF files (Snowden et al. 2008) and the instrumental Al K-$\alpha$ line at 1.49 
keV was modeled by adding Gaussian components separately for MOS1, MOS2 and PN (Snowden et al. 2008). The powerlaw 
indices were found to be negative for most cases and hence were frozen to the value of 0.3 (i.e. the minimum recommended in Snowden 
et al. 2008), while the powerlaw normalizations were left as free parameters. The centre position and the widths of the Gaussian 
components were frozen to 1.49 keV and 0.02 keV respectively in all the spectra and their normalizations were left as free 
parameters. The resulting spectra, along with the histograms of best fit model spectra are shown in 
Figure~\ref{fig:XMM_Chandra_NE_SW_W_bridge_average_spectra} and the best fit values of the temperature, abundance and \textbf{apec} 
normalizations from the XMM-Newton and Chandra spectral analyses are given in 
Tables~\ref{tab:XMM_A3395_total_regions_spectral_results} and ~\ref{tab:Chandra_A3395_total_regions_spectral_results} respectively. 
The confidence contours at the 68.3\%, 90\% and 99\% confidence levels for the free parameters are shown in 
Figure~\ref{fig:XMM_Chandra_NE_SW_S_W_chisq_cont}.

  From the confidence contours produced from the spectral analysis using XMM-Newton, it can be seen that the values of the 
temperatures and abundances for the NE and the SW regions; the NE and W regions and the NW and W regions are different only at a 
confidence level of 68.3\%. However, the temperature and abundance values for the SW and W regions hardly differ. The temperature and 
abundance values for the filament appear to be distinct from the values for all other components except for the W region, which are 
distinct only at a confidence level of 68.3\%. The best fit parameters from Chandra have errors much larger than those from 
XMM-Newton because of fewer counts and hence, all four sets of confidence contours have large overlaps. While the results for all 
other regions are in agreement with those from XMM-Newton (within errors), the results from Chandra for the NE region show a 
temperature slightly higher than that from XMM-Newton. This is possibly because the part of NE which is not covered by Chandra has a 
lower than average temperature, as is indicated by the temperature map produced from XMM-Newton data in 
\S\ref{sec:box_thermodynamic_maps}. This was also confirmed by using the same polygon shaped region for the extraction of spectrum 
from the NE with MOS1, MOS2 and PN, as was used for Chandra spectral analysis, and the results were in mutual agreement. The 
background systematic errors (assuming 10\%) do not affect the results significantly.\\

\subsection{Azimuthally Averaged spectrally determined radial profiles of thermodynamic quantities}
\label{sec:Azimuth_spec_analys}
  We also produced azimuthally averaged profiles of temperature, density, entropy, and pressure for the cluster by extracting spectra 
in elliptical annuli using both XMM-Newton and Chandra data. The NE, SW, and W subclusters were divided into 7, 5 and 4 elliptical 
annuli respectively (shown in Figure~\ref{fig:A3395_unsharp_mask_comb_MOS1_MOS2_and_tot_spec_regions}). The centres of the ellipses 
were at the peak of the X-ray emission for all three  regions. Ellipses in the NE regions had semi-major axis lengths of 84$^{\prime\prime}$, 
132$^{\prime\prime}$, 184$^{\prime\prime}$, 240$^{\prime\prime}$, 300$^{\prime\prime}$, 364$^{\prime\prime}$, and 432$^{\prime\prime}$. Similarly, ellipses in the SW region had semi-major axis 
lengths of 93.6$^{\prime\prime}$, 140.4$^{\prime\prime}$, 187.2$^{\prime\prime}$, 234$^{\prime\prime}$, and 280.8$^{\prime\prime}$ and W region had semi-major axis lengths of 
64.8$^{\prime\prime}$, 108$^{\prime\prime}$, 151.2$^{\prime\prime}$, and 194.4$^{\prime\prime}$. Spectra were extracted 
from all the annuli for all three detectors of XMM-Newton. However, for Chandra data, only the innermost 4 annuli in the NE and 
innermost two annuli in the W region were used as Chandra did not cover the NE and W parts completely, while for the SW subcluster, 
spectra were extracted from all 5 annuli. All the spectra were extracted in the energy band of 0.5-9.0 keV. Both the projected and
 deprojected profiles of temperature, density, entropy, and pressure were obtained.
 
\subsubsection{2-D Projected Profiles}
\label{sec:projection_analysis} 
  The details of spectral analyses performed are the same as given in \S\ref{sec:Total-Spec-Analy} except that the elemental 
abundances were fixed to the respective average abundance value obtained from the spectral fitting of the NE, SW, and W regions using 
XMM-Newton data i.e., 0.42, 0.35, and 0.3 times the solar value ($Z_{\odot}$) 
(Table~\ref{tab:XMM_A3395_total_regions_spectral_results}.) respectively, for the annuli belonging to the respective region. The 
temperature profile was obtained directly from the spectral analysis and was used to derive the density, entropy, and pressure 
profiles. To derive the electron density n$_{\rm e}$, we used the \textbf{apec} normalization, 
K$=$10$^{-14}$EI/(4$\pi$[D$_{\rm A}$(1+z)]$^{2}$) (expressed in units of cm$^{-5}$ in cgs system), where EI is the emission integral 
$\int \rm n_{\rm e} \rm n_{\rm p}$dV. We assume $\rm n_{\rm p}= 0.855 \rm n_{\rm e}$ (Henry et al. 2004), which under the assumption 
of constant density within each ellipsoidal shell gives, EI$=0.855\rm n_{\rm e}^{2}$V, where V$=$volume of ellipsoidal shell. 
For the volume estimation of the ellipsoidal shells we have assumed oblate ellipsoids (i.e., line of sight axis = major-axis of the 
ellipse). Then the volume of a shell, V$=(4 \pi /3)(\rm a_{\rm out}^{2} \rm b_{\rm out} - \rm a_{\rm in}^{2} \rm b_{\rm in})$, where 
$\rm a_{\rm out}$ and $\rm b_{\rm out}$ are the semi-major and semi-minor axis of the outer shell and $\rm a_{\rm in}$ and 
$\rm b_{\rm in}$ are those of the inner shell. From the commonly adopted definition, entropy is given by S$=\rm kTn_{\rm e}^{-2/3}$ 
and electron pressure P$=\rm n_{\rm e}\rm kT$ (Gitti et al. 2010). If the ions have the same temperature as that of the electrons then
 the total pressure is twice as large. We admit that the volume estimates may introduce errors $\sim20\%$ (Finoguenov et al. 2004, and
 references therein; Henry et al. 2004), but even these errors are quite small as compared to the errors in temperatures obtained from
 the deprojection analysis because of poor statistics of the data. For confirmation we also estimated the entropy and pressure for 
prolate ellipsoidal shells (i.e., line of sight axis = minor-axis of the ellipse) but the differences were very small ($\sim20\%$).

\subsubsection{Deprojected Profiles}
\label{sec:deprojection_analysis}
  Projection effects along the line of sight can smooth out the variations in the measured quantities. To correct for this we
have performed the deprojection analysis on the same annular regions shown in 
Fig.~\ref{fig:A3395_unsharp_mask_comb_MOS1_MOS2_and_tot_spec_regions} by using the \textbf{Xspec projct} model, which estimates the 
parameters in 3-D space from the 2-D projected spectra of ellipsoidal shells, using the \textbf{wabs*apec} model. The model calculates
 the geometrical weighting factor according to which the emission is redistributed amongst the projected annuli. The elemental 
abundances for the annuli belonging to the NE, SW, and W regions were fixed to 0.42, 0.35, and 0.3 times the solar value 
($Z_{\odot}$), same as in \S\ref{sec:projection_analysis}. The residual soft proton contamination and instrumental Al lines were 
modeled by adding powerlaws and Gaussian components to the models as described in \S\ref{sec:Total-Spec-Analy} with the only 
difference that a single model for all MOS1, MOS2 and PN was used. This was because the \textbf{projct} model requires all the spectra
 belonging to the same annulus to be part of the same group and therefore have to be modeled using the same model. Electron density 
n$_{\rm e}$, entropy (S), and electron pressure (P) have been calculated using the same relations as given in 
\S\ref{sec:projection_analysis}.

\begin{figure}
\centering
\subfigure[]
{
$\begin{array}{ccc}
\hspace{0.6cm}
\includegraphics[width=1.5in]{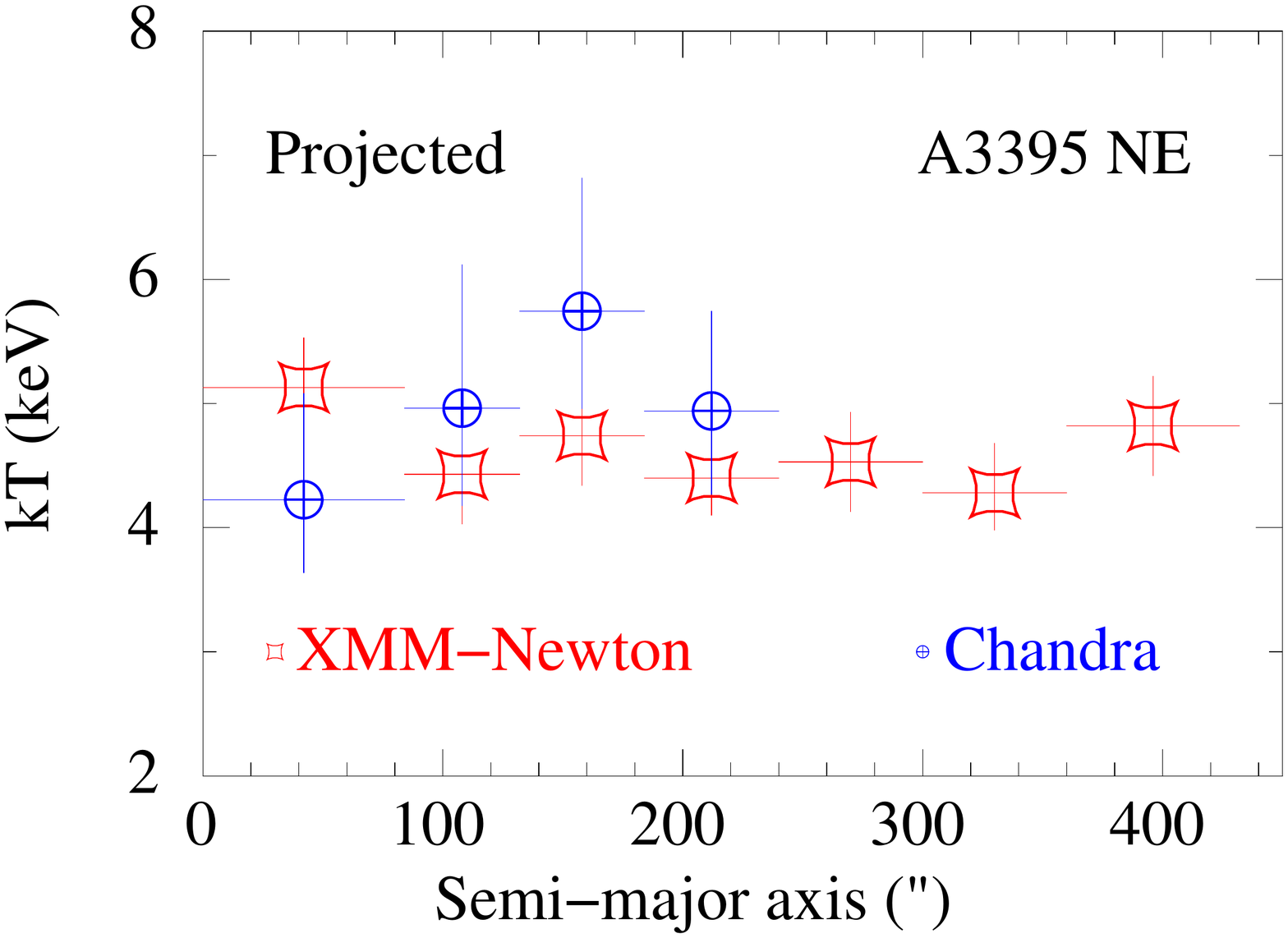} &
\includegraphics[width=1.5in]{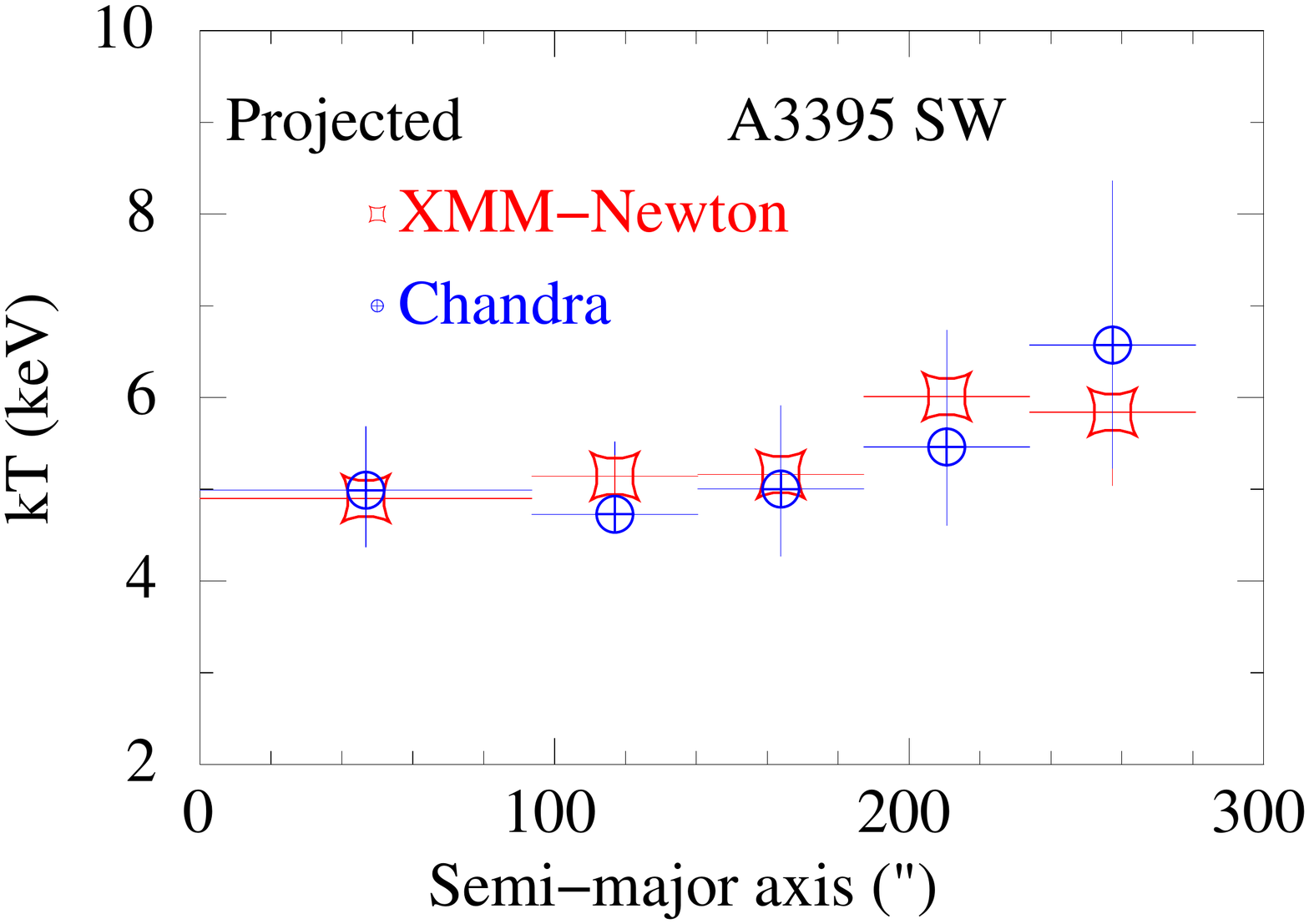} &
\includegraphics[width=1.5in]{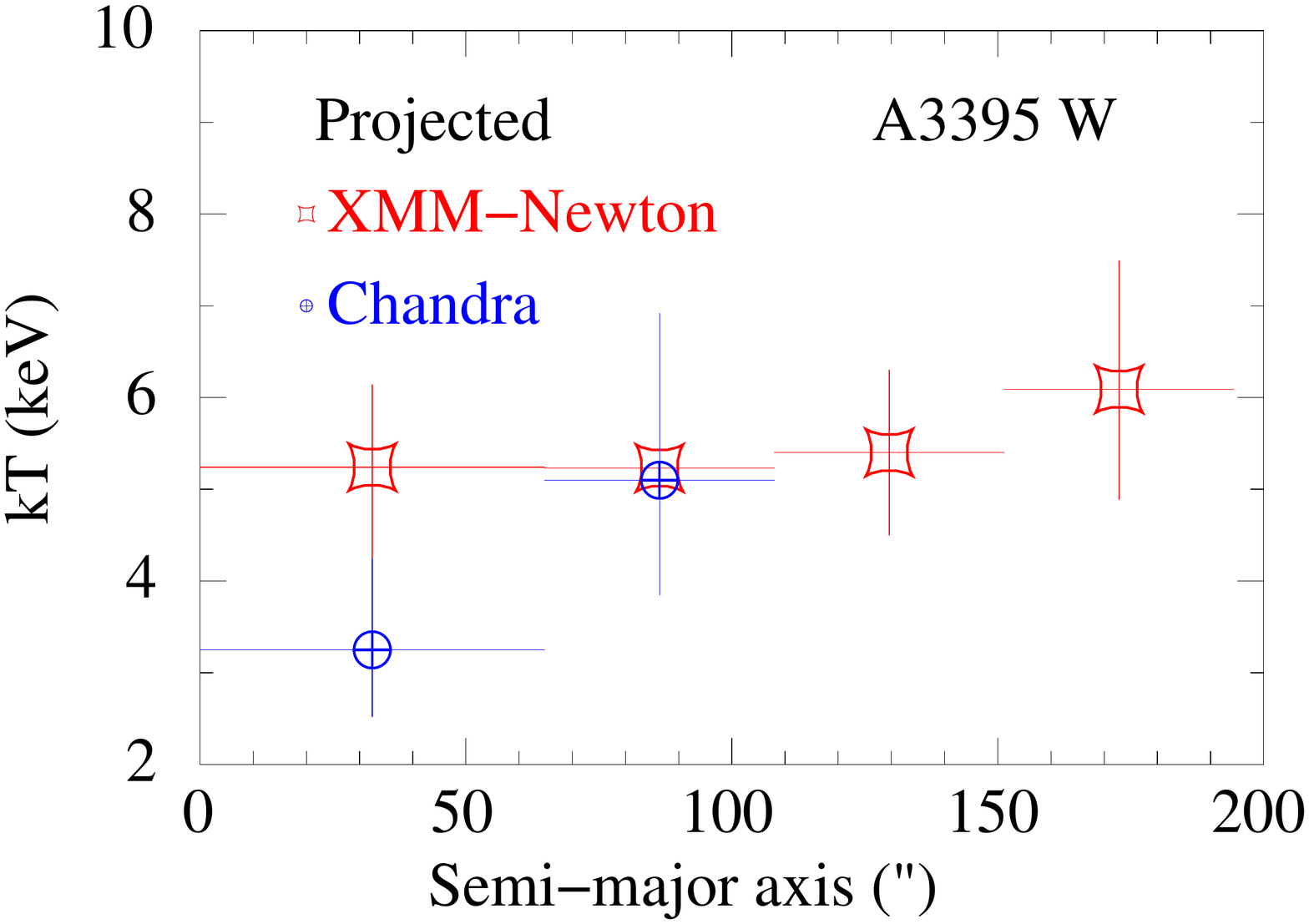} \\
\hspace{0.6cm}
\includegraphics[width=1.5in]{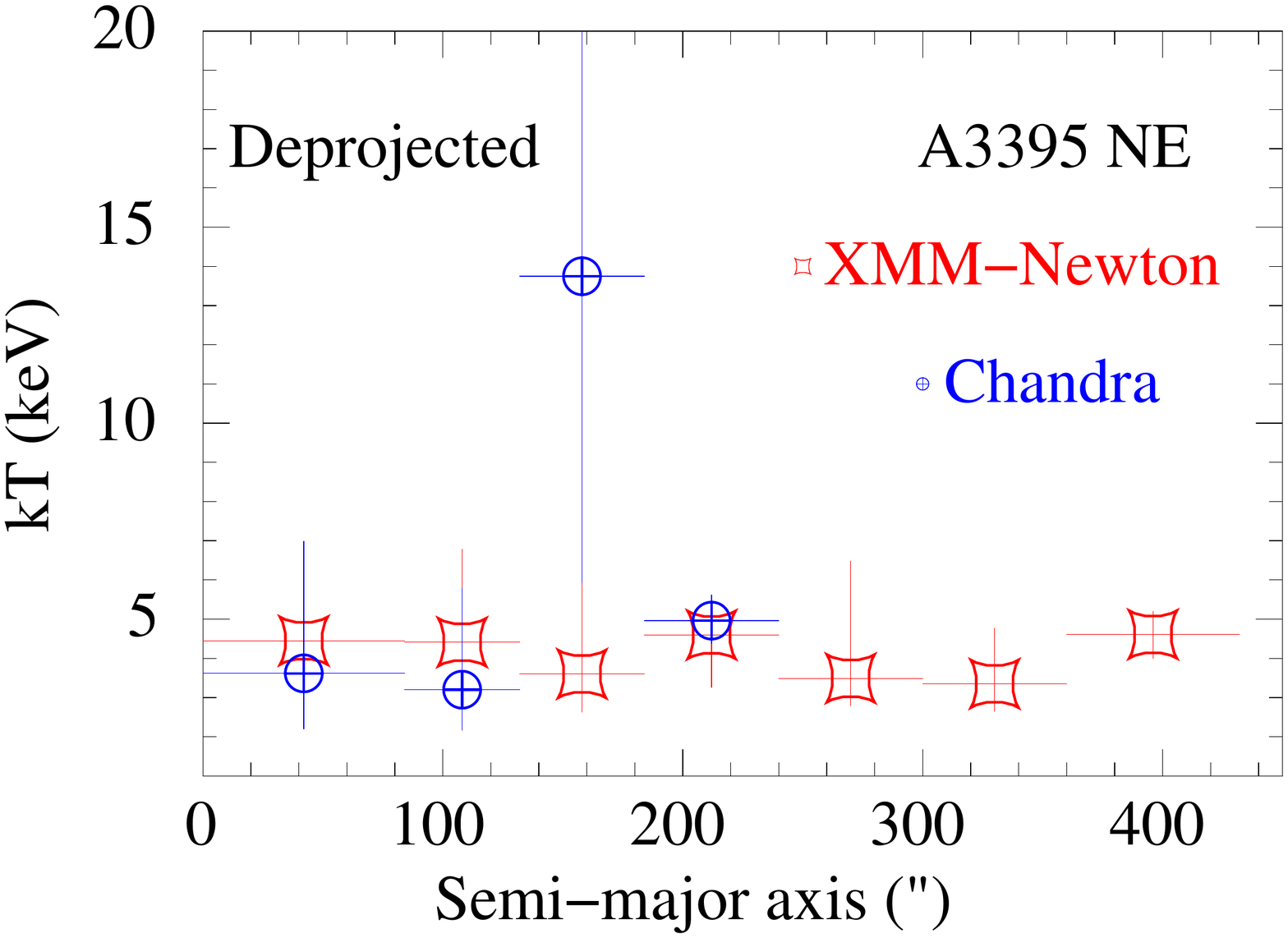} &
\includegraphics[width=1.5in]{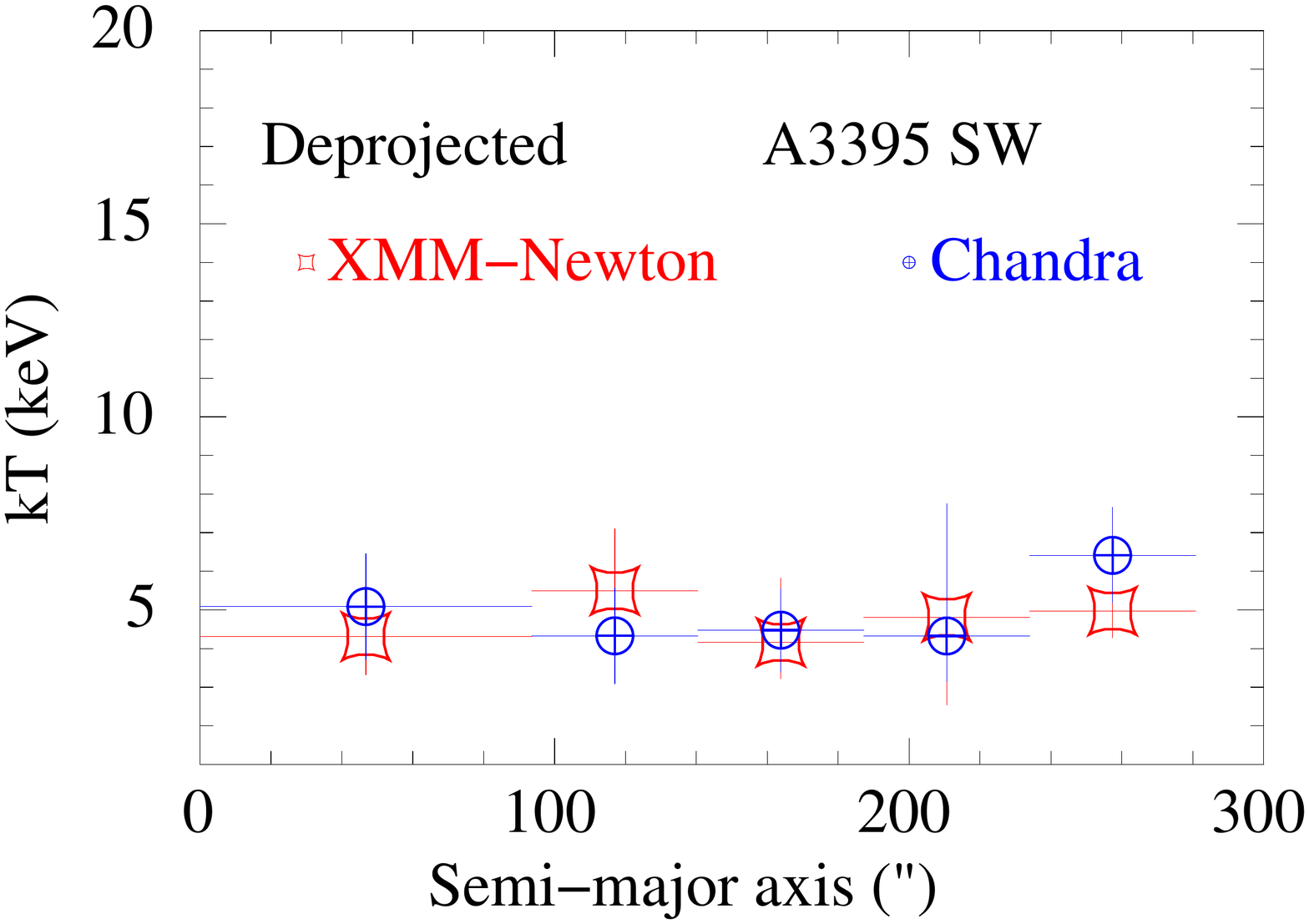} &
\includegraphics[width=1.5in]{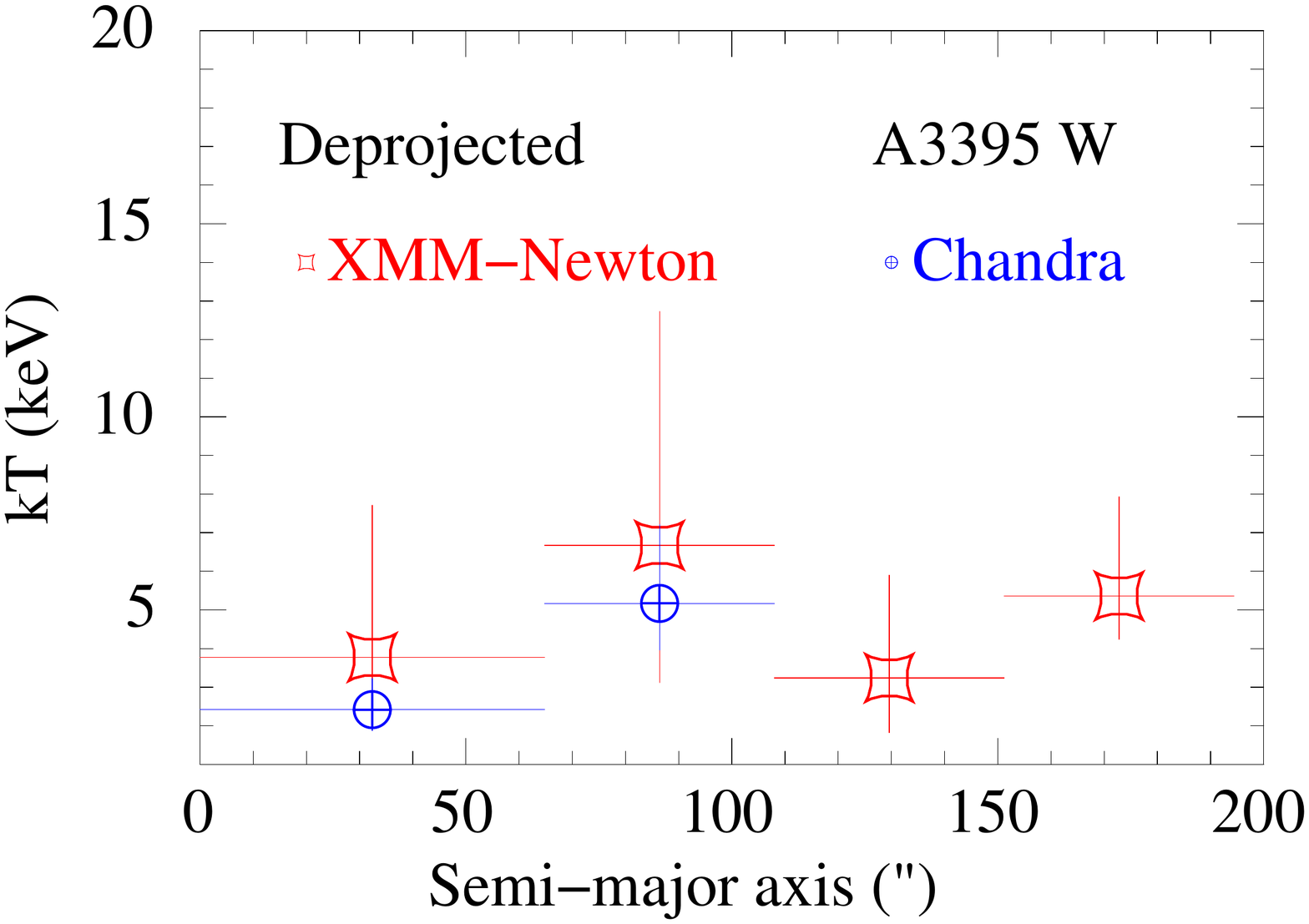}\\
\end{array}$
\label{fig:annuli_temp}
}
\subfigure[]
{
$\begin{array}{ccc}
\includegraphics[width=1.5in]{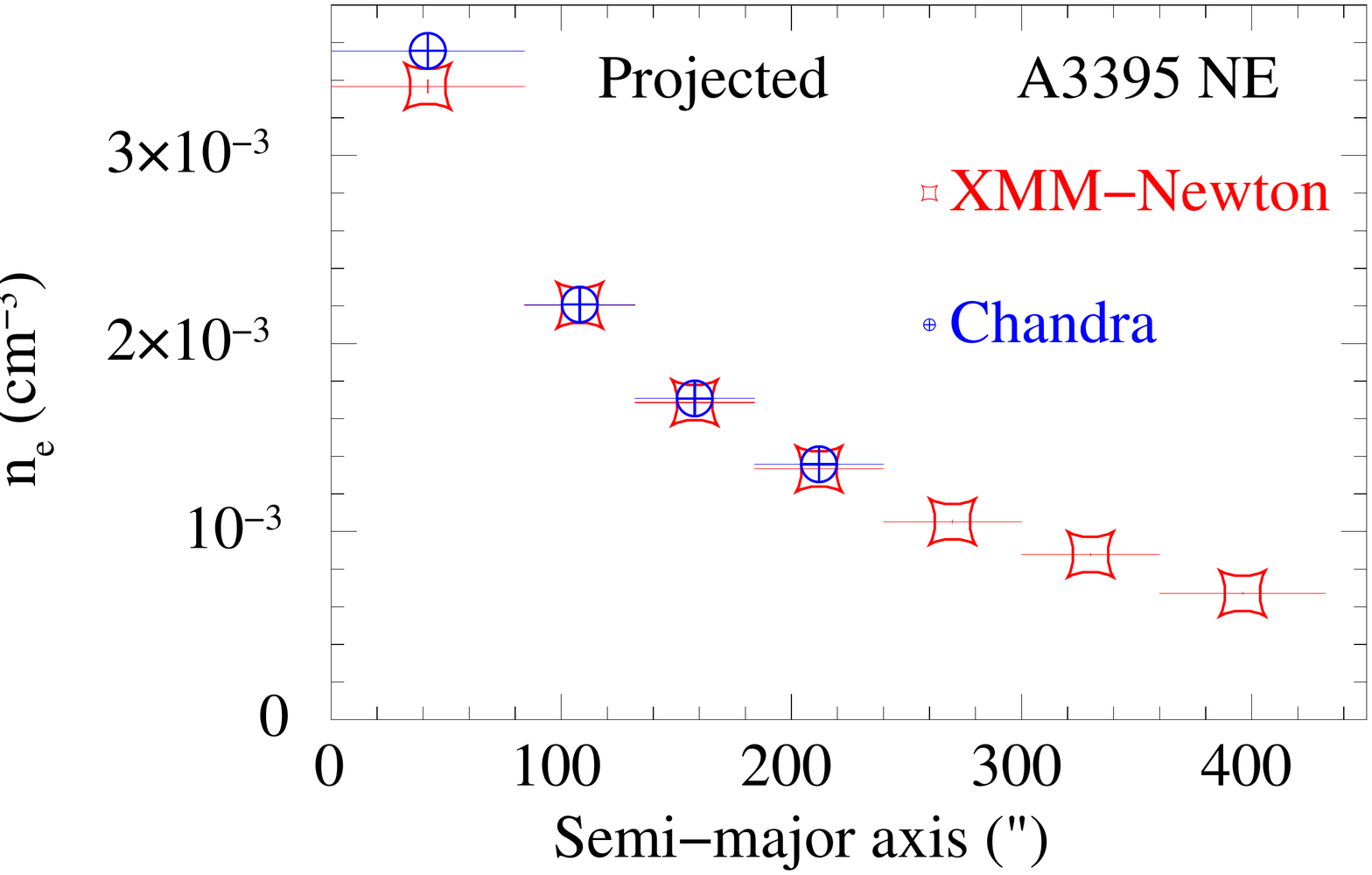} &
\includegraphics[width=1.5in]{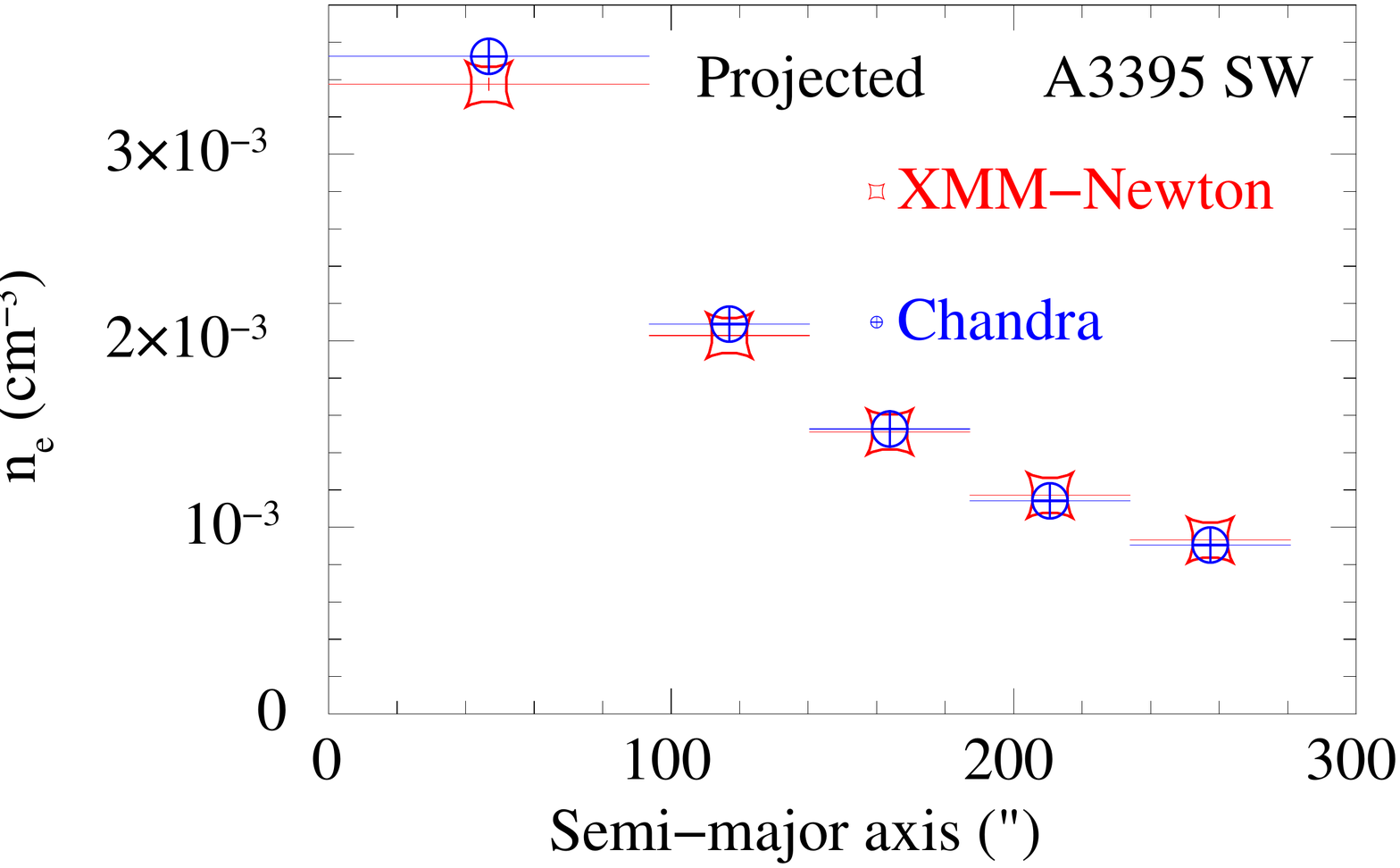} &
\includegraphics[width=1.5in]{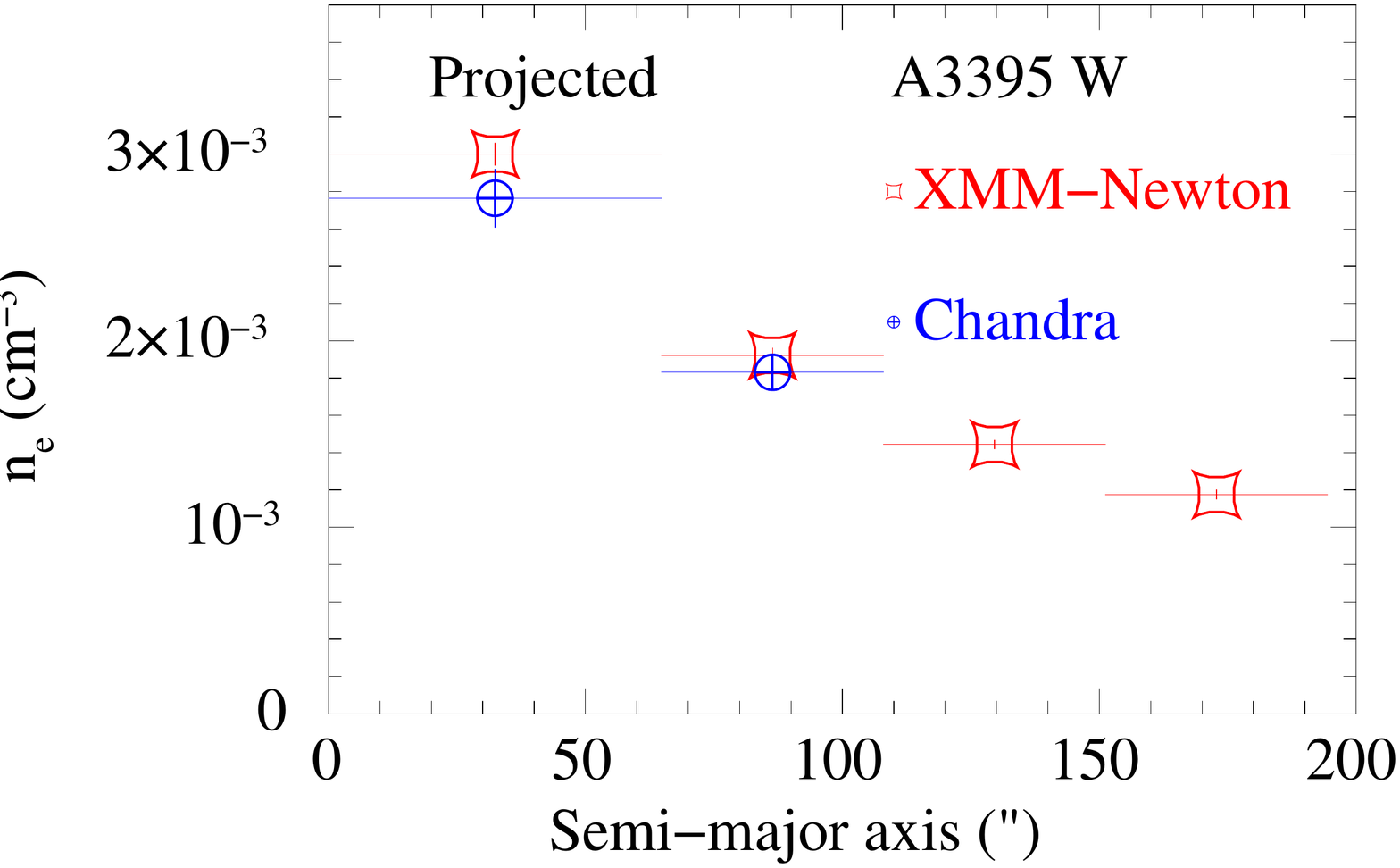}\\
\includegraphics[width=1.5in]{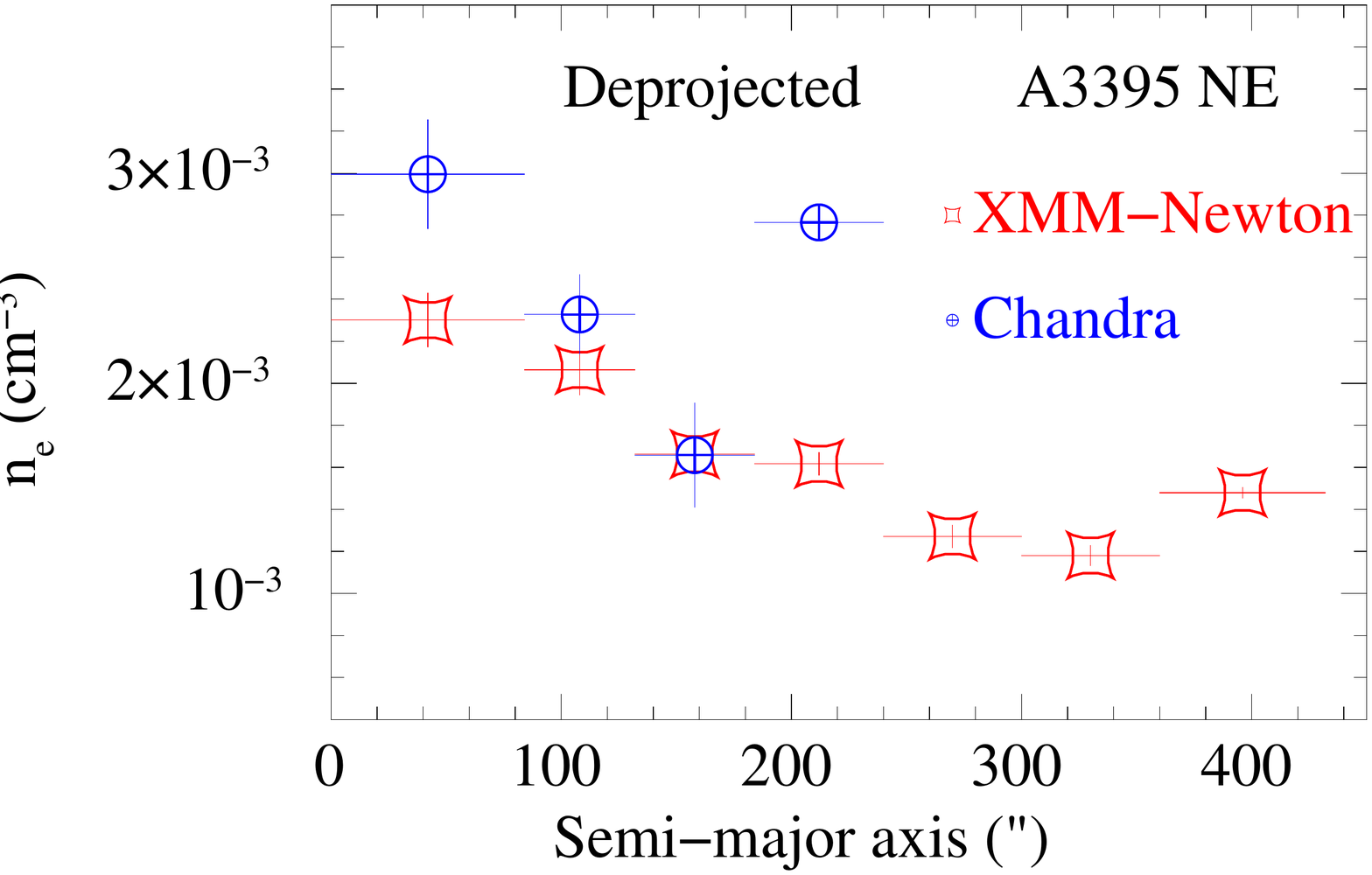} &
\includegraphics[width=1.5in]{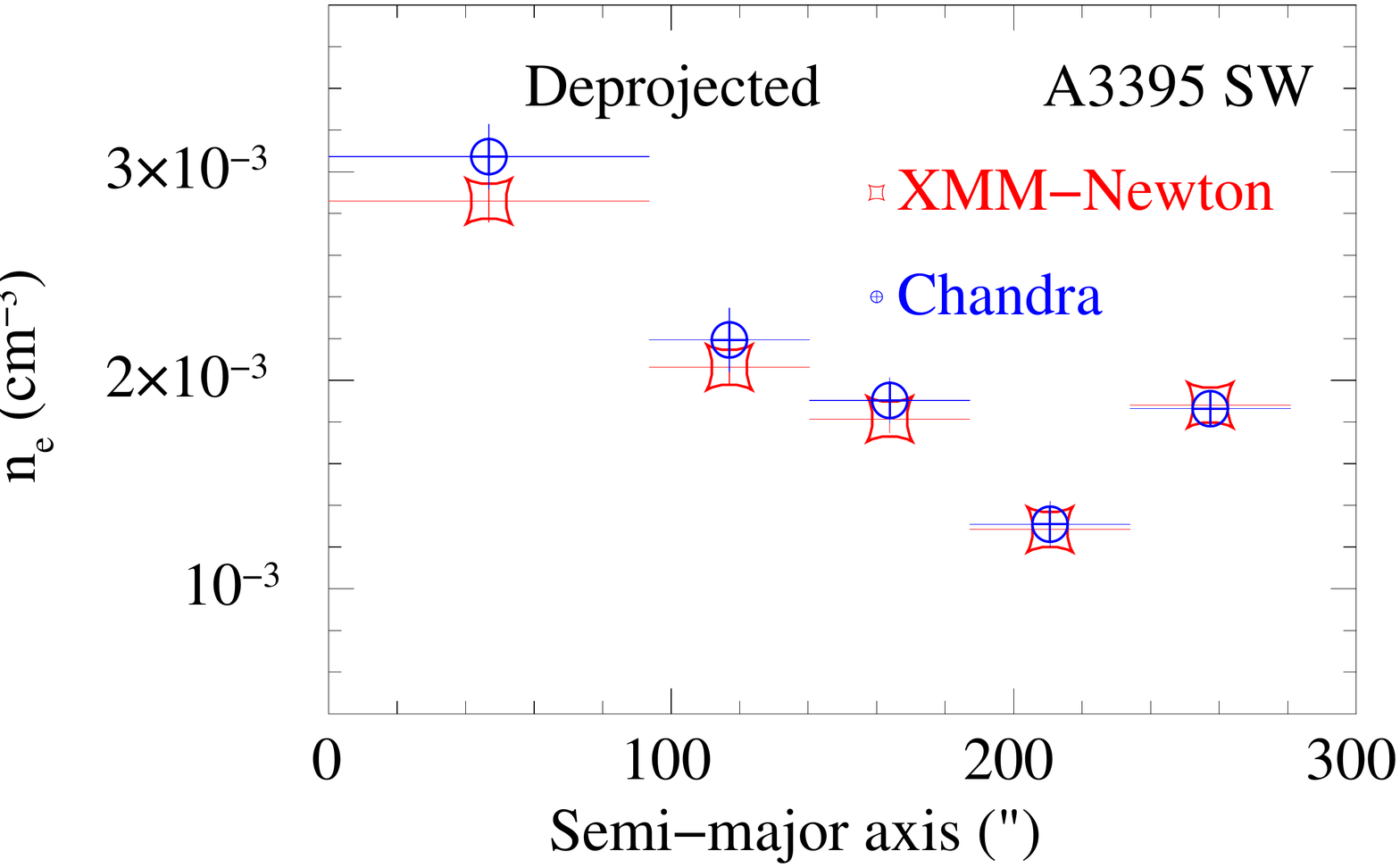} &
\includegraphics[width=1.5in]{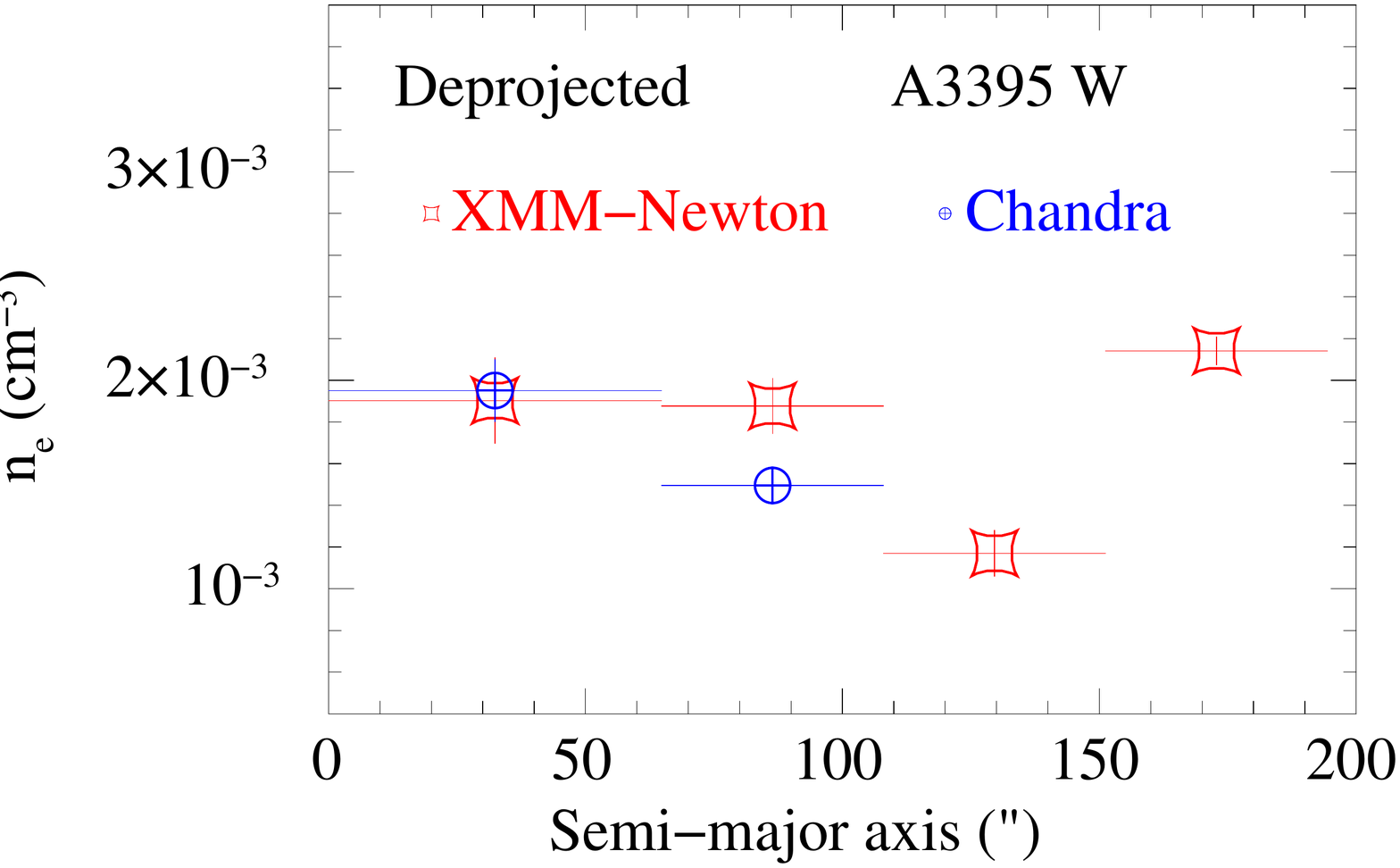}\\
\end{array}$
\label{fig:annuli_density}
}
\caption{Projected and deprojected, temperature (kT) (\ref{fig:annuli_temp}) and electron density (\ref{fig:annuli_density})
(n$_{\rm e}$) profiles from the spectral analysis done using XMM-Newton (Red) and Chandra (Blue) data on elliptical annuli (shown in 
Fig.~\ref{fig:A3395_unsharp_mask_comb_MOS1_MOS2_and_tot_spec_regions}) of NE, SW, and W regions.  For projected spectral analysis, 
the spectra for all annuli in each of the regions were fitted using the model \textbf{wabs*apec} while for deprojected spectral 
analysis the spectra for all annuli in each of the regions were fitted together using the model \textbf{projct(wabs*apec)} for a 
fixed Galactic absorption. Powerlaw and Gaussian components were also added to model the residual soft proton contamination and the 
instrumental Al line at 1.49 keV. Temperature, and \textbf{apec} normalizations were left as free parameters and the best fit values
 of the \textbf{apec} normalizations were used for deriving the n$_{\rm e}$.}
\label{fig:NE_SW_W_projected_deprojected_temp_dens_profile}
\end{figure}

\begin{figure}
\centering
\subfigure[]
{
$\begin{array}{ccc}
\hspace{0.6cm}
\includegraphics[width=1.5in]{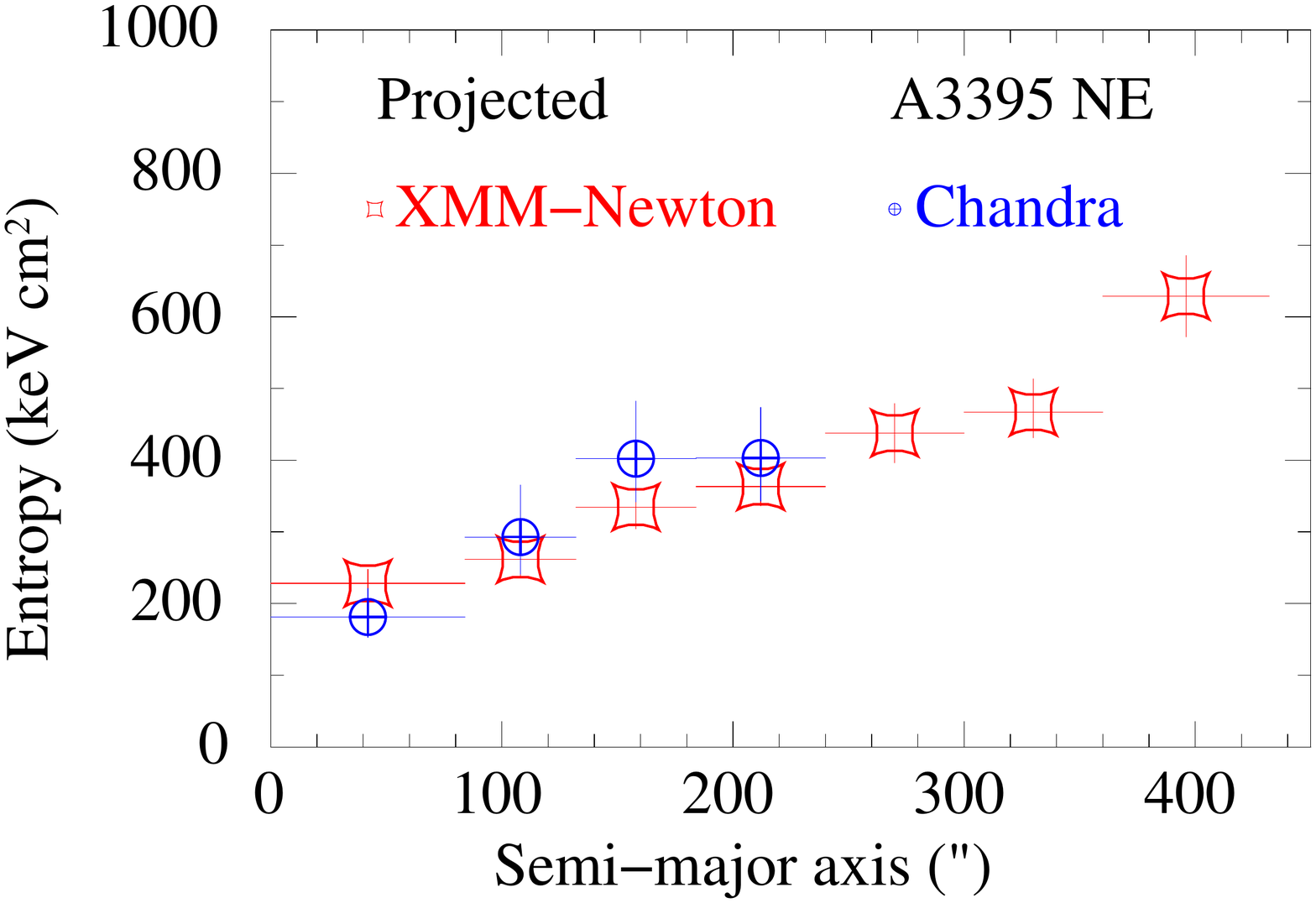} &
\includegraphics[width=1.5in]{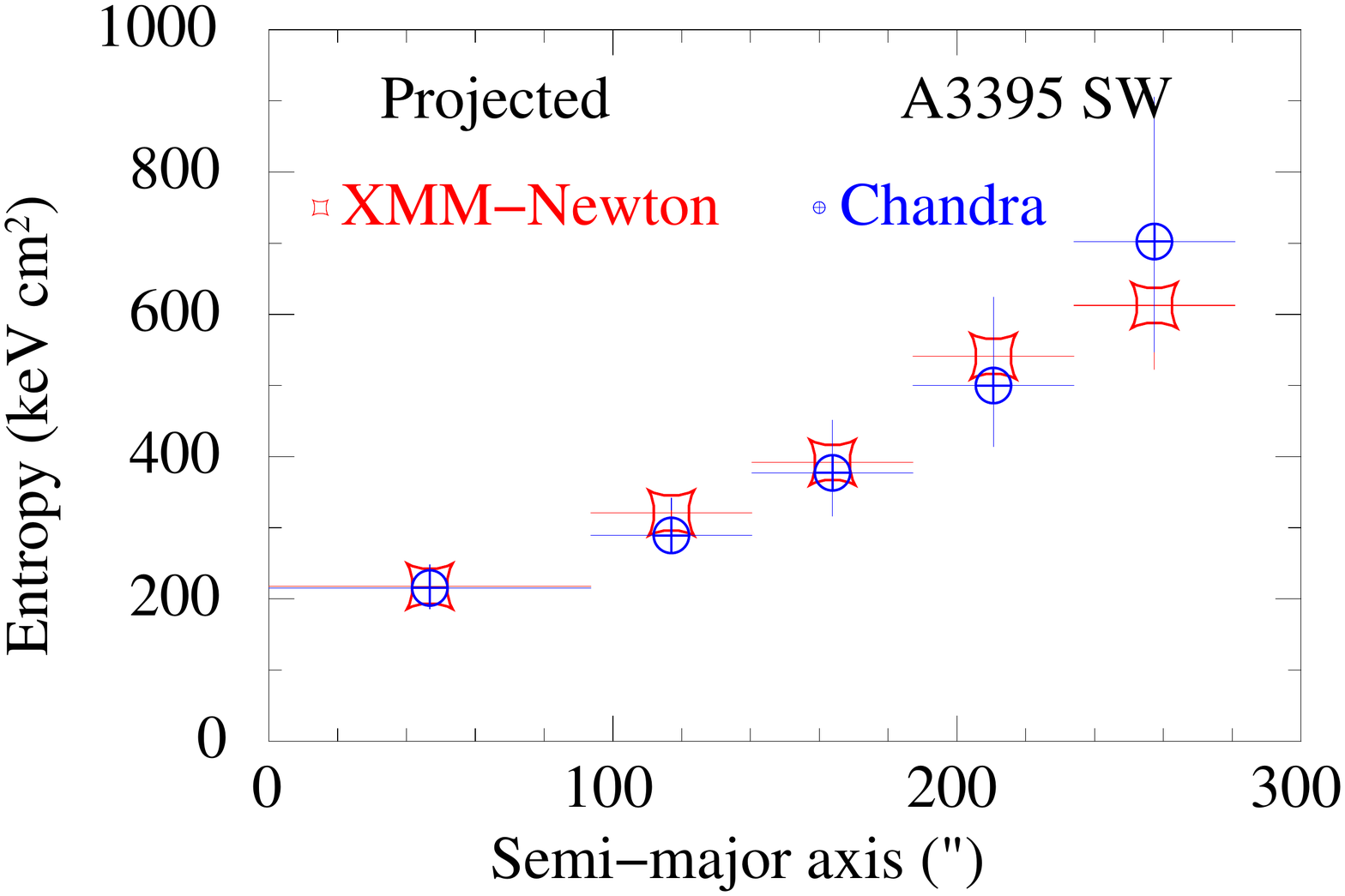} &
\includegraphics[width=1.5in]{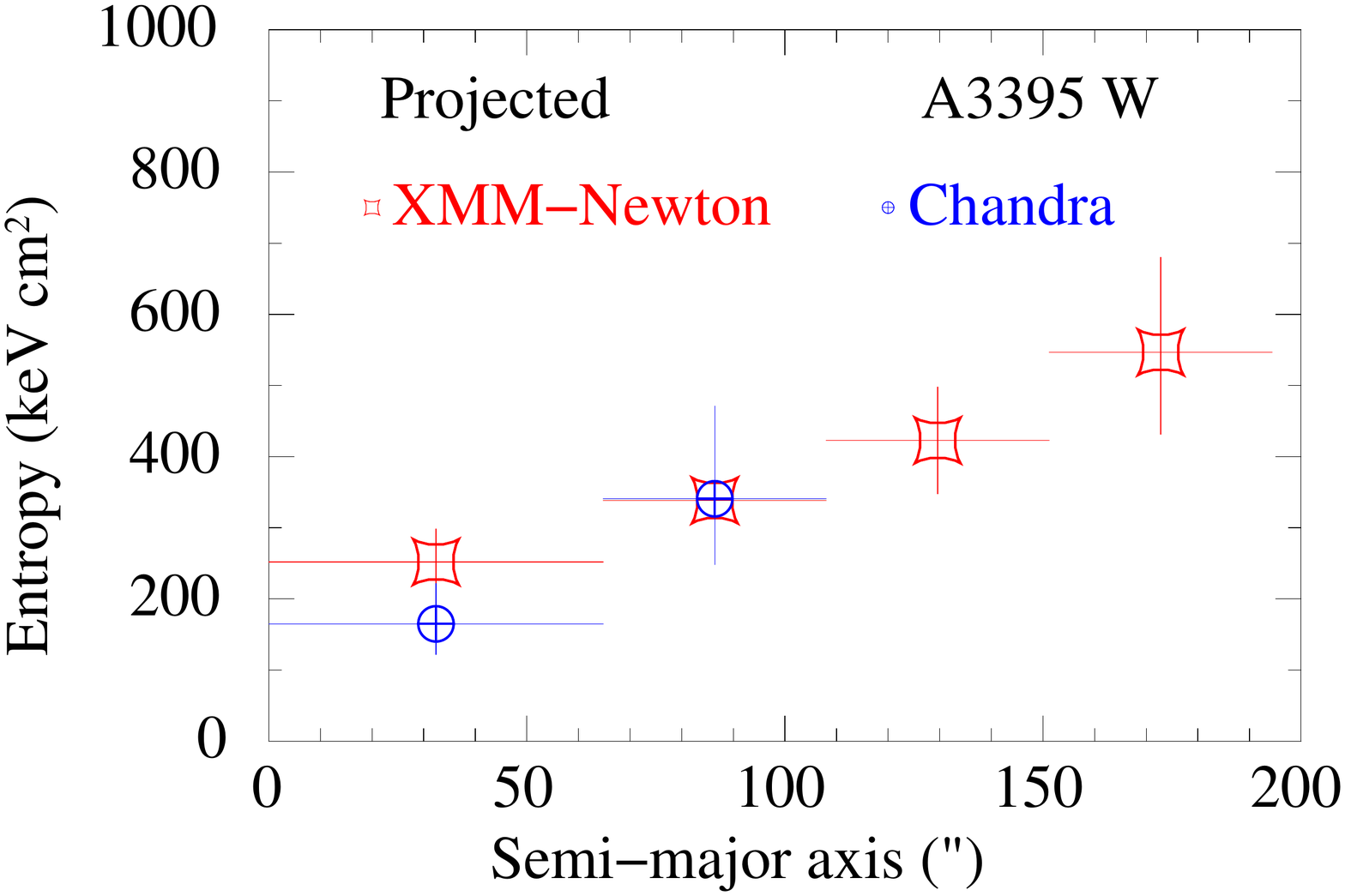}\\
\hspace{0.6cm}
\includegraphics[width=1.5in]{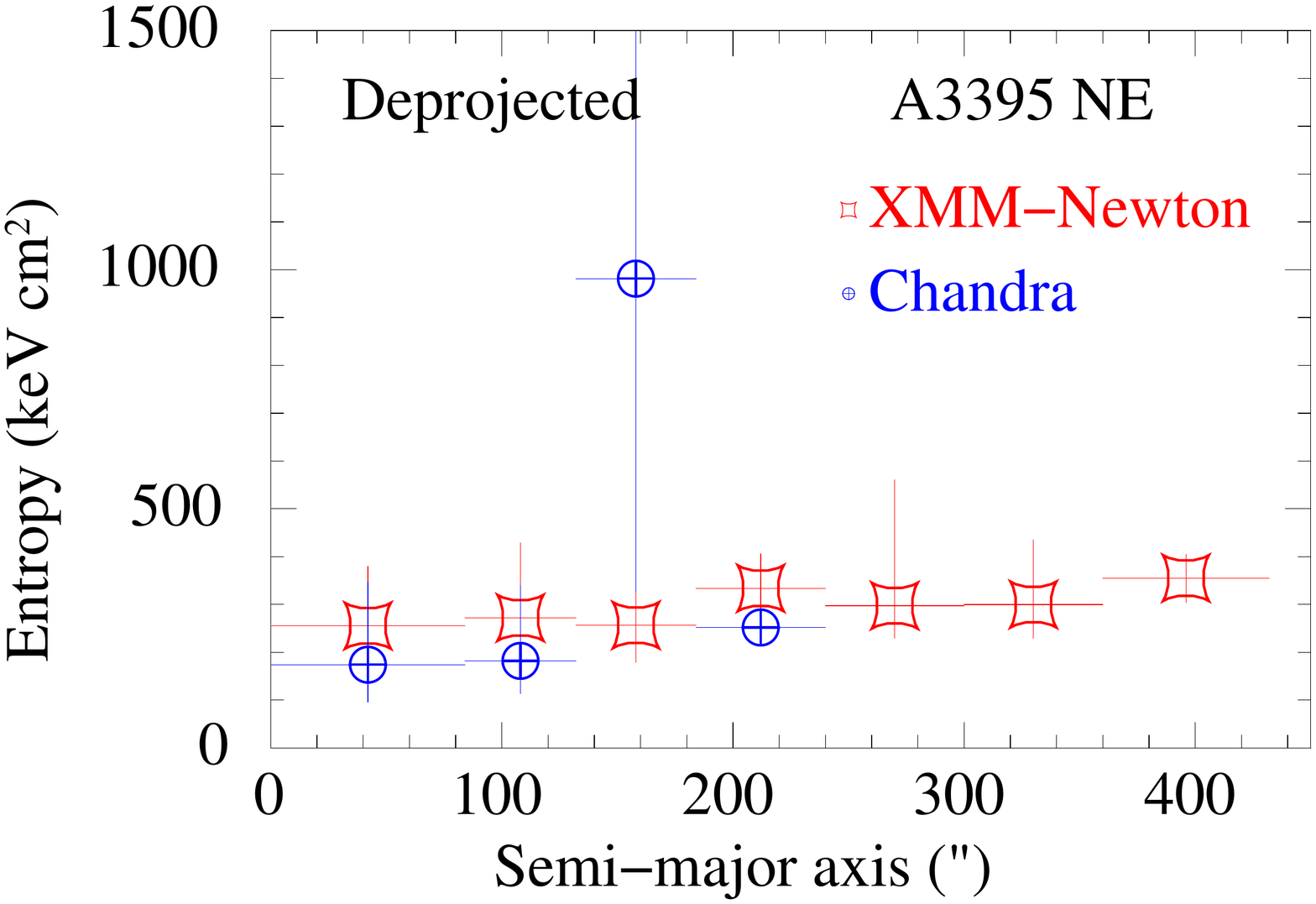} &
\includegraphics[width=1.5in]{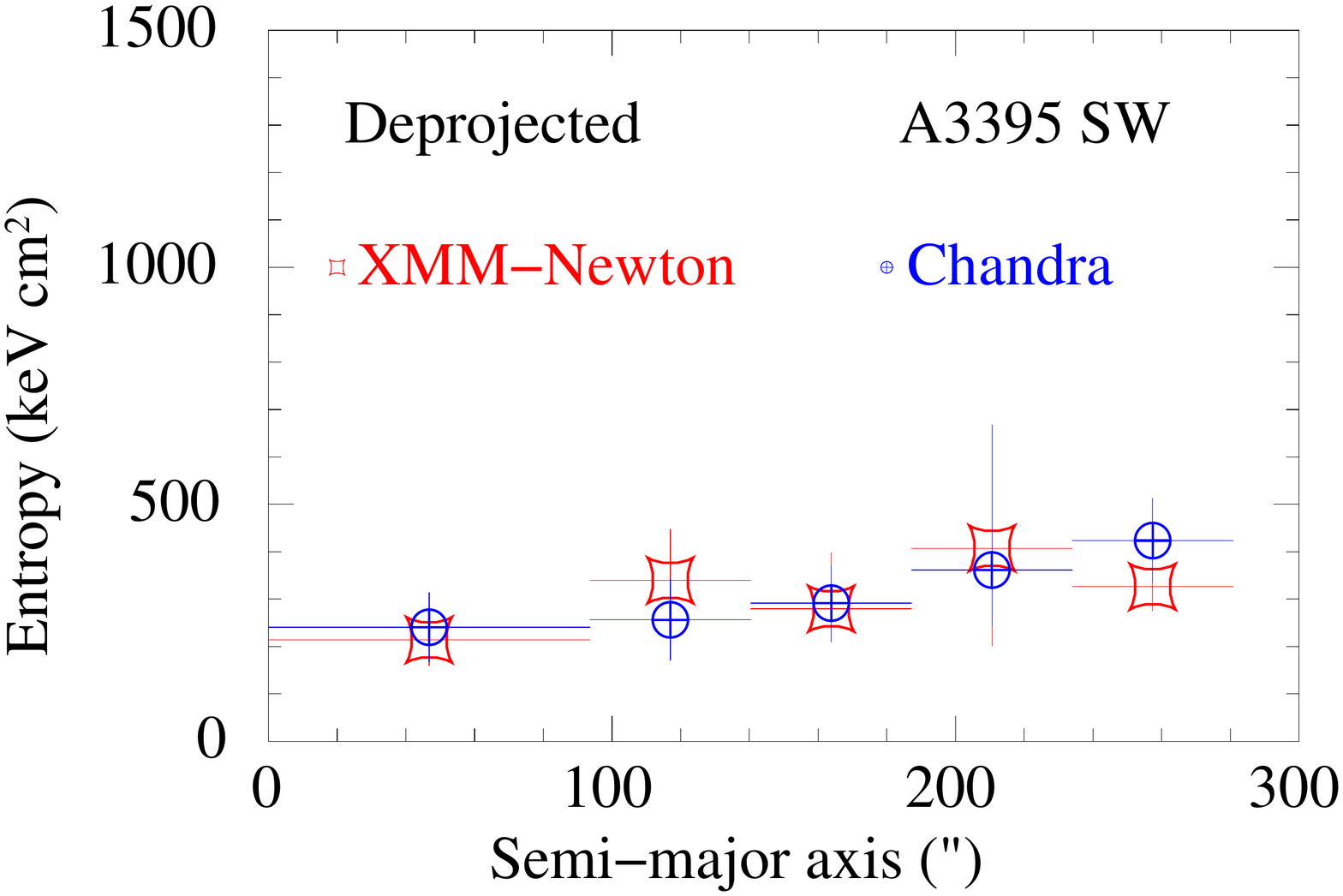} &
\includegraphics[width=1.5in]{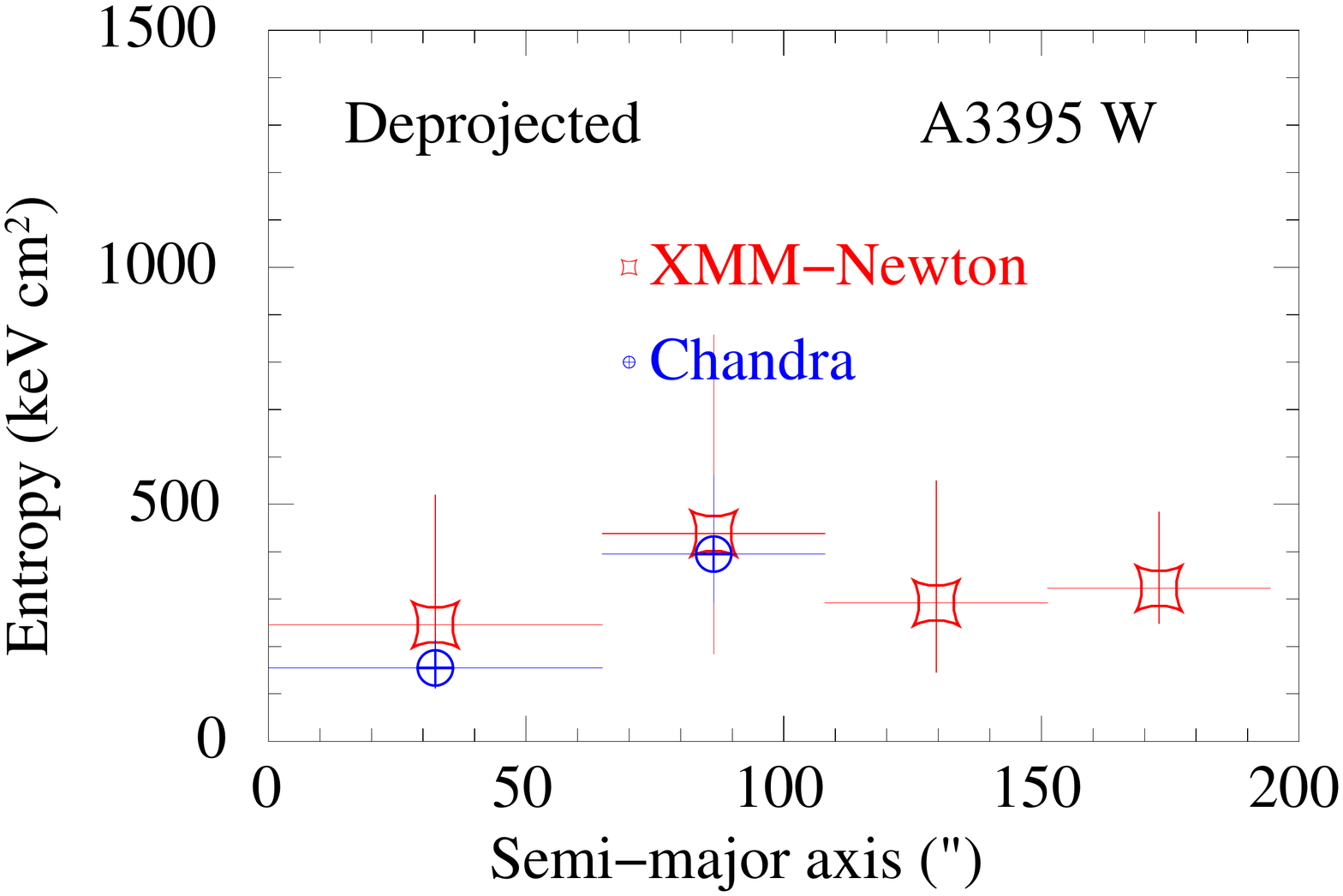}\\  
\end{array}$
\label{fig:annuli_entropy}
}
\subfigure[]
{
$\begin{array}{ccc}
\includegraphics[width=1.5in]{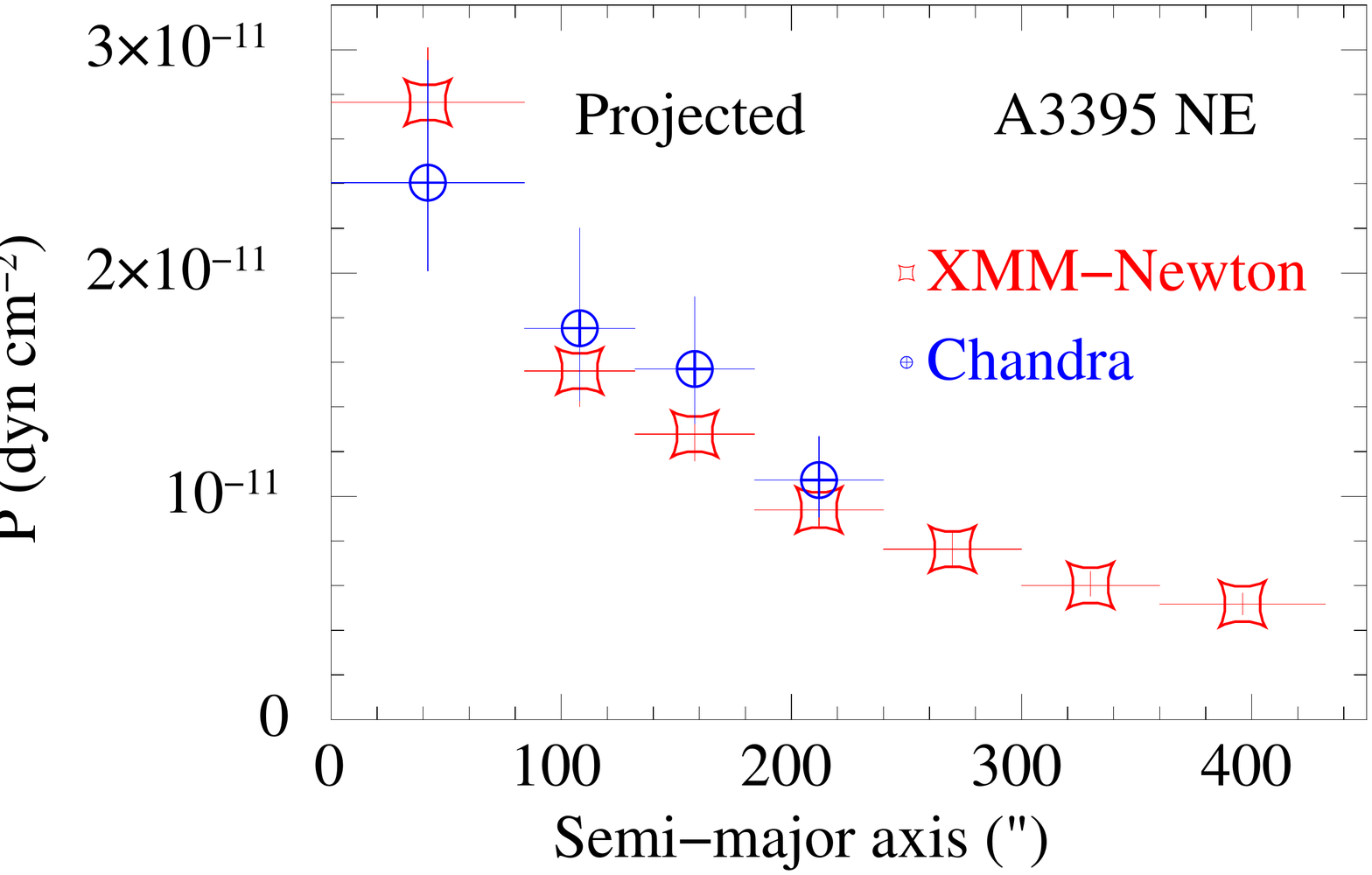} &
\includegraphics[width=1.5in]{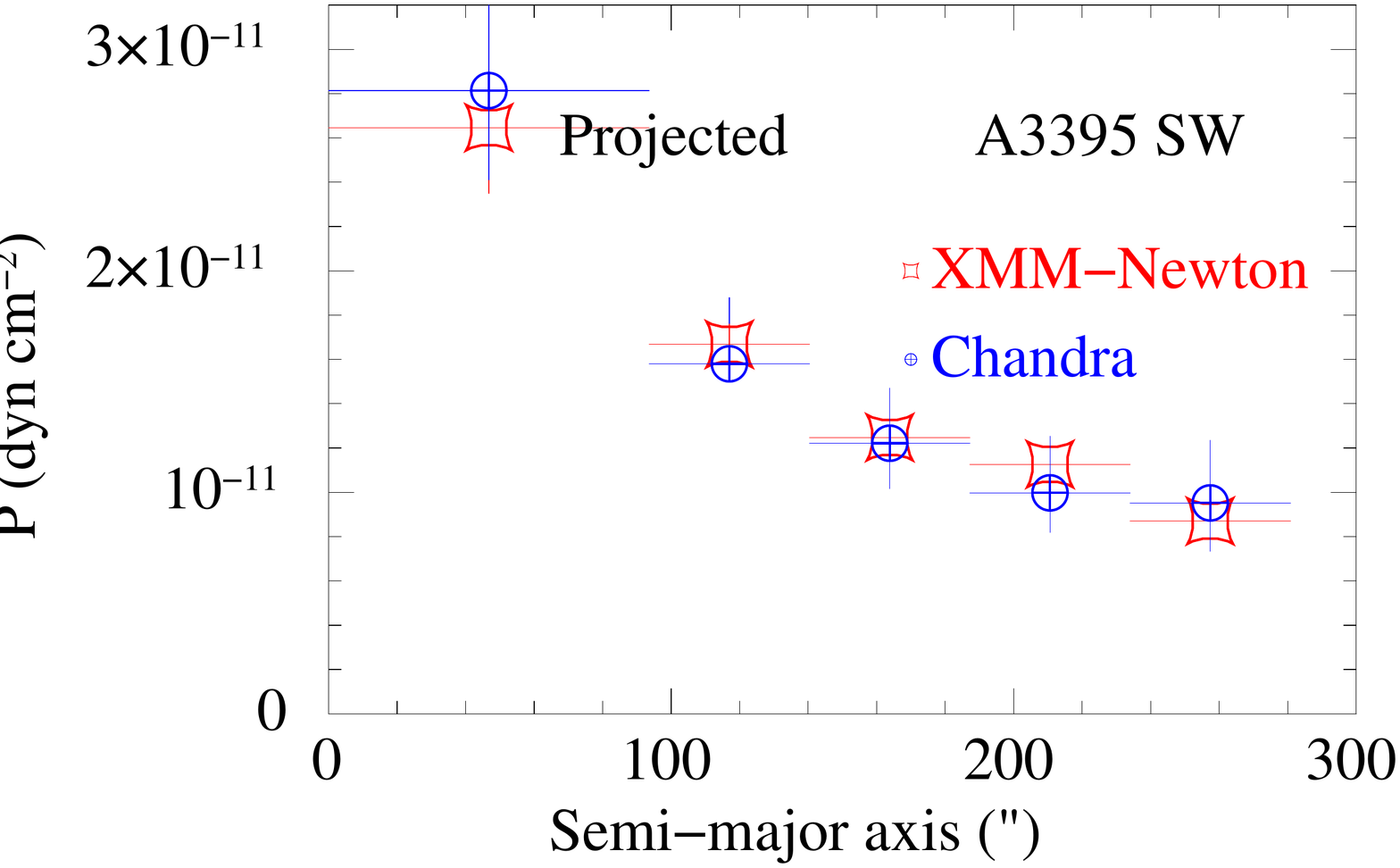} &
\includegraphics[width=1.5in]{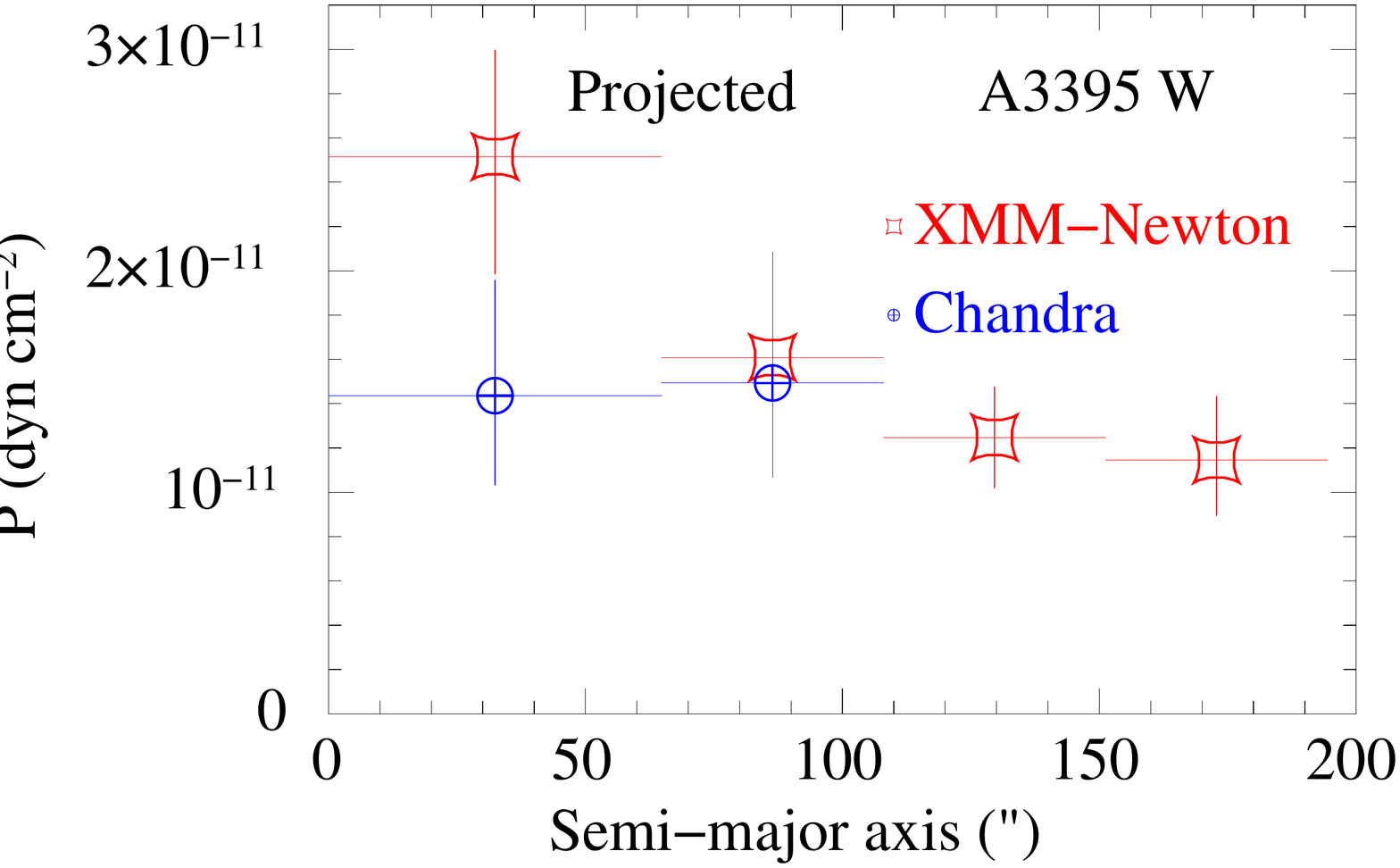}\\
\includegraphics[width=1.5in]{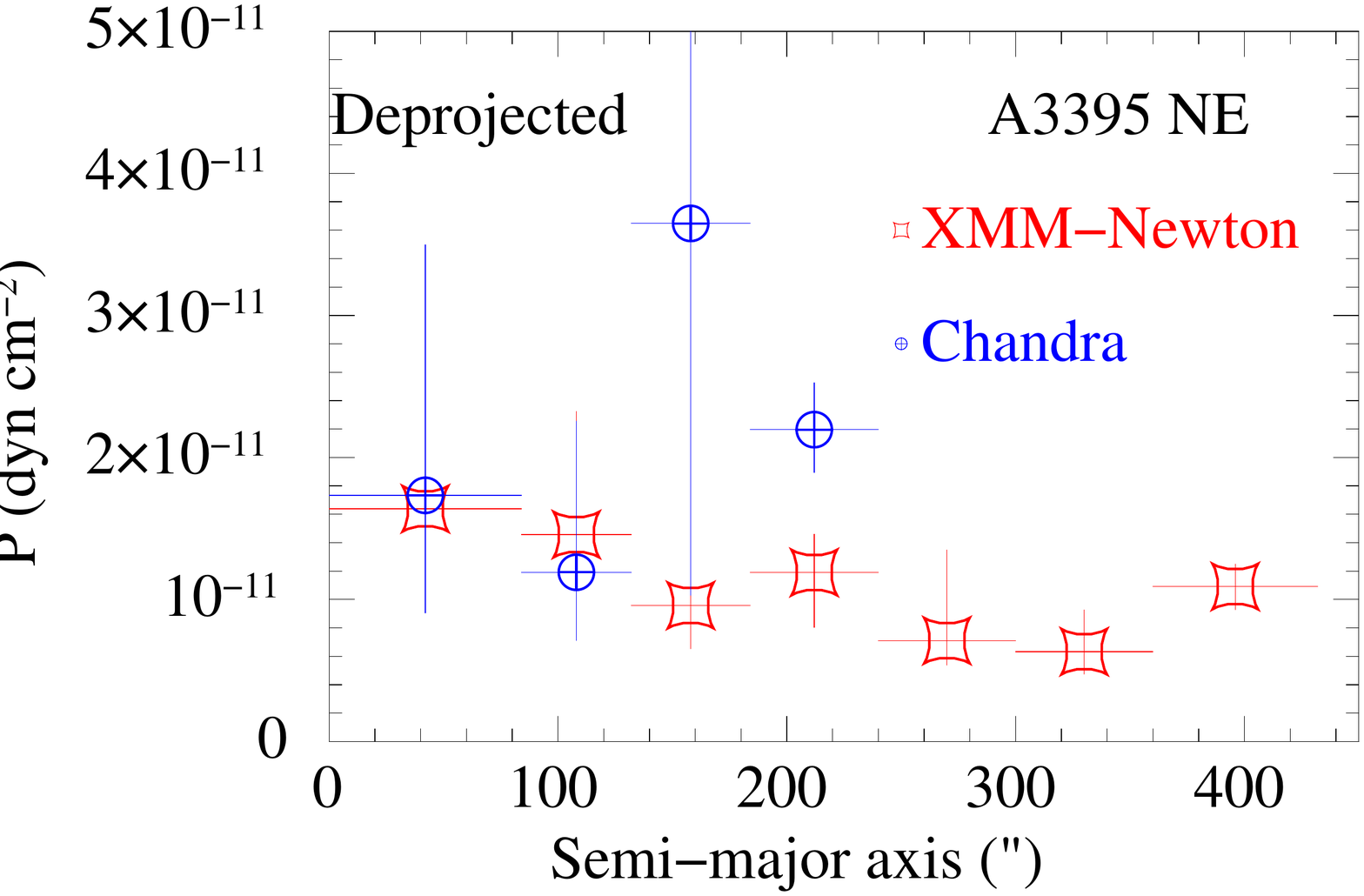} &
\includegraphics[width=1.5in]{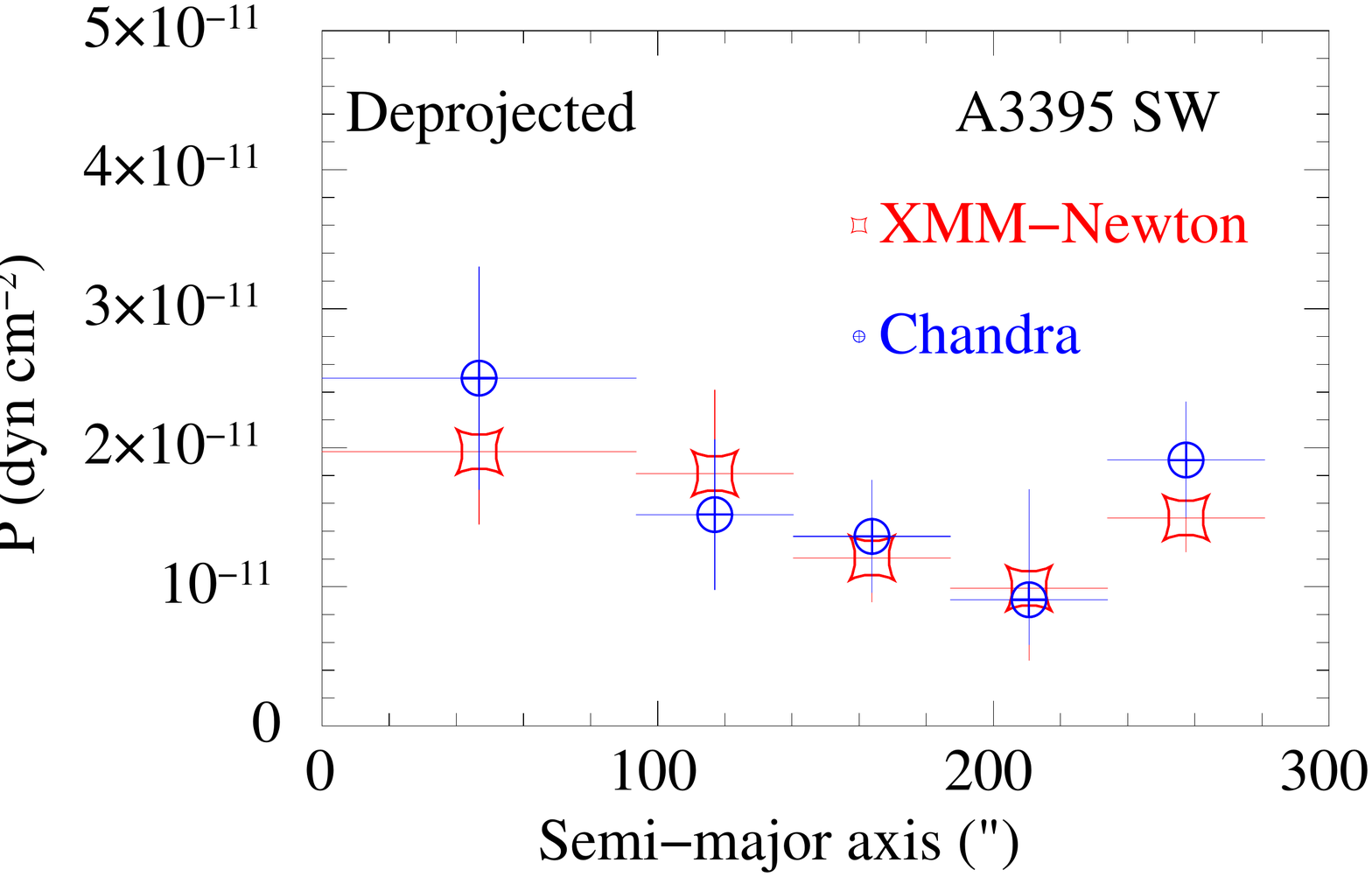} &
\includegraphics[width=1.5in]{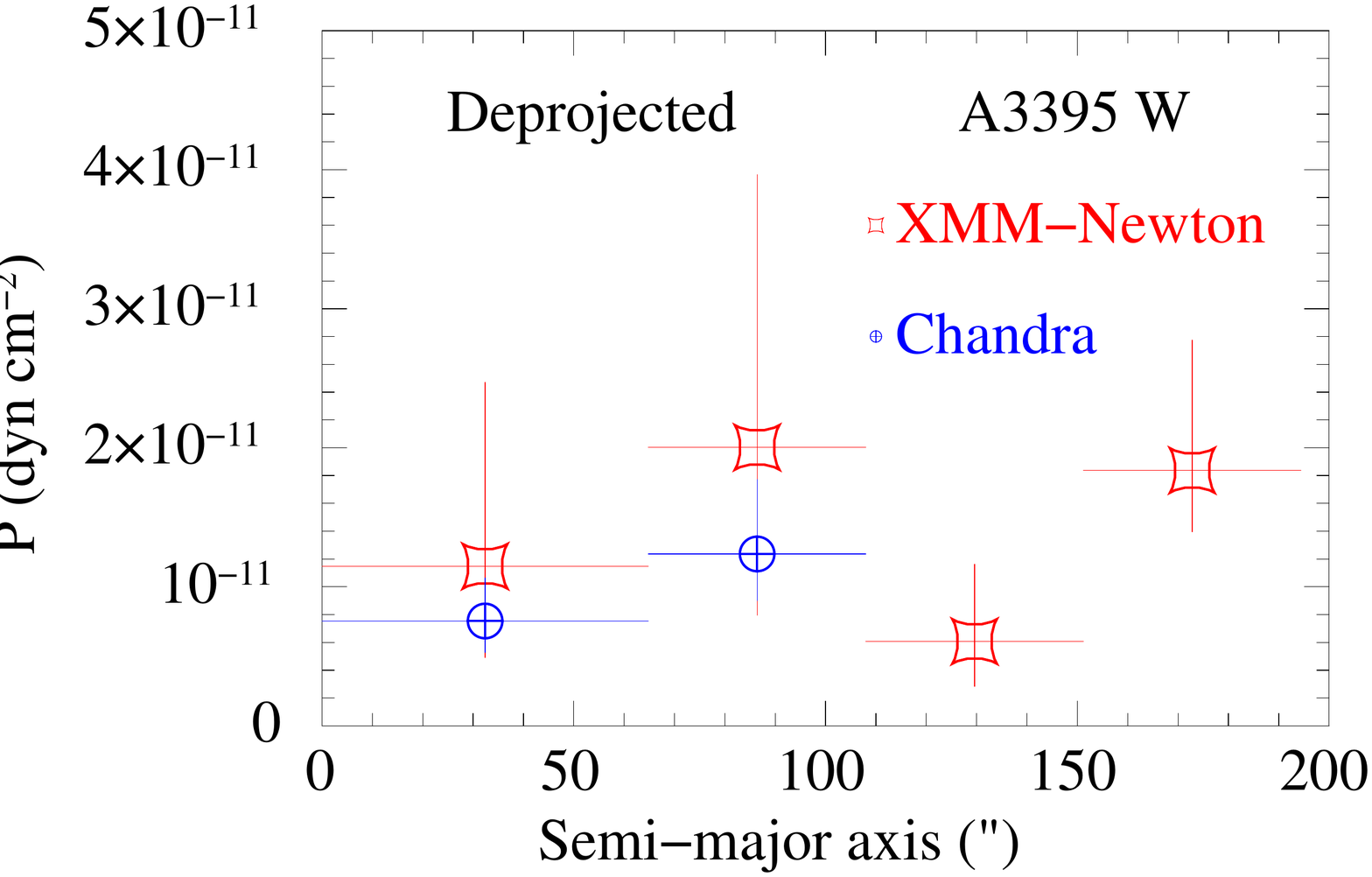}
\end{array}$
\label{fig:annuli_pressure}
}
\caption{Projected and deprojected, entropy (\ref{fig:annuli_entropy}) and pressure (P) (\ref{fig:annuli_pressure}) profiles from the 
spectral analysis done using XMM-Newton (Red) and Chandra (Blue) data on elliptical annuli (shown in 
Fig.~\ref{fig:A3395_unsharp_mask_comb_MOS1_MOS2_and_tot_spec_regions}) of NE, SW, and W regions. For projected spectral analysis, the 
spectra for all annuli in each of the regions were fitted using the model \textbf{wabs*apec} while for deprojected spectral analysis 
the spectra for all annuli in each of the regions were fitted together using the model \textbf{projct(wabs*apec)} for a fixed 
Galactic absorption. Powerlaw and Gaussian components were also added to model the residual soft proton contamination and the 
instrumental Al line at 1.49 keV. Best fit values of the temperature, and \textbf{apec} normalizations were used for deriving the 
entropy and pressure.}
\label{fig:NE_SW_W_projected_deprojected_entr_press_profile}
\end{figure}

  The resulting projected and deprojected temperature and density profiles are shown in 
Figure~\ref{fig:NE_SW_W_projected_deprojected_temp_dens_profile}, and the projected and deprojected entropy and pressure profiles are 
shown in Figure~\ref{fig:NE_SW_W_projected_deprojected_entr_press_profile}. Best fit parameters and the derived values of other 
dependent parameters are given in Tables \ref{tab:XMM_projected_annuli_spectral_results}-
\ref{tab:Chandra_deprojected_annuli_spectral_results}. The projected spectral analysis resulted in a nearly constant 
temperature profile for all three subclusters, while the projected density, entropy, and pressure 
profiles show a uniform decrease, increase, and decrease respectively for all three subclusters from the innermost to the outermost 
annulus. The deprojected temperatures, entropies and pressures of all the three subclusters show a nearly uniform profile. The 
deprojected density profiles from all three regions show a uniform decrease except for the outermost annulus where a slight increase 
in the density is seen. This is a common artifact of deprojection analysis because the excess emission from shells with radii larger 
than that of the outermost annulus, gets added to the outermost annulus. The projected profiles from Chandra and XMM-Newton are in 
complete agreement with each other although the former has larger errors as compared to the latter. For the deprojected profiles also,
 there is a good agreement in the results between XMM-Newton and Chandra with a few anomalies. For example, the deprojected density 
for the innermost annulus in the NE region from Chandra is greater while the density in the second annulus in the W region from 
Chandra is lower as compared to those from XMM-Newton. However, the higher deprojected density and pressure from Chandra in the 
fourth annulus (outermost annulus for Chandra) in NE as compared to those from XMM-Newton can easily be explained as the artifact of 
deprojection described earlier in this section.

\subsection{Spectrally determined 2-D projected thermodynamic maps at a higher resolution}
\label{sec:box_thermodynamic_maps}
  We have made projected temperature, abundance, density, entropy, and pressure maps of A3395, using box shaped regions for spectral 
analysis to improve the spatial resolution of the spectral parameters and to look for anisotropy in their spatial distribution. To 
make the results more robust, both XMM-Newton and Chandra data have been used and the whole cluster was divided into 139 boxes for 
XMM-Newton and only 42 boxes for Chandra. The number of boxes used for Chandra were less because of fewer counts and also because 
Chandra did not cover the outer parts of the cluster significantly due to smaller FOV. The sizes for the boxes were chosen adaptively 
to get sufficient counts in each region. For XMM-Newton, large size boxes ($\sim 3.1^{\prime}\times1.4^{\prime}$ or $\sim 1.6^{\prime}\times2.9^{\prime}$)
 for the outermost parts, small size boxes ($\sim 47^{\prime\prime}\times43^{\prime\prime}$) for the innermost brightest parts, and medium sized boxes 
($\sim 1.6^{\prime}\times1.4^{\prime}$) for the regions in between were made to get more than 700 total counts from all three detectors in each 
box. For Chandra, large size boxes ($\sim 3.1^{\prime}\times1.6^{\prime}$) for outer parts and small size boxes ($\sim 1.6^{\prime}\times1.6^{\prime}$) 
for the inner parts were made to get more than 700 counts in each box. Hence, spatial resolution of the thermodynamic maps obtained 
with XMM-Newton is better than that from Chandra. Spectra from all boxes were fitted using \textbf{wabs*apec} model with fixed 
Galactic absorption. The residual soft proton contamination and instrumental Al lines were modeled by adding powerlaws and 
Gaussian components to the models as described in \S\ref{sec:Total-Spec-Analy}. For the boxes where MOS1 
data could not be used (i.e., boxes lying in regions of MOS1 missing CCD6 and anomalous CCD5), only MOS2 and PN spectra were used.
 The electron density, entropy, and electron pressure were calculated using the same relations as in \S\ref{sec:projection_analysis}.
 The volume calculation for the box regions, however, was not as straightforward as in the case of ellipsoidal shells. We had to 
assume spherical geometry which can be justified because the volumes considered here are very small as compared to those for the 
ellipsoidal shells. We assumed the 139 box regions as projections of parts of spherical shells (centred at the X-ray intensity peak of
 the subcluster nearest to the box) with inner and outer radii (R$_{\rm in}$, R$_{\rm out}$) equal to the smallest and largest 
distance from the centre of their respective spheres. The volume for each box region was estimated as 
D$_{\rm A}^{3} \rm \Omega (\rm \theta_{\rm out}^{2} - \rm \theta_{\rm in}^{2})^{1/2}$ (Ehlert et al. 2010; Henry et al. 2004), where 
$D_{A}$ is the angular diameter distance and $\rm \Omega$ is the solid angle subtended by the region. $\rm \theta_{\rm in}$ and 
$\rm \theta_{\rm out}$ are equal to the distances R$_{\rm in}$ and R$_{\rm out}$ expressed in angular units respectively.\\

\begin{figure}[!ht]
\begin{center}
\includegraphics[width=3.0in,height=3.5in]{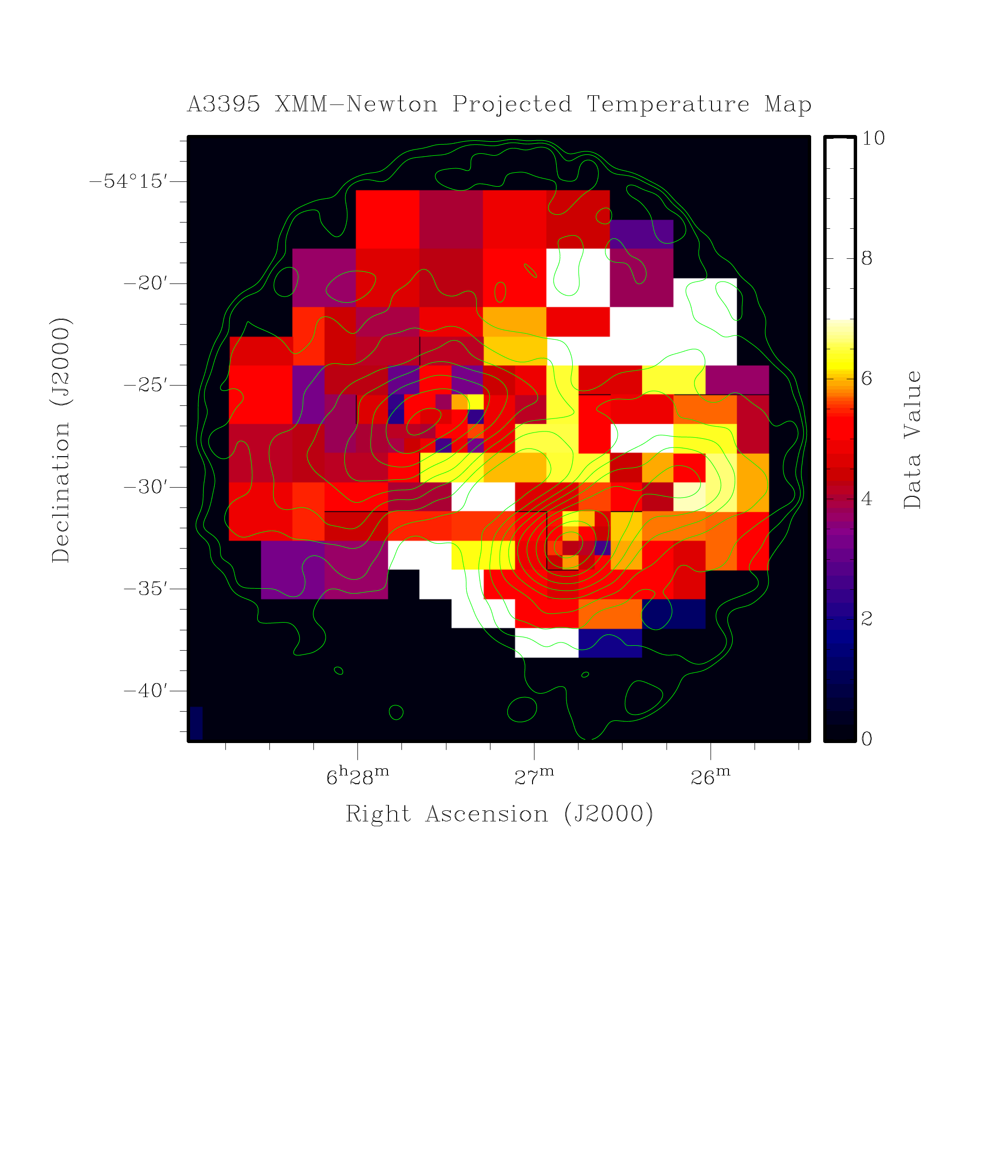}
\includegraphics[width=2.5in,height=3.2in]{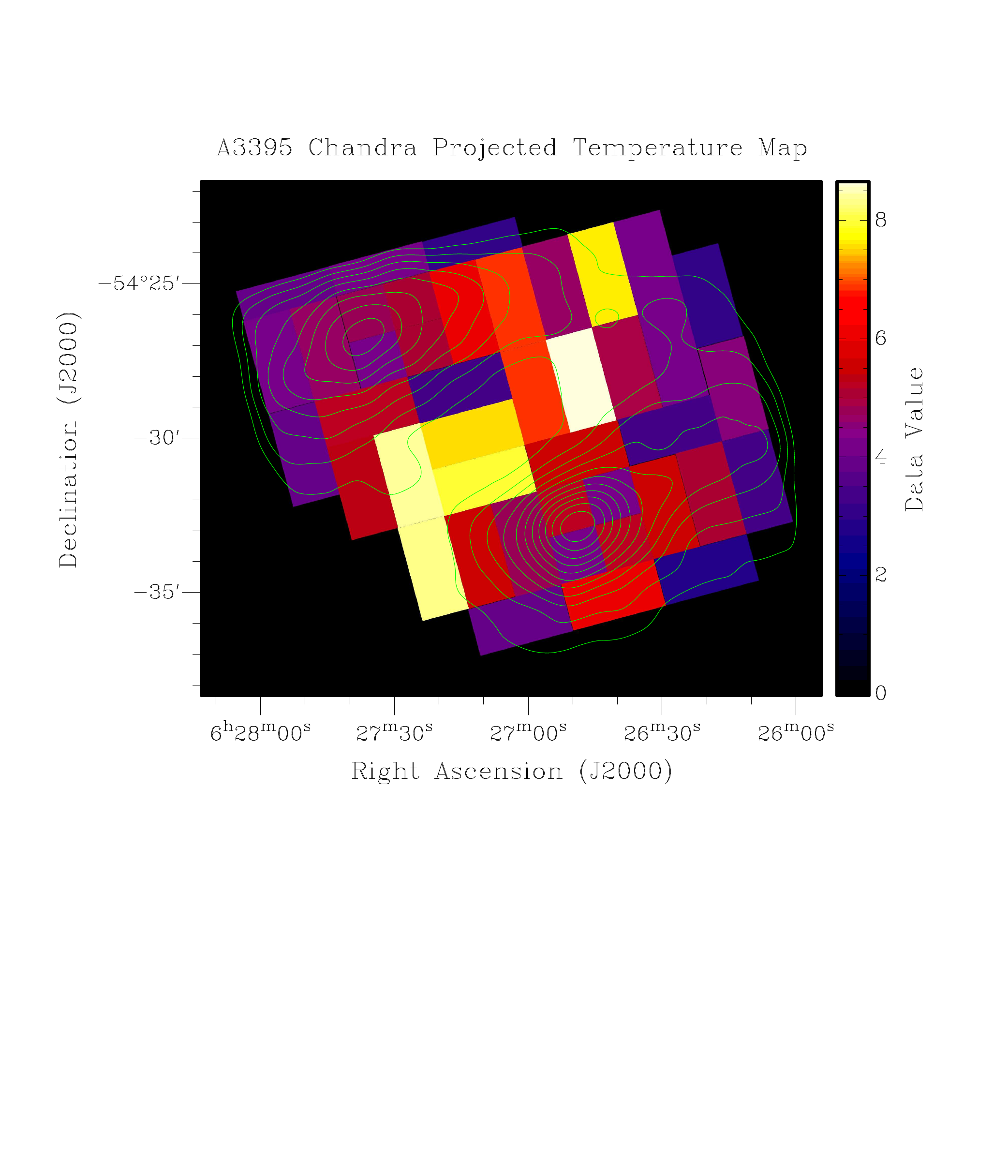}  
\end{center}
\caption{Projected temperature maps from 139 box regions using XMM-Newton data (top) and 42 box regions using Chandra data (bottom) 
with overlaid X-ray surface-brightness contours having levels same as in Figs.~\ref{fig:Combined_MOS1_MOS2_smoothedimage} and 
\ref{fig:Chandra_smoothedimage} respectively. The scales are expressed in keV units shown in the bars alongside. Details of the 
spectral fittings are provided in \S\ref{sec:box_thermodynamic_maps}.}
\label{fig:A3395_box_temp_map}
\end{figure}

\begin{figure}[!ht]
\centering
\includegraphics[width=3.0in,height=3.5in]{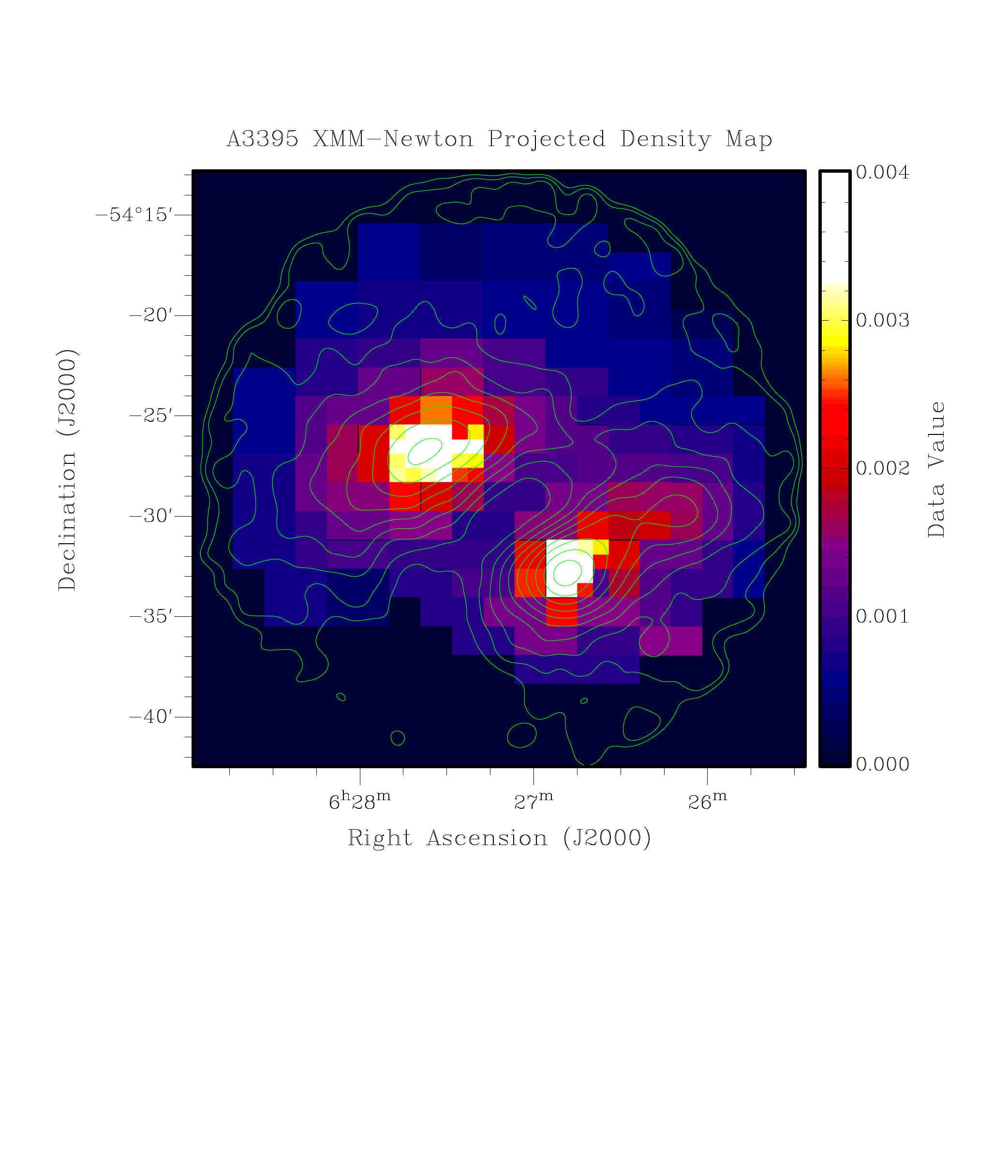}
\includegraphics[width=2.5in,height=3.2in]{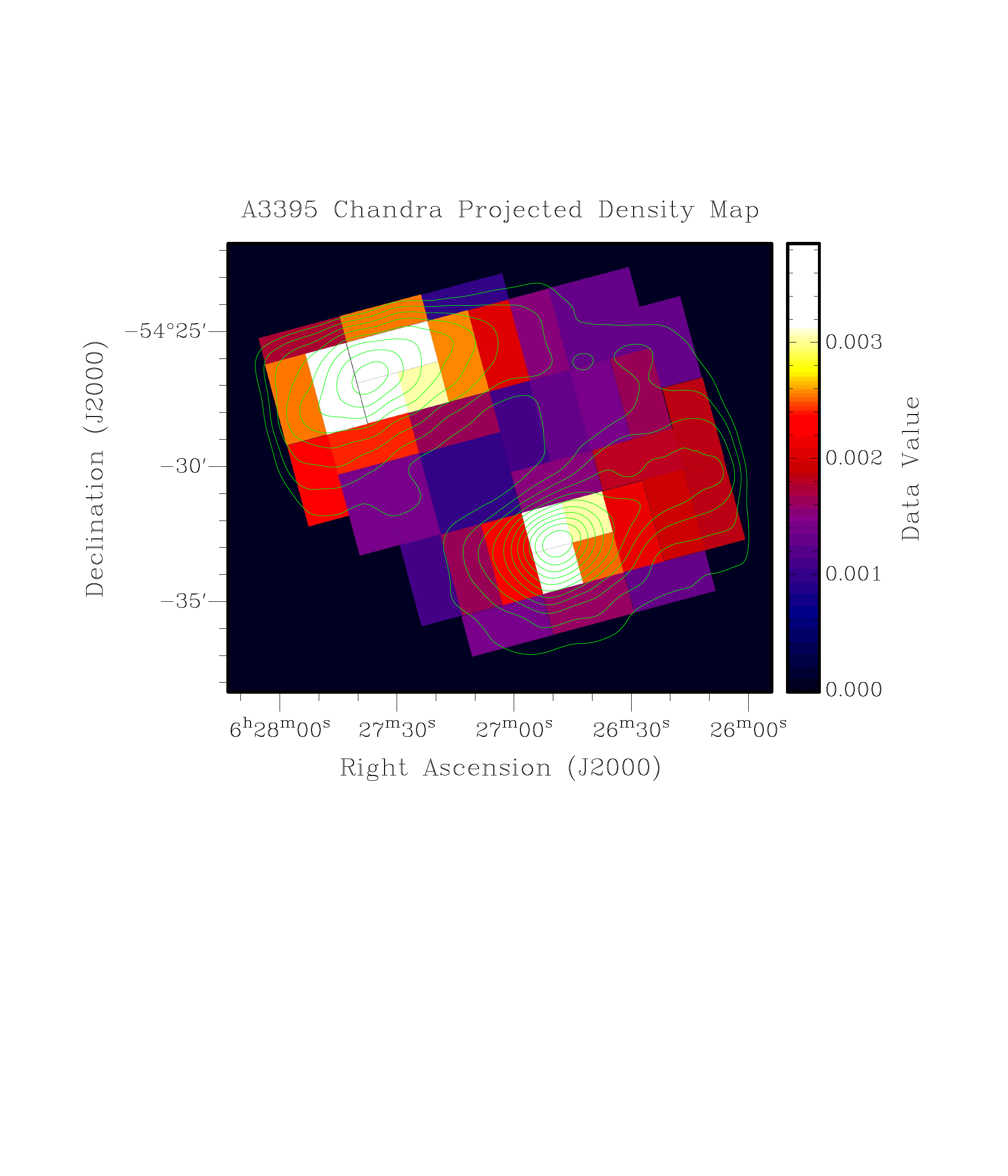}  
\caption{Projected density maps from 139 box regions using XMM-Newton data (top) and 42 box regions using Chandra data (bottom) 
with overlaid X-ray surface-brightness contours having levels same as in Figs.~\ref{fig:Combined_MOS1_MOS2_smoothedimage} and 
\ref{fig:Chandra_smoothedimage} respectively. The scales are expressed in units of cm$^{-3}$ shown in the bars alongside. Details of 
the spectral fittings are provided in \S\ref{sec:box_thermodynamic_maps}.}
\label{fig:A3395_box_density_map}
\end{figure}

\begin{figure}[!ht]
\centering
\includegraphics[width=3.0in,height=3.5in]{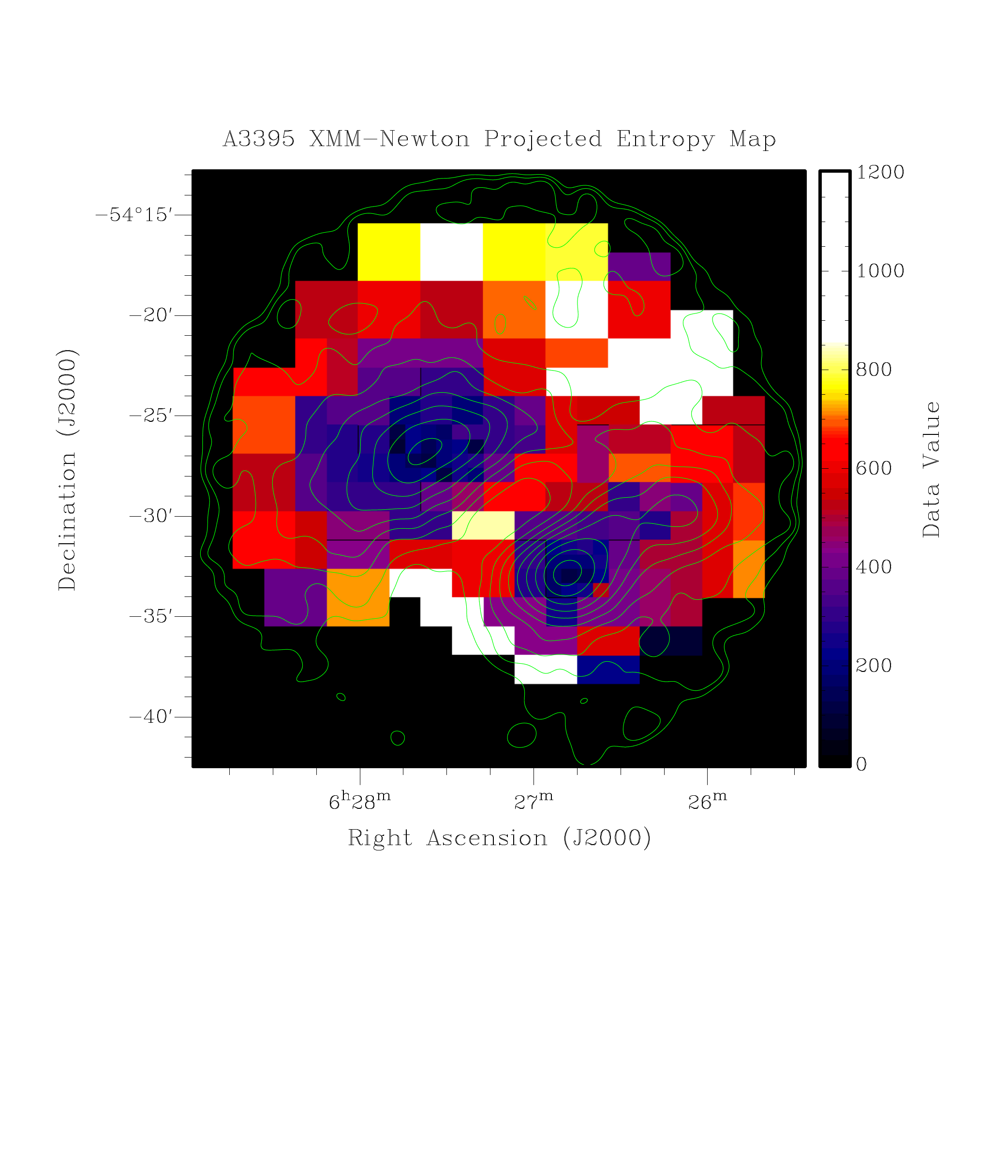}
\includegraphics[width=2.5in,height=3.2in]{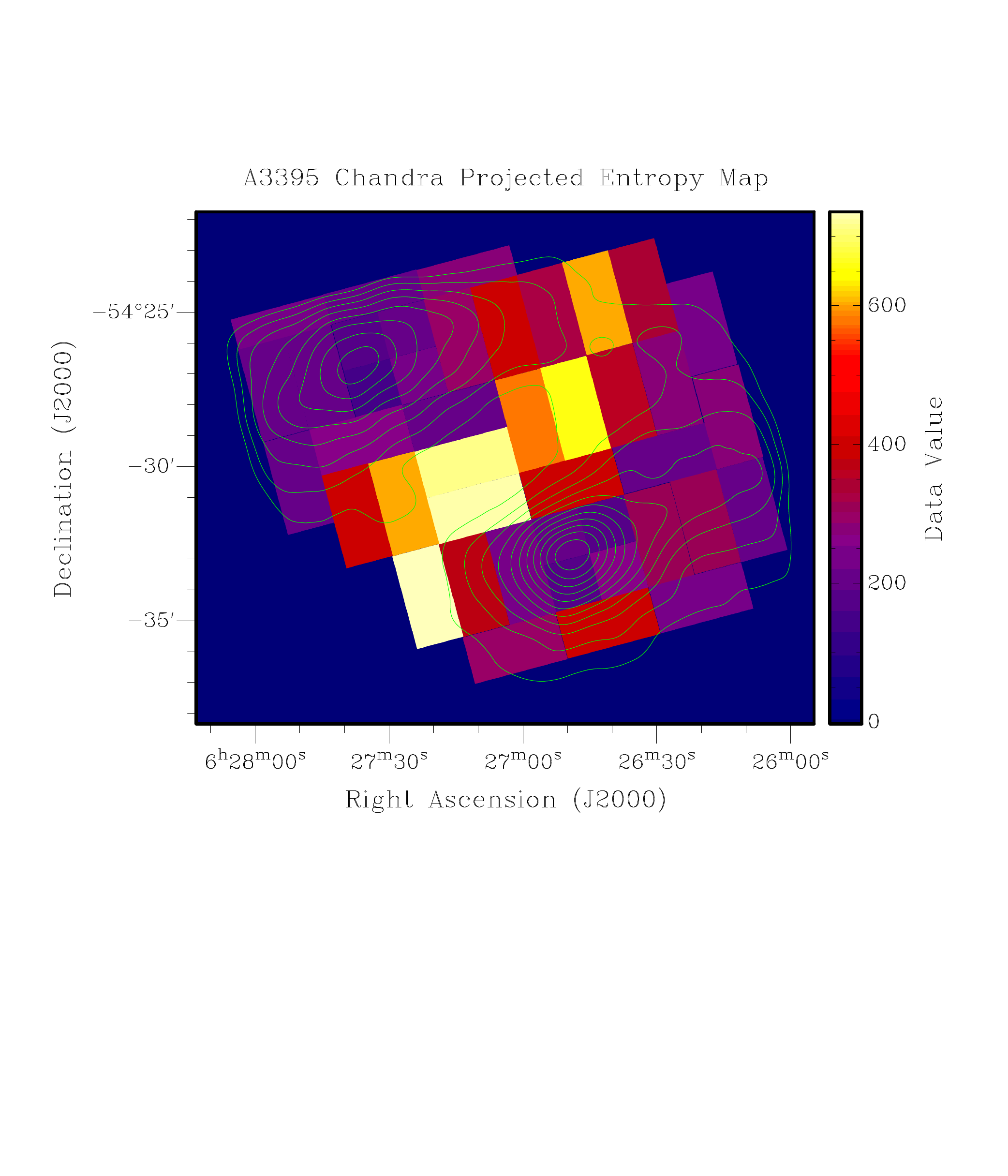}  
\caption{Projected entropy maps from 139 box regions using XMM-Newton data (top) and 42 box regions using Chandra data (bottom) 
with overlaid X-ray surface-brightness contours having levels same as in Figs.~\ref{fig:Combined_MOS1_MOS2_smoothedimage} and 
\ref{fig:Chandra_smoothedimage} respectively. The scales are expressed in units of keV cm$^{2}$ shown in the bars alongside. Details 
of the spectral fittings are provided in \S\ref{sec:box_thermodynamic_maps}.}
\label{fig:A3395_box_entropy_map}
\end{figure}

\begin{figure}[!ht]
\centering
\includegraphics[width=3.0in,height=3.5in]{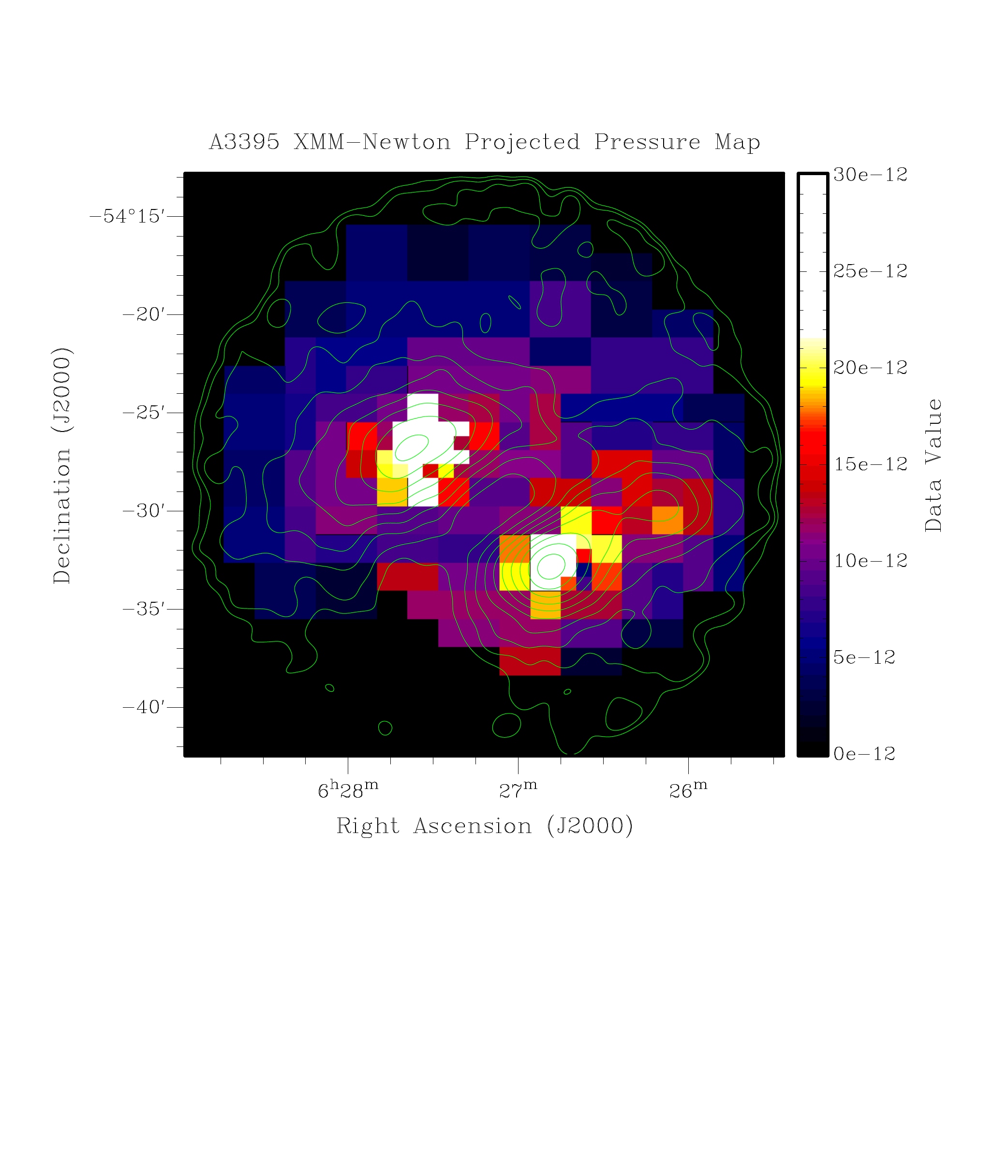}
\includegraphics[width=2.5in,height=3.2in]{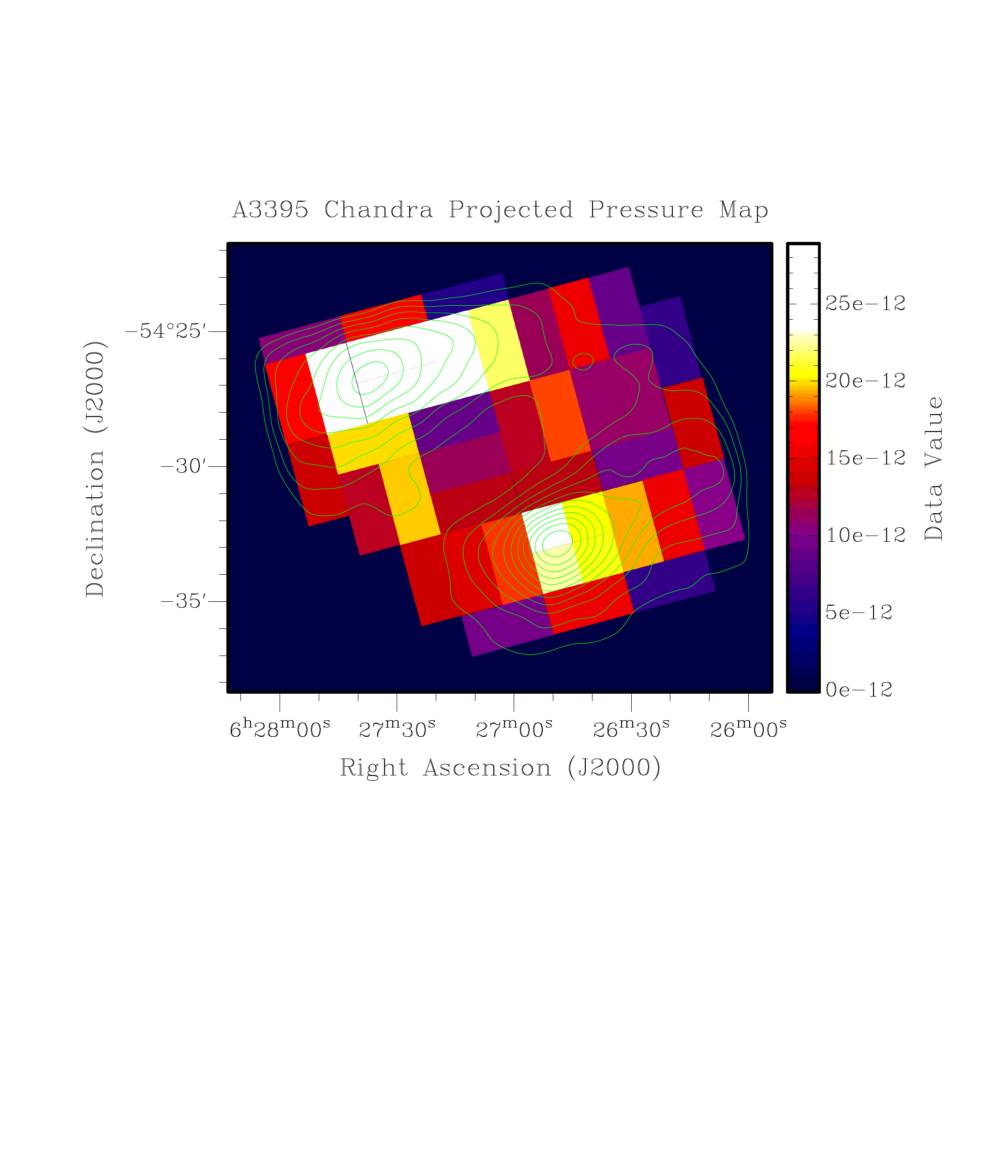}  
\caption{Projected pressure maps from 139 box regions using XMM-Newton data (top) and 42 box regions using Chandra data (bottom) 
with overlaid X-ray surface-brightness contours having levels same as in Figs.~\ref{fig:Combined_MOS1_MOS2_smoothedimage} and 
\ref{fig:Chandra_smoothedimage} respectively. The scales are expressed in units of erg cm$^{-3}$ shown in the bars alongside. 
Details of the spectral fittings are provided in \S\ref{sec:box_thermodynamic_maps}.}
\label{fig:A3395_box_pressure_map}
\end{figure}

  The temperature, density, entropy, and pressure maps produced (from both XMM-Newton and Chandra) are shown in 
Figures~\ref{fig:A3395_box_temp_map}, ~\ref{fig:A3395_box_density_map}, ~\ref{fig:A3395_box_entropy_map}, and 
~\ref{fig:A3395_box_pressure_map} respectively. Maps obtained from Chandra had larger errors as compared to those from XMM-Newton, 
and in all the maps fainter regions had larger errors. The errors range from as low as $\sim$15\% for the innermost regions of NE and
 SW subclusters to as high as $\sim$60\% for the extreme outermost box regions and the highest temperature regions in the NW and 
towards the southeastern edge of the SW. Abundance values had large errors in the maps obtained from both XMM-Newton and Chandra and 
for most of the boxes only upper limit of abundance could be obtained, hence the abundance maps have not been shown. Maps from
 both XMM-Newton and Chandra show a high temperature and high entropy region where the NE and SW regions meet. The SW, on an average 
shows a temperature slightly higher than the average temperature of the NE and even more higher temperature regions are seen in the W 
component, the region joining the NE and SW parts and the filament regions. The NE and SW components have similar average entropies 
and the W component has slightly higher average entropy. The filament and the region between the NE and SW components have even 
higher entropies. The highest temperatures and entropies are seen at the south-eastern edge of the SW subcluster and in a few regions 
in the NW component lying above the filament, to the west of the NE subcluster. Although, the single \textbf{apec} model provided a 
good fit for the source spectra of all the regions, for some of the regions (for e.g., the region between the NE and SW) they could 
also be very well fitted using powerlaw model (in addition to the fixed index powerlaws used for modeling the residual soft proton 
contamination), thus pointing towards a completely non-thermal emission from shock accelerated particles. None of the two models 
seemed to be statistically preferred. Fitting the spectra with \textbf{apec+power-law} or two \textbf{apec} models resulted in the 
values of the free parameters being insensitive to the fits. In addition, the normalizations of the non-thermal (power-law) components
 in the \textbf{apec+power-law} fits were found to be negligible as compared to those of the thermal (apec) components.

\subsection{Radio sources}
\begin{figure}
\centering
\subfigure[]
{
\vspace{3.0cm}
\includegraphics[width=2.5in]{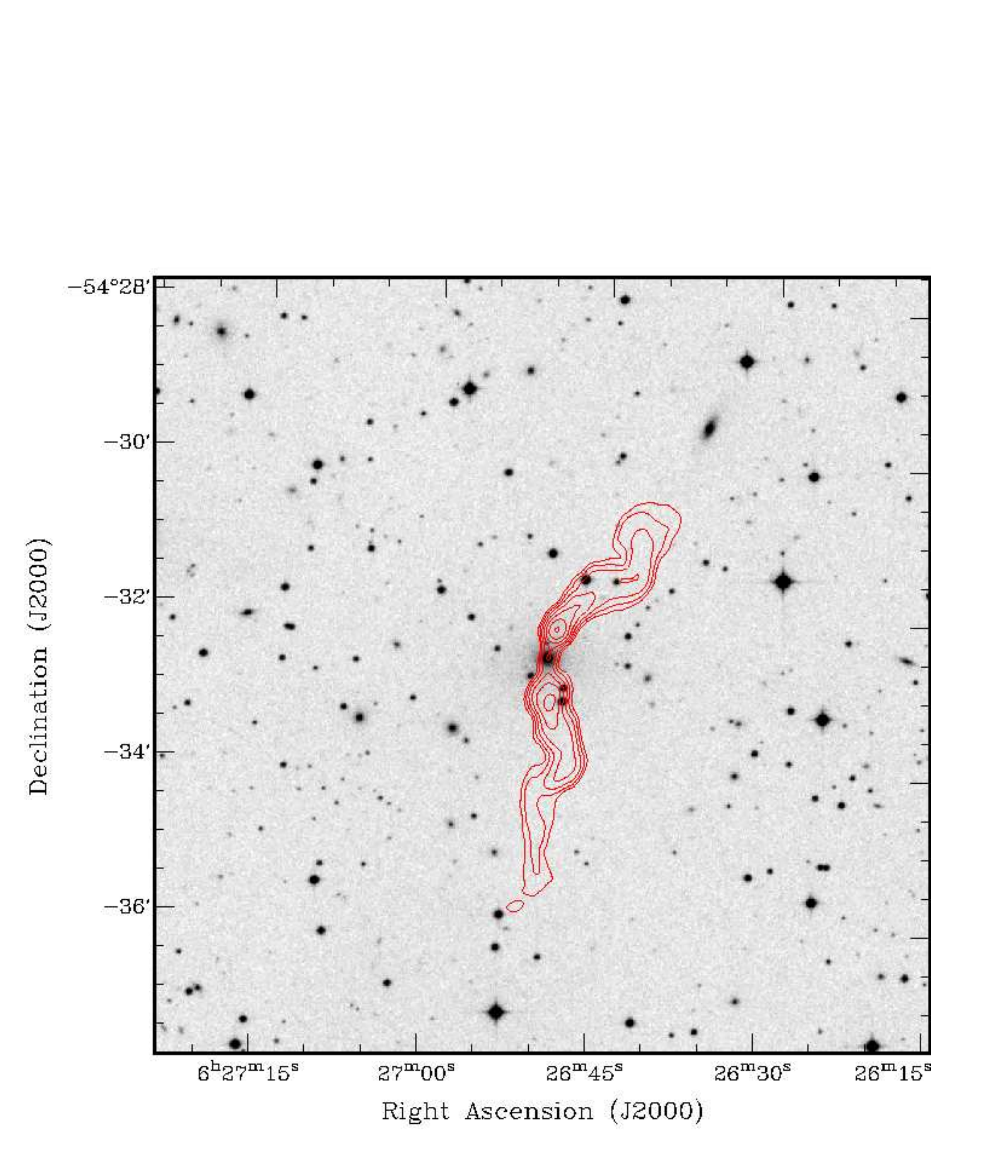}
\label{fig:A3395_WAT_SuperCOSMOS_overlaid_ATCA_radio_contours}
}
\subfigure[]
{
\includegraphics[width=3.3in]{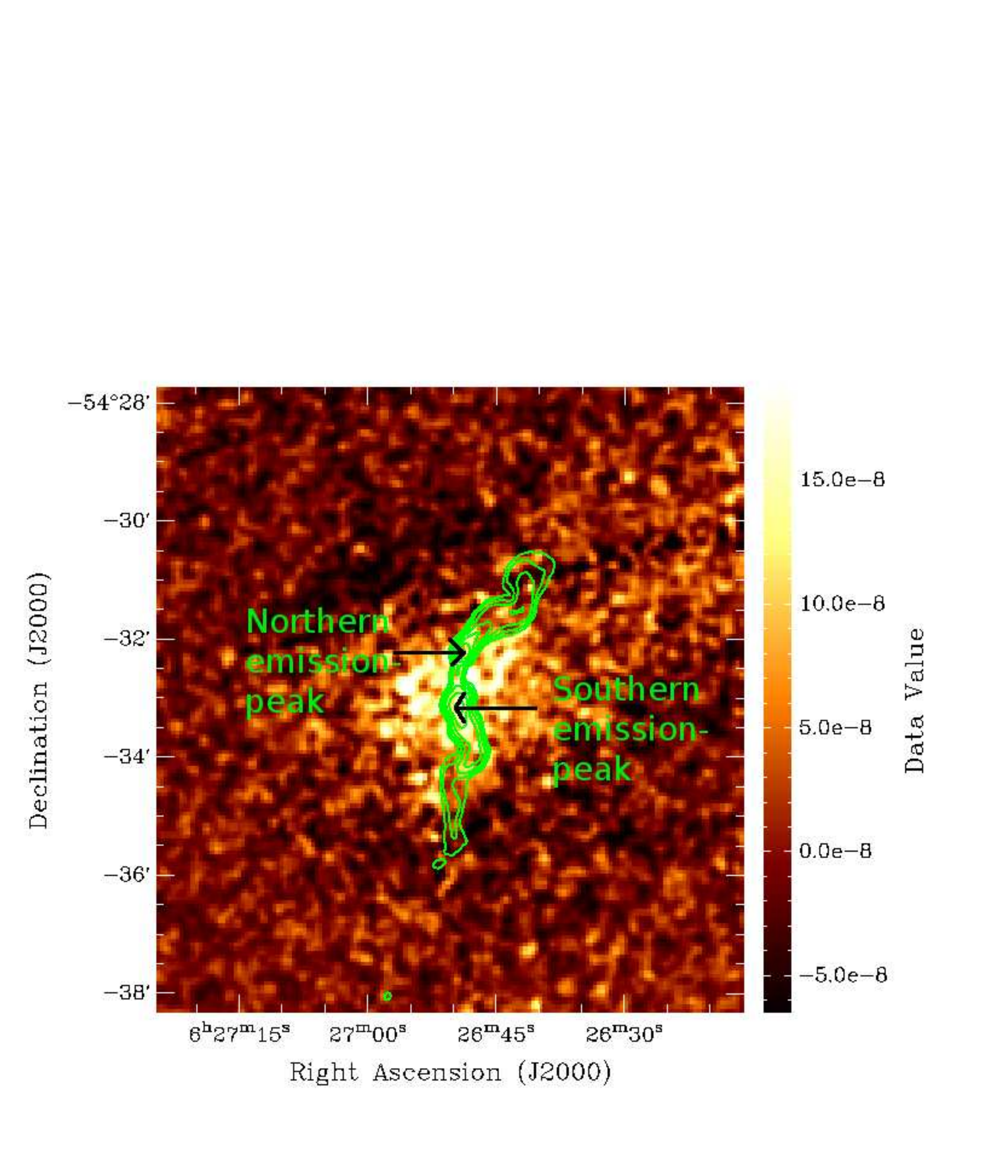}
\label{fig:A3395_WAT_unsharp_mask_Chandra_overlaid_ATCA_radio_contours}
}
\caption{\ref{fig:A3395_WAT_SuperCOSMOS_overlaid_ATCA_radio_contours} : Optical image from the SuperCOSMOS survey in the $B_J$ band 
of the region surrounding the WAT source in A3395 SW overlaid by ATCA 1348 MHz radio continuum contours. Contour levels are at 0.005 
times 1, 2, 4, 8, 16, and 32 Jy beam$^{-1}$. 
\ref{fig:A3395_WAT_unsharp_mask_Chandra_overlaid_ATCA_radio_contours} : Chandra unsharp-masked image produced by subtracting a large 
scale (80$^{\prime\prime}$) smoothed image from a small scale (4$^{\prime\prime}$) smoothed image of the same region as 
\ref{fig:A3395_WAT_SuperCOSMOS_overlaid_ATCA_radio_contours} overlaid by the same ATCA 1348 MHz radio continuum contours (green). 
Positions of the northern and southern emission peaks have also been shown. 
The colour scale shown as a bar is expressed in units of counts s$^{-1}$arcsec$^{-2}$.}
\label{fig:A3395_WAT_SuperCOSMOS_and_unsharp_mask_Chandra_overlaid_ATCA_radio_contours}
\end{figure}

  The ATCA images at 1348 and 2374 MHz show that the most prominent source in the field is the wide-angle tailed (WAT) source PKS 
B0625$-$545 (MRC B0625$-$545) which is identified with an elliptical galaxy at a redshift of 0.05174 (Teague, Carter, \& Gray 1990) 
(Figure~\ref{fig:A3395_WAT_SuperCOSMOS_overlaid_ATCA_radio_contours}). Both the oppositely-directed jets bend initially towards the 
west and then again bend sharply further down the flow at $\sim$90 kpc from the host galaxy so that the emission is approximately 
along the north-south axis. Douglass et al. (2008) have observed similar bends in the WAT source in Abell 562 and have interpreted the
 bends further down the flow as the regions  where the effects of buoyancy become important. Figure 
\ref{fig:A3395_WAT_unsharp_mask_Chandra_overlaid_ATCA_radio_contours} shows the unsharp-masked Chandra image of the cluster (zoomed in
 to show the WAT source) produced by subtracting a large scale (80$^{\prime\prime}$) smoothed image from a small scale 
(4$^{\prime\prime}$) smoothed image, overlaid with the 1348 MHz radio continuum contours. 
It may be worth noting that there is no evidence of X-ray cavities associated with 
the regions of radio emission. 

\begin{figure}
\centering
\includegraphics[height=5.0in]{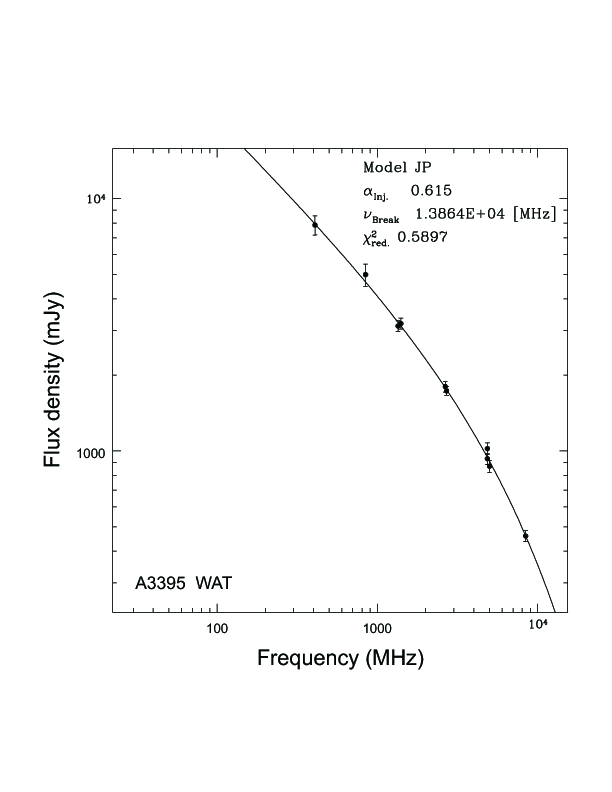}
\caption{The radio spectrum of the complete WAT source in A3395 SW, as described in the text. The flux density 
measurements have been fitted for the \citep{1973A&A....26..423J} model using the {\tt SYNAGE} package (\citep{1999A&A...345..769M}). }
\label{A3395_radio_spectrum}
\end{figure}

  The flux densities of the WAT galaxy at a number of frequencies from low-resolution observations are listed in Table 
\ref{tab:radio_flux}, including the values listed by Burgess \& Hunstead (2006), and estimates from the ATCA images presented in this 
paper. The flux density estimated from the ATCA image at 1348 MHz is very close to the measurement at 1410 MHz, suggesting that the 
interferometric observations at this frequency have not missed any significant amount of flux density. However the expected flux 
density at 2374 MHz from the derived spectrum using the high-frequency points ($>$800 MHz) is higher than our measured value by 
$\sim20\%$. Although the high-frequency points are consistent with a straight spectrum with a spectral index of $\sim$1.05, a 
good fit to the spectrum is obtained for the Jaffe \& Perola (1973) model using the {\tt SYNAGE} package (Murgia et al. 1999). 
Excluding the ATCA estimate at 2374 MHz where some flux density is missing, the spectrum is well fitted with an injection spectral 
index of 0.62 and a break frequency of $\sim$13.9 GHz (Figure \ref{A3395_radio_spectrum}). Higher frequency observations would be 
useful to determine this more reliably.
    
\begin{figure}
\centering
\includegraphics[width=7.0in]{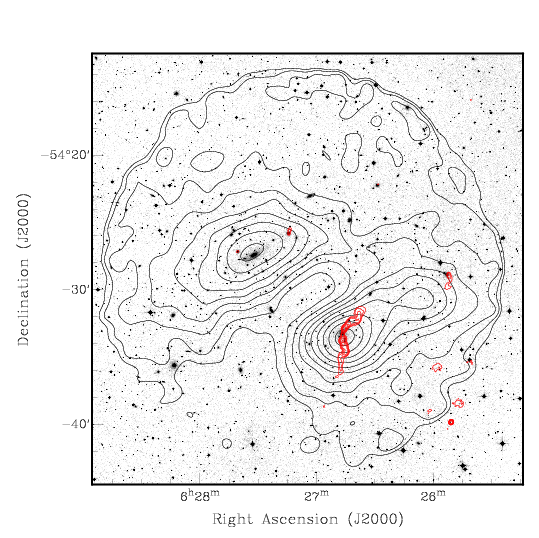}
\caption{A3395 image from the SuperCOSMOS survey in the $B_J$ band with overlaid X-ray contours from 
Figure~\ref{fig:Combined_MOS1_MOS2_smoothedimage} (shown with black) and ATCA 1348 MHz radio continuum contours 
(shown with red; levels are at 0.005 times 1, 2, 4, 8, 16, and 32 Jy beam$^{-1}$).}
\label{fig:A3395_SuperCOSMOS_optical_image}
\end{figure}

  Besides the WAT galaxy, there is a head-tail (HT) source located at RA(J2000): $06^{\rm h}25^{\rm m}56.7^{\rm s} $ and Dec.(J2000): 
$-54^{\rm d} 27^{\prime} 50^{\prime\prime}$ which is associated with a galaxy at a redshift of 0.05955 (Teague, Carter, \& Gray 1990) and is shown
 in Figure~\ref{fig:A3395_SuperCOSMOS_optical_image}. The extent of the HT galaxy is smaller than the WAT, with an angular size of
 $\sim80^{\prime\prime}$ including the diffuse emission, which corresponds to a linear size of $\sim$90 kpc at the redshift of the 
galaxy. The peak and total flux densities of the HT source at 1348 MHz are 25 mJy beam$^{-1}$  and $\sim$150 mJy. The tail is not as 
well imaged at 2374 MHz  where the peak and total flux densities are 8 mJy beam$^{-1}$ and $\sim$70 mJy respectively. MOST 
observations show that the peak and total flux densities of the HT source are 135 mJy beam$^{-1}$ and  299 mJy respectively at 843 MHz
 (Reid 2000). The HT galaxy is located at the periphery of the W region of A3395. HT galaxies are often detected in non-relaxed 
cluster but can also be found in the periphery of clusters and near regions of enhanced X-ray emission (cf. Klamer, Subrahmanyan \& 
Hunstead 2004; Mao et al. 2009, and references therein). 

  In addition, Reid (2000) lists two radio sources towards the NE region, namely J0627$-$5425 (RA (J2000): $ 06^{\rm h} 27^{\rm m} 
18.8^{\rm s} $, Dec.(J2000): $-54^{\rm d} 25^{\prime} 10^{\prime\prime}$) identified with a dumbbell system at a redshift of 0.04512 (Teague, 
Carter \& Gray 1990), and J0627$-$5426 (RA (J2000): $ 06^{\rm h} 27^{\rm m} 44.9^{\rm s} $ and Dec (J2000): $-54^{\rm d} 26^{\prime} 
46^{\prime\prime}$), associated with a galaxy at a redshift of 0.04348. Both sources have been shown
 in Figure~\ref{fig:A3395_SuperCOSMOS_optical_image}. The peak and total flux densities estimated from the ATCA image at 1348 
MHz are 14 mJy beam$^{-1}$ and 46 mJy respectively for J0627$-$5425. The corresponding values for J0627$-$5426 are 6.9 mJy 
beam$^{-1}$ and 17 mJy respectively. The redshifts of these sources are much below the average redshifts 
(z=$\overline{\rm v}/c$) of the main components of A3395, hence these sources are most probably not cluster members.

\subsection{X-ray Luminosity estimates}
\label{sec:luminosity_estimates}
  X-ray luminosities for the NE, SW, W, NW and the filament regions using XMM-Newton and for the NE, SW, and W and the filament 
regions using Chandra in the energy range 0.5-9.0 keV were estimated from the flux values obtained from the spectral analysis of these
 regions described in \S\ref{sec:Total-Spec-Analy}. The fluxes ($\rm F_{\rm X}$) were estimated by convolving the model used in 
\S\ref{sec:Total-Spec-Analy} with the \textbf{Xspec} convolution model, \textbf{cflux} after freezing the \textbf{apec} normalization.
 The X-ray luminosities ($\rm L_{\rm X}$) were then obtained from the fluxes using the formula: 
\begin{equation}
\rm L_{\rm X} = 4 \pi \rm D_{\rm L}^{2} \rm F_{\rm X}
\end{equation}
where $\rm D_{\rm L}$ is the luminosity distance to the source. 
 The luminosities (L$_{\rm x}$), derived from the flux values obtained from the spectral analysis done using XMM-Newton and 
Chandra data are given in Tables~\ref{tab:XMM_A3395_total_regions_spectral_results} and 
~\ref{tab:Chandra_A3395_total_regions_spectral_results} respectively. We also obtained the bolometric luminosities of the NE, SW, W and 
NW components. For this purpose, we fitted the $0^{\circ}-360^{\circ}$ surface brightness profiles of the NE, SW and W subclusters using
 $\beta$-profiles and the average count rates were obtained from the same regions as in \S\ref{sec:Total-Spec-Analy}. For the NW 
component the average count rate was obtained from an elliptical region (an approximation to the polygon shaped region used for the 
NW in section \S\ref{sec:Total-Spec-Analy})  with semi-major and semi-minor axis lengths of $11.11^{\prime}$ and $4.45^{\prime}$ respectively. 
By using the HEASARC tool \textbf{ Web Portable, Interactive, Multi-Mission Simulator (WebPIMMS)}, these count rates were converted 
to fluxes in the energy range of 0.01-100 keV from which bolometric X-ray luminosities were obtained using the relation given above 
in this section. The results obtained from the $\beta$-model fitting along with the estimated X-ray bolometric luminosities of the 
NE, SW, W and NW components have been given in Table~\ref{tab:bol_xray_luminosity}. Xue and Wu (2000) obtained the L$_X$-kT relations 
based on 274 clusters and 66 groups as $\rm L_{X} = 10^{-0.032\pm0.065}T^{2.79\pm0.08}$ and $\rm L_{X} = 
10^{-0.27\pm0.05}T^{5.57\pm1.79}$ for clusters and groups of galaxies respectively, where L$_{X}$ is in units of 10$^{43}$ ergs 
s$^{-1}$ and T is in units of keV. On comparing the bolometric X-ray luminosities and temperatures of the NE, SW, W and NW 
components of A3395 with those obtained from these relations, we find that the NE, SW and W subclusters have bolometric X-ray 
luminosities within a factor of 1.5-2 of each other, and are close to the L$_X$-kT relation for the rich clusters obtained by Xue and 
Wu (2000). The temperatures of these regions are almost twice those of the rich clusters of similar luminosities. However, the NW 
component is sub-luminous by about an order of magnitude, as compared to the rich clusters of similar temperatures and is hotter as 
compared to the isolated groups of similar luminosities. From these observations, it seems highly probable that all the components of 
the cluster have been heated up due to the ongoing merger processes.
\subsection{Gas Mass}
\label{sec:gas_mass_estimates}
The projected gas densities obtained from \S\ref{sec:projection_analysis} for different annuli in the NE, SW, and W regions were 
fitted using a $\beta$-model i.e., 
\begin{equation}
\rm n_{\rm e}(\rm r)=\rm n_{\rm e}\left(0\right)\left(1+\frac{\rm r^{2}}{\rm r_{\rm c}^{2}}\right)^{(3/2)\rm \beta},
\end{equation}
 where $\rm n_{\rm e}(0)$ is the central density and $\rm r_{\rm c}$ is the 
core radius. The gas mass $\rm M_{\rm gas}(r)$ out to radii 0.5 Mpc and 1 Mpc for the 
NE, SW, and W regions were obtained by using the following formula (see Donnelly et al. 2001) : 
\begin{equation}
\rm M_{\rm gas}(\rm r)=4 \pi \rho_{0} \int_{0}^{\rm r} \rm s^{2} \left[1+ \left( \frac{\rm s}{\rm r_{\rm c}} \right)^{2} \right] ^{(3/2)\rm \beta} ds
\end{equation} 
where $\rho_{0}= \rm \mu \rm n_{\rm e}(0) \rm m_{\rm p}$ ; $\rm m_{\rm p}$ is the mass of a proton, and $\mu$= 0.609 is the average 
molecular weight for a fully ionized gas (Gu et al. 2010). The values of $\rm \beta$, $\rm r_{\rm c}$, 
$\rm \rho_{0}$, and $\rm M_{\rm gas }$ based on fitting the density profiles with the above model are listed in 
Table~\ref{tab:gas_mass_estimate}. The deprojected densities have not been used for determining the gas masses as they had 
very poor $\beta$-model fits, even after ignoring the outermost annuli densities, which were mostly an overestimate. The results, 
therefore, had large errors, especially for the W subcluster. A rough estimate for the gas mass of the NW component was also 
made by assuming a constant density in an oblate ellipsoid made from the ellipse used for the NW region in 
\S\ref{sec:luminosity_estimates}. The density was derived from the \textbf{apec} normalization obtained from the spectral analysis of 
NW region in \S\ref{sec:Total-Spec-Analy} using the relation given in \S\ref{sec:projection_analysis} and the gas mass was estimated 
to be $\sim(1.1\pm0.2)\times10^{12}$ M$_{\odot}$.

\subsection{Cooling time}
\label{sec:cool_time_estimate}
  Using the central gas temperatures ($\rm T_{\rm g}$) and densities (n) derived from the deprojection analysis in 
\S\ref{sec:deprojection_analysis} using XMM-Newton data and by using the equations from Sarazin (1988), we calculate the cooling 
times for the three regions NE, SW, and W of A3395, assuming them to be subclusters, as :
\begin{equation}
\rm t_{\rm cool}= 8.5 \times 10^{10} \rm yr \left[ \frac{\rm n}{10^{-3} \rm cm^{-3}}\right]^{-1} \left[\frac{\rm T_{\rm g}}{10^{8}K} \right]^{1/2}
\end{equation}
  The cooling times for the NE, SW, and W subclusters are $2.9\times 10^{10}$y, $2.7\times 10^{10}$y, and 
$4.3\times 10^{10}$y respectively, which are much longer than the Hubble time. This shows that none of the subclusters 
has any ongoing cooling flow in it. 

\section{Discussion}
\label{sec:discussion}
  The present X-ray observations of A3395 show that the cluster morphology is much more complex than the simple bimodal 
structure reported earlier. We identify four distinct regions of strong diffuse X-ray emission, namely, the NE, SW, W, and the 
filament connecting the W to the NE part of the cluster. In addition, a fifth component A3395 NW is seen as a weak excess emission in 
the northwest of the cluster, which was also detected by Tittley \& Henriksen (2001) and is most probably a part of the supercluster 
filament connecting the clusters A3395 and A3391. Because of a larger FOV, better sensitivity, and larger energy band coverage of the
 XMM-Newton, we have been able to better constrain the values of the X-ray temperature and luminosities for the NE and SW subclusters,
 as compared to their values from earlier observations that used ROSAT and ASCA data. The temperatures determined by us 
for the NE ($4.8\pm0.1$ keV) and SW ($5.1\pm0.1$ keV) subclusters are very close (within uncertainties) to the values reported earlier 
by Markevitch et al. (1998) ($\sim$5.8$\pm$0.8 keV for the NE and $\sim5.5\pm0.8$ keV for the SW subcluster) based on ASCA data. 
We have estimated the X-ray luminosities of all the components of the cluster in the energy range of 0.5-9.0 keV and the 
values are given in Tables~\ref{tab:XMM_A3395_total_regions_spectral_results} and 
\ref{tab:Chandra_A3395_total_regions_spectral_results}.  A3395 NE is clearly the dominant subcluster with the highest X-ray luminosity
 and extent of emission. Our estimates of the bolometric X-ray luminosity for the NE subcluster is equal to, while that for the SW 
subcluster is slightly higher than, those from Donnelly et al. 
(2001), who estimated the bolometric X-ray luminosity within 1 Mpc of the NE and SW subcluster cores based on the surface 
brightness profiles using ROSAT data. In addition to the previously identified SW subcluster, there are strong indications for a 
separate clump W in the west that is either a clump of gas stripped-off the SW subcluster or is probably a separate subcluster 
merging with the A3395 SW. This W region is particularly delineated in the unsharp-masked MOS2 image shown in 
Fig.~\ref{fig:A3395_unsharp_mask_comb_MOS1_MOS2_and_tot_spec_regions}. Note that, in contrast to Donnelly et al. (2001), our estimate 
of the luminosity for the SW subcluster does not include the W region.

  We now examine whether the W clump of X-ray emission represents a smaller group of galaxies in the process of merging in A3395.

\subsection{Subclustering Analysis}
\begin{figure}
\centering
\includegraphics[width=7.0in,height=4.2in]{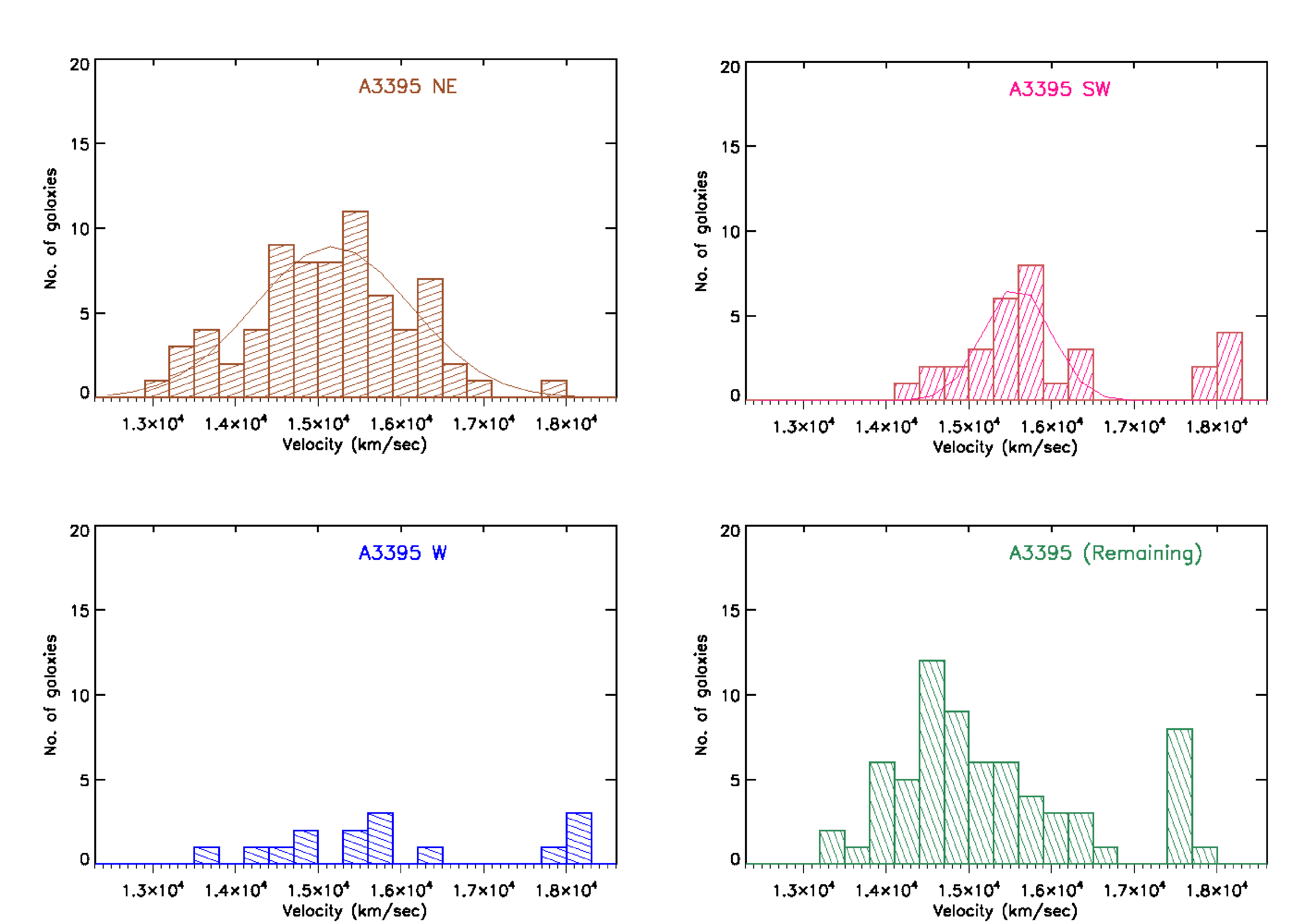}
\caption{Velocity histograms of the galaxies belonging to the NE, SW, W subclusters, and the remaining set of galaxies that did not
 belong to any of the subclusters. Bins are 300 km s$^{-1}$ wide. Overlaid are the Gaussian fits to the velocity 
distribution of the NE and SW subclusters.}
\label{fig:A3395_gala_vel_histogram}
\end{figure}

 A3395 W appears to be a distinct region of X-ray emission (Figs.~\ref{fig:Combined_MOS1_MOS2_smoothedimage} 
to~\ref{fig:A3395_unsharp_mask_comb_MOS1_MOS2_and_tot_spec_regions}). The detection of an HT source at the periphery of the W 
subcluster, which is indicative of enhanced X-ray emission and a deep potential well, points towards a distinct atmosphere of the W 
subcluster. The surface brightness profile of the W subcluster (Fig. \ref{fig:W_surface_brightness_profiles_full}) which shows a peak at the 
centre, also strengthens the idea of a distinct identity of the W subcluster. We have also re-examined the galaxy velocity sample for 
A3395 from Donnelly et al. 2001, which had velocity information for 157 cluster member galaxies to look for dynamical evidence for 
A3395 W being a separate group. We divided the galaxies into the three subclusters by considering ellipses with major axes 0.4, 0.3, 
and 0.2 times $R_{180}$ (=34.6$^{\prime}$) for the NE, SW, and W subclusters respectively and counting the number of galaxies in these regions. 
The major-axis position angles and ellipticities for the ellipses in NE, SW, and W regions were the same as given in section 
\S\ref{X-ray_morphology}. This led to 71 galaxies in the NE,  32 galaxies in the SW, and 15 galaxies in the W subcluster, leaving 69 
galaxies which could not be clearly identified as belonging to any particular subcluster. Also, there were some slight overlaps in the
 three sets of galaxies. For example, NE and SW subclusters had 13 galaxies in common ; NE and W subclusters had 9 galaxies in common,
 and SW and W subclusters had 12 galaxies in common. The velocity histograms obtained for the four sets are shown in Figure 
\ref{fig:A3395_gala_vel_histogram}, where the  bin-size was chosen to be 300 km $\mathrm{s}^{-1}$ wide. Donnelly et al. (2001) had 
reported a high velocity hump in A3395. From Fig. \ref{fig:A3395_gala_vel_histogram}, we observe that the galaxies in this hump belong
 to either the SW or the W parts or to none of the subclusters. The average subcluster velocities and their dispersions were obtained 
by fitting Gaussians to their velocity distributions and are given in Table \ref{tab:A3395_gala_vel_distbn_results}. The velocity data
 in the W region shows a very flat and wide distribution and is not sufficient for fitting a Gaussian. In addition, the galaxies in 
the W subcluster seem to follow the velocity distribution of the SW subcluster.

  Though the velocity distribution of the W subcluster had a very poorly fitted Gaussian, we have force-fitted a Gaussian to 
it for getting an estimate for the velocity dispersion and a lower limit of the virial mass used in the bound system analysis in 
\S\ref{sec:bound_sys_anlys}. We estimated the virial mass for each of the subclusters assuming that the galaxies included in each 
subcluster are bound and the velocity dispersions are isotropic. We used the following formula (see Beers, Geller, \& Huchra (1982)):
\begin{equation}
 \rm M_{\rm virial} = \frac{3 \pi}{\rm G} \sigma_{\rm r}^{2} \left\langle \frac{1}{\rm r_{\rm p}} \right\rangle^{-1} 
\end{equation}
where $\sigma_{\rm r}$ is the velocity dispersion along the line of sight and $\langle 1/\rm r_{\rm p}\rangle^{-1}$ is the harmonic 
mean projected separation between galaxy pairs. The mean velocity ($\bar{\rm v}$), velocity dispersion ($\sigma_{\rm v}$), and the
 virial masses of the three subclusters thus estimated, are given in Table \ref{tab:A3395_gala_vel_distbn_results}. Compared with 
Donnelly et al. (2001), our estimate of the virial mass for the NE subcluster ($=8.1\pm1.1\times10^{14}\rm M_{\odot}$) is slightly 
larger, while that for the SW subcluster ($=1.4\pm0.4\times10^{14}\rm M_{\odot}$) is almost equal (see \S\ref{sec:A3395}) within 
uncertainties. A possible reason for a higher mass estimate for the NE region than Donnelly et al. is that our area for including 
subcluster member galaxies in the NE region ($\sim1$Mpc$^{2}$) is $\sim$25\% larger than the area used by them ($\sim0.8$Mpc$^{2}$), 
which leads to a larger velocity dispersion and harmonic mean distance between galaxy pairs.

  Application of wavelet transform techniques to the positions of galaxies in A3395 by Flin \& Krywult (2006) led to the 
detection of  two subclusters which most probably correspond to the bimodal structure of the cluster at a scale of $\sim$527 kpc. It 
should be noted that if the W component is indeed a possible subcluster then its length scale is only $\sim$250 kpc. However, 
considering the velocity distribution of galaxies in the W region, the W component is most probably either a clump that has been 
stripped-off the SW subcluster by the ram pressure or a subcluster that has merged and relaxed into the SW subcluster (also see 
\S\ref{sec:disc_evid_fo_merg}).

\subsubsection{Bound System Analysis}
\label{sec:bound_sys_anlys}
\begin{figure}
\begin{center}$
\begin{array}{ccc}
\includegraphics[width=2.3in]{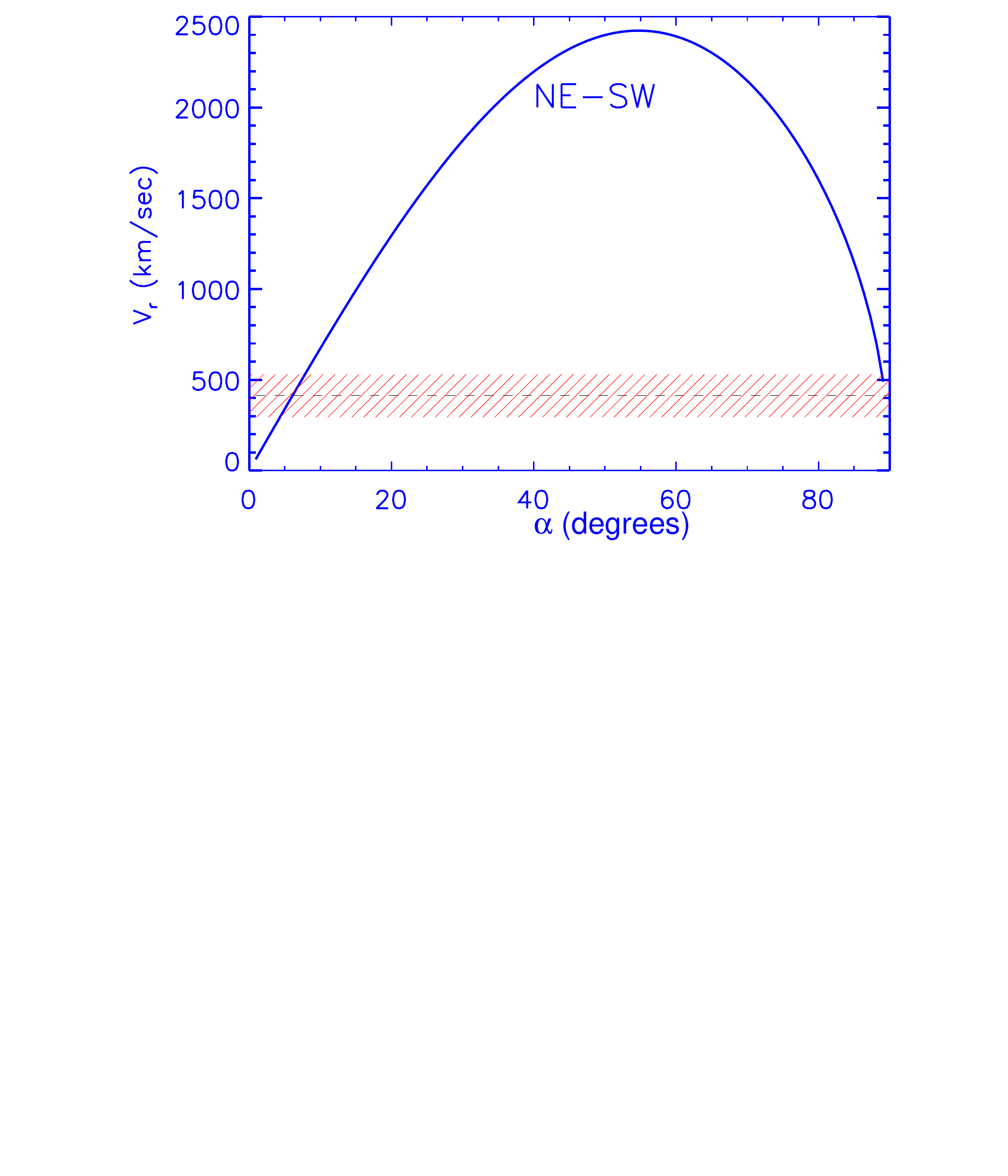} &
\includegraphics[width=2.3in]{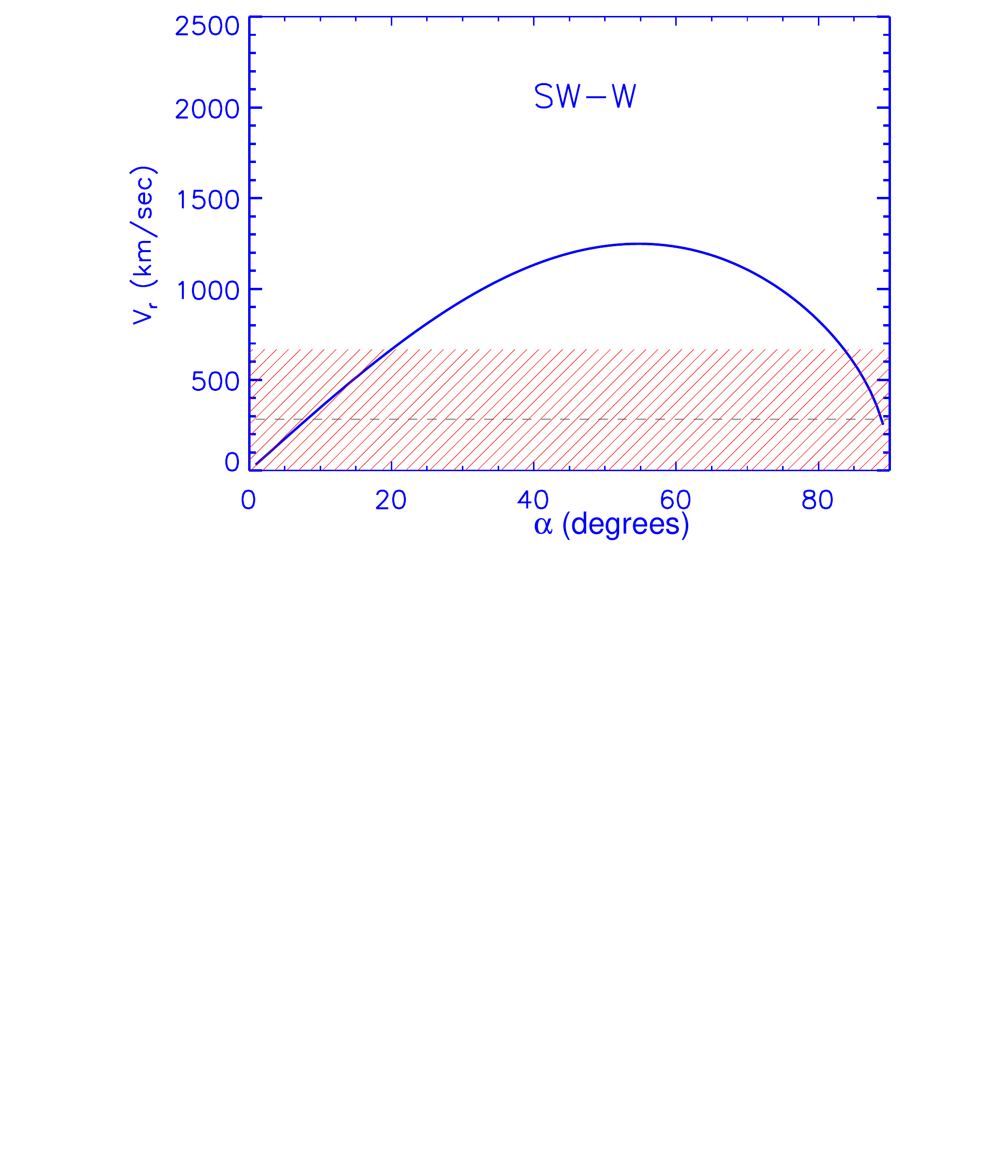} &
\includegraphics[width=2.3in]{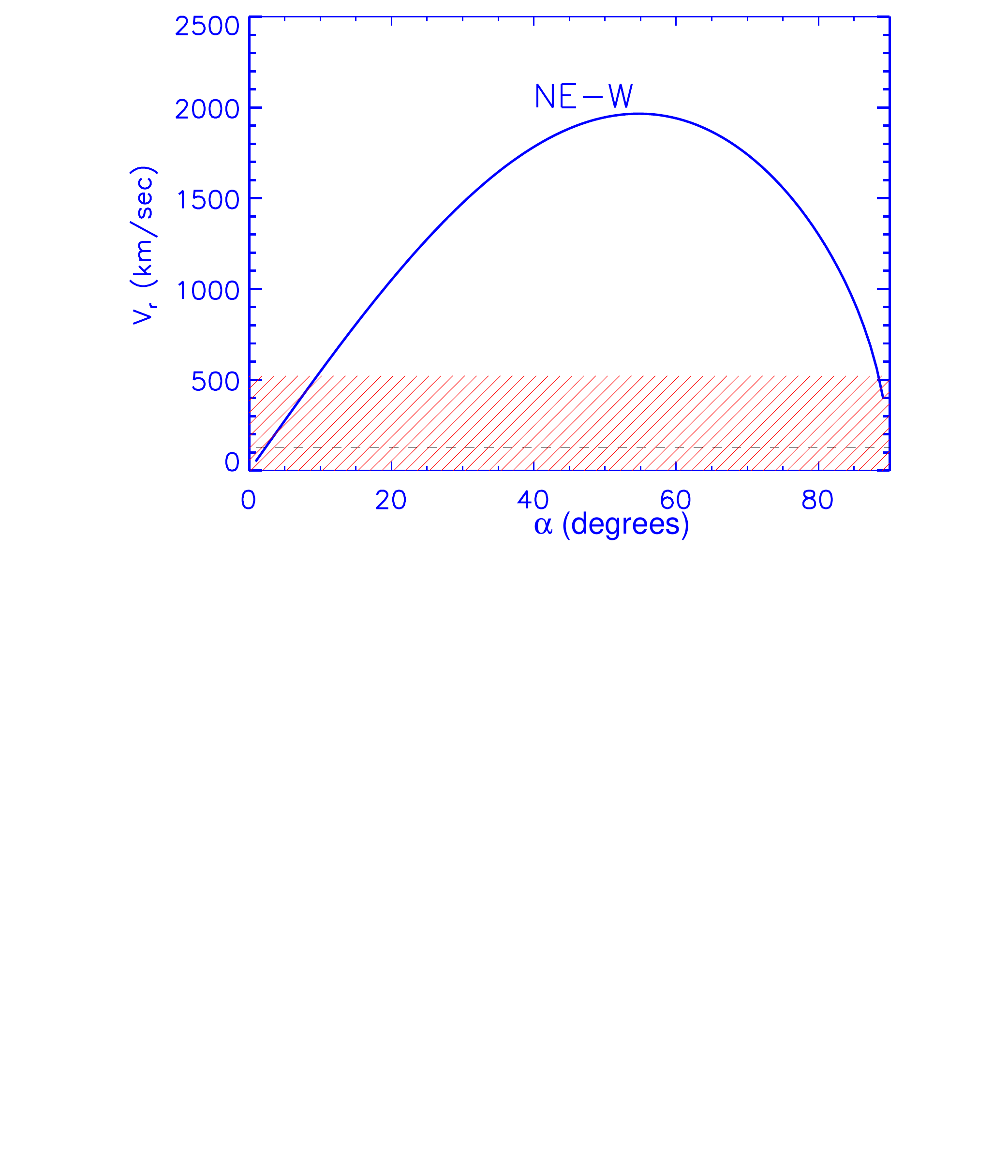}
\end{array}$
\end{center}
\caption{Plots of the Newtonian energy binding conditions as a function of measured relative velocity ($V_{r}$) and 
projection angle from the plane of the sky ($\alpha$). The hyperbolic curve is the limiting case for bound systems i.e., all 
orbit solution above it are unbound while those below it are bound. The horizontal dashed line indicates the measured 
relative velocities of the subcluster pair and the cross-hatching shows its 68\% confidence region.}
\label{fig:A3395_bnd_system_analysis}
\end{figure}

  In this section, we have tested whether the NE, SW, and W subclusters (pairwise) make a bound system or not. From simple Newtonian 
energy considerations for a bound system (see Donnelly et al. 2001):\\
\begin{equation}
\rm V_{\rm r}^{2} \rm R_{\rm p} \leq \rm 2GM \rm \sin^{2} \alpha \rm \cos \alpha
\end{equation}
where $\rm V_{\rm r}$ is the relative radial velocity between the two subclusters (the difference in mean velocities of the two 
subclusters from Table~\ref{tab:A3395_bnd_system_analysis}), $\rm R_{\rm p}$ is the projected separation between the centres of the 
two subclusters, M is the sum of the masses of the two subclusters (we used the lower limit of the sum of the two masses), and 
$\alpha$ is the projection angle from the plane of the sky. The values of $\rm V_{\rm r}$ and  $\rm R_{\rm p}$ for the NE-SW, SW-W, 
and NE-W subcluster pairs are given in Table~\ref{tab:A3395_bnd_system_analysis}. Plots are shown in 
Figure~\ref{fig:A3395_bnd_system_analysis}. The hyperbolic curve represents the quantity (2GM$\rm \sin^{2} \alpha \rm \cos \alpha/\rm 
R_{\rm p}$)$^{1/2}$ while the horizontal dashed line represents V$_{\rm r}$ (with its 68\% confidence region shown with the cross 
hatching). All orbit solutions below the hyperbolic curve are bound while those above it are unbound. Thus, all three pairs of 
subclusters are most probably bound systems.

\subsection{Evidence for Mergers}
\label{sec:disc_evid_fo_merg}
  Amongst the three regions, the NE region is the most luminous and most massive 
(Tables~\ref{tab:XMM_A3395_total_regions_spectral_results}, \ref{tab:Chandra_A3395_total_regions_spectral_results}, 
\ref{tab:gas_mass_estimate} and \ref{tab:A3395_gala_vel_distbn_results}). The region A3395 W is connected to A3395 NE via a 
filamentary structure which is at an average temperature of $6.5^{+0.8}_{-0.6}$ keV. The 2-D thermodynamic maps show a high 
temperature of $\sim7.5^{+2.2}_{-1.2}$ keV in a region between the NE and SW components along with a high entropy. As the surface 
brightness profiles of the NE and SW subclusters did not show any abrupt discontinuity, direct evidence for any shock heating could 
not be obtained. Possibly, the high temperature is not due to shock heating but due to viscous dissipation, which has also been 
invoked as the possible reason for heating in the merging binary cluster A115 by Gutierrez \& Krawczynski (2005). The temperature map 
also shows high temperature regions in the W component. A part of the NW component just above the filament shows the highest 
temperature and entropy in the 2-D thermodynamic maps. Also, from \S\ref{sec:luminosity_estimates}, we find indications of a 
possible heating of all the components of the cluster due to the ongoing mergers. Taking into account the facts that non-detection of
 the shock front can also be due to projection effects and that some of the high temperature regions could also be fitted by purely 
non-thermal models, the possibility of X-ray emission in these regions being purely non-thermal from shock accelerated particles can 
not be ignored.

  Two different merging scenarios for the cluster A3395 are possible. If the W region is indeed a separate subcluster, it is 
possible that the cluster is going through its first merger and SW and W subclusters are most probably falling and merging with the 
more massive and luminous NE subcluster. However, under this assumption, the origin of the filament joining the W to the NE subcluster
 can not be explained. Another possibility is that the two main subclusters NE and SW have already gone through their first 
merger and the filament and the W region are likely results of two different phases of ram-pressure stripping from the SW subcluster. 
The observed strong gradient in the surface-brightness profile of the SW subcluster along the southeast direction is possibly due to 
an orbital motion of the SW subcluster around the NE subcluster. Under this assumption, the outer hot and diffuse layers of the SW 
subcluster were perhaps stripped off by the cooler and denser gas of the NE subcluster during the first phase of ram-pressure 
stripping, resulting in the formation of the filament and the W region was stripped off in a later (more recent) phase of 
ram-pressure stripping where a relatively colder clump of the SW subcluster was stripped off the main subcluster by the hotter 
surrounding gas. Randall et al. (2008) have analyzed a similar merging scenario of the M86 galaxy with the ICM of the Virgo cluster 
in which a hotter tail and a colder plume were formed in two different phases of ram-pressure stripping.\\

  Clusters hosting a WAT with bent lobes are known to be the sites of ongoing mergers. Thus, the bent lobes of the WAT source in 
A3395 SW are consistent with its being a merging cluster. The WAT sources in merging clusters have been known to show some additional 
interesting characteristics, viz., the offset of the X-ray centroid from the position of the bright central galaxy hosting the WAT 
(Sakelliou \& Merrifield, 2000). In A3395 SW, we have found an offset of about $\sim 16$ kpc ($\sim$16.5$^{\prime\prime}$) between the WAT hosting 
BCG and the X-ray centroid. Merging clusters hosting a WAT also show an elongation of the ICM distribution along the line that bisects
 the WAT (G\'{o}mez et al. 1997), which we don't find in the ICM distribution of the SW subcluster. Following these observations we 
propose a scenario where the subclusters SW and W are falling into and merging with the subcluster A3395 NE, although the 
actual merger geometries seem to be much more complex.

\subsection{Radio and X-ray comparison}
\subsubsection{WAT Radio galaxy}
  The radio properties of the WAT radio galaxy have been well determined over a large frequency range. We estimate the equipartition 
magnetic field and the minimum energy densities in the northern and southern peaks of emission which are closest to the core of the 
WAT radio galaxy, by assuming a value of unity for the particle to electron energy density ratio and also for the filling factor of 
the relativistic plasma. The formulae used to calculate the equipartition magnetic field (B(U$_{\rm min})$) and the minimum energy 
densities (U$_{\rm min}$) are (see Moffet 1975, p. 211):
\begin{equation}
 \rm B(\rm U_{\rm min}) = 2.3 (\rm aAL/V)^{2/7}
\end{equation}
and
\begin{equation}
 \rm  U_{\rm min} = 0.5 (\rm aAL)^{4/7} V^{3/7}
\end{equation} respectively, where, 
\begin{equation}
\rm  A=\frac{\rm C_{1}^{1/2}}{\rm C_{2}} \frac{2 \alpha + 2}{2 \alpha + 1} \frac{\nu_{2}^{\alpha + 1/2} - \nu_{1}^{\alpha + 1/2}}{\nu_{2}^{\alpha + 1} - \nu_{1}^{\alpha + 1}}
\end{equation}
\begin{equation}
\rm  L=4 \pi \rm D_{\rm L}^{2} \rm S
\end{equation}
 $\rm C_{1} = 6.266 \times 10^{18}$ and $\rm C_{2} = 2.368 \times 10^{-3}$ in cgs units, L is the total luminosity calculated from 
the total flux S, (obtained by integrating the flux density ($\rm S_{\nu} \propto \nu^{-\alpha}$) from $\nu_{1}=10$MHz to 
$\nu_{2}=100$MHz), $\alpha$ is the radio spectral index, V is the volume of the source region used, a is the ratio of the total 
particle energy to the energy in the electrons (assumed to be 1), and $\rm D_{\rm L}$ is the luminosity distance to the source. 
The spectral index of the northern emission peak of the WAT is $\sim$0.7 and its deconvolved size has been estimated to be 
$\sim$16$\times$8 arcsec$^2$ from two-dimensional Gaussian fits using the AIPS task JMFIT. Assuming a cylindrical geometry, the 
equipartition magnetic field is $\sim$20 $\mu$G and the minimum energy density is $\sim$3.8$\times$10$^{-11}$ erg cm$^{-3}$, implying 
a pressure of $\sim$1.3$\times$10$^{-11}$ dynes cm$^{-2}$. For the southern feature, $\alpha$ is again $\sim$0.7, and for a 
deconvolved size of $\sim$26$\times$12 arcsec$^2$, the equipartition magnetic field is 14$\mu$G and the minimum energy density is 
$\sim$1.9$\times$10$^{-11}$ erg cm$^{-3}$, implying a pressure of 
$\sim$0.6$\times$10$^{-11}$ dynes cm$^{-2}$. The deprojected pressure near the centre of the SW subcluster is 
$\sim$4$\times$10$^{-11}$ dynes cm$^{-2}$. Although an estimate of the volume will also be affected by projection effects, a change in
 the volume by $\sim$20\% will affect the pressure by only $\sim$8\%. Although the X-ray pressure appears somewhat larger than the 
internal pressure of the emission peaks of radio-emitting plasma, a ratio of particle to electron energy of 50 will increase the 
energy density and pressure by a factor of $\sim$9.4. If this is the case the northern emission peak would be overpressured, while the
 southern emission peak would approximately be in pressure equilibrium with its environment. The pressure of the emission-peaks would 
also increase if the radiating particles extend to lower energies and hence lower frequencies than have been assumed here. 

  We also estimate an average value of the equipartition magnetic field and pressure for the whole source with similar assumptions. 
The total length of the WAT radio source along its ridge line is $\sim6^{\prime}$, and the average deconvolved width is 
$\sim16^{\prime\prime}$, although this varies significantly along the axis of the source. Here, the radio spectrum has been integrated
 from 10 MHz to 1 GHz using the injection spectral index of 0.62, and from 1 to 100 GHz using an estimated value of 1.05 for the 
high-frequency spectral index resulting from the steepening of spectra due to aging. A cylindrical geometry has been assumed. This 
yields an equipartition magnetic field of $\sim$6$\mu$G, which indicates a minimum energy density of 3$\times$10$^{-12}$ erg cm$^{-3}$
 and a minimum pressure of $\sim$10$^{-12}$ dynes cm$^{-2}$. Again, a contribution from heavier particles and/or integration to lower 
energies appears necessary to achieve pressure balance with the external environment where the deprojected pressure drops to 
$\sim$10$^{-11}$ dynes cm$^{-2}$ at a distance of $\sim3^{\prime}$. As suggested by O'Sullivan et al. (2010), a possible source of 
additional pressure could be provided by the gas entrained and heated by the jets, which are mostly found in FRI jets (Worrall 2009). 
For a field strength of 5.7 $\mu$G, a spectral break at $\sim$13.9 GHz as suggested by {\tt SYNAGE} (Murgia et al. 1999) fit for the 
Jaffe \& Perola (1973) model, the spectral age of the WAT is $\sim$10 Myr. 

  There appear to be no cavities in the X-ray maps near the location of the WAT radio source (Figure 
\ref{fig:A3395_WAT_unsharp_mask_Chandra_overlaid_ATCA_radio_contours}). This is either because the observations are not deep 
enough to detect the cavities or because the surrounding hot thermal plasma has had time to leak into and fill the cavities.

\subsubsection{HT galaxy}
  The properties of the HT source are less well determined compared with the WAT. Such HT sources are produced by galaxies moving at
 a high relative velocity into a high density of the ICM. The integrated flux densities discussed earlier suggest a rather steep 
spectral index of $\sim$1.4. Although there are considerable uncertainties in this value, the equipartition magnetic field with the 
same assumptions as those for the WAT is $\sim$4 $\mu$G, while the minimum energy density is $\sim$1.6$\times$10$^{-12}$ erg 
cm$^{-3}$, implying a pressure of $\sim$0.5$\times$10$^{-12}$ dynes cm$^{-2}$. This is again smaller than the deprojected pressure 
of the W subcluster where the pressure is $\sim$1.5$\times$10$^{-11}$ dynes cm$^{-2}$. The deprojected electron density in the central
 region of the W subcluster is close to 2$\times$10$^{-3}$ cm$^{-3}$, indicating a ram pressure, $\sim\rho \rm v_{\rm g}^2$ where 
$\rm v_{\rm g}$ is the velocity of the host galaxy relative to the ICM, of $\sim$3$\times$10$^{-11}$ dynes cm$^{-2}$.  Considering the
 galaxies in the W subcluster, the HT host galaxy is moving at $\sim$2000 km s$^{-1}$ relative to the median velocity 
of this group  (Figure \ref{fig:A3395_gala_vel_histogram}) as would be expected for the tails to be bent by ram pressure of the ICM. 
  
\section{Summary}
  X-ray observations of the cluster A3395 have revealed five main components in its X-ray morphology: the NE, SW, W subclusters, the 
NW region and a filament connecting the NE subcluster to the W subcluster. The surface brightness profiles of the various components 
of the subcluster do not show any shock fronts in the cluster. The 2-D thermodynamic maps of the cluster, however, provide evidence for
 high temperature regions at the interfaces of the various components of the cluster. The very high temperature and entropy regions in
 the NW component, which is most probably a part of the supercluster filament connecting the clusters A3395 and A3391, point to the 
merging environment in A3395 possibly being affected by a much larger supercluster network. The X-ray bolometric luminosities 
of the NE, SW and W components, and the NW component are similar to those of the rich clusters and the isolated groups of galaxies 
respectively. However, the temperatures of all the components of A3395 are unusually higher as compared to those of the rich 
clusters and groups of galaxies and suggest a possible heating of the whole cluster resulting from the ongoing mergers. None of the 
NE, SW, and W subclusters shows any cooling flows, as it is very likely that the cooling flows have been disrupted by the mergers in 
the cluster. The galaxy velocity distributions of the NE and SW subclusters have well defined Gaussian fits and hence are confirmed as
 well defined subclusters in the process of merging. Although the morphology, the surface brightness profile and the presence of an HT
 galaxy in the W region show it as a separate subcluster, the galaxy velocity distribution of A3395 W could not be fitted with a 
Gaussian because of very few galaxies in it. Another possibility is that the cluster may have already gone through its first 
merger and the filament and the W components have been stripped off the SW subcluster in two different phases of ram-pressure 
stripping. However, if the W component is indeed a separate subcluster, it is possible that the SW and W subclusters are falling into 
and merging with the more massive and luminous NE subcluster.

  We have estimated the equipartition magnetic field and the minimum energy pressure for the WAT and the HT source seen in the ATCA 
radio images near the centre of the SW subcluster and at the periphery of the W subcluster respectively. Neither source shows 
any cavities associated with it in the unsharp masked X-ray images, which could be either because the observations are not deep enough
 to detect the cavities or because the hot thermal plasma has leaked into and filled the cavities. Although the minimum energy 
pressure of both the radio sources is somewhat less than the external X-ray pressure, pressure equilibrium can be achieved 
by considering a larger contribution from heavier particles, integration to much lower energies, and additional pressure due to gas 
entrained and heated by the jets. From the spectral analysis of the source based on various radio observations, we have estimated a 
spectral age of $\sim10$Myr for the WAT source.

  Deeper and higher resolution X-ray observations are required to properly understand the merging scenario and the geometry of the 
cluster, to quantify the structure of the shock fronts between the subclusters, and to understand the filamentary region in 
detail. Deeper radio observations of the HT source are required to better constrain the results from the imaging and spectral analysis
 of the source. Also, deeper radio observations at low frequencies ($\sim$300-500 MHz) are required to look for the extended sources 
of diffuse radio emission such as radio halos which are mostly found in merging clusters. The galaxy velocity distribution of the W 
region requires observations of redshifts from a larger sample of galaxies in this region and its surroundings in order to derive 
the velocity dispersion and mass estimates with better precision, and to confirm the distinct identity of the W subcluster. 
Pointed observations from XMM-Newton in the NW region and also along the filament between the clusters A3391 and A3395 
(discussed in Tittley \& Henriksen 2001) can help in a better understanding of the merging environment of A3395 and its connection 
with the supercluster network.

\section{Acknowledgement}
  The data for this research have been obtained from the High Energy Astrophysics Science Archive Research Center (HEASARC), 
provided by NASA's Goddard Space Flight Center. This work is based on observations obtained with XMM-Newton, an ESA science mission 
with instruments and contributions directly funded by ESA Member States and the USA (NASA). We have also made use of data from the 
Chandra X-ray Observatory, managed by NASA's Marshall Center. We thank the XMM helpdesk for their assistance on XMM-Newton data 
analysis. Data were also obtained from the Australia Telescope Compact Array which is a part of the Australia
 Telescope National Facility, funded by the Commonwealth of Australia for operation as a National Facility managed by CSIRO. We would 
like to thank Keith Arnaud for his help on various issues related to XSPEC. Finally, we thank the anonymous referee for his detailed 
comments and suggestions, which have helped us in improving the analyses and the presentation.



\begin{table}
 \caption{X-ray Observations table for A3395}
\label{tab:observation_table}
\vskip 0.5cm
\centering
{\small
\begin{tabular}{c c c c c c c}
\hline
Satellite & Detector & $\alpha$ (J2000) & $\delta$ (J2000) & Observation ID & Date of Obs. & Exp. time \\
\hline
\hline
XMM-Newton & MOS1, MOS2, PN & 06 27 11 & -54 29 06 & 0400010301 & 2007 Jan 24 & 29.9 ks \\
Chandra & ACIS-I & 06 26 50 & -54 32 35 & 4944 & 2004 Jul 11 & 22.2 ks\\
\hline
\end{tabular}}
\end{table}

\begin{table}
 \caption{Best fit parameters obtained from the spectral analysis of the full NE, SW, W and NW regions plus the filament connecting 
the NE and W regions using \textbf{XMM-Newton} data (all regions as shown in 
Fig.~\ref{fig:A3395_unsharp_mask_comb_MOS1_MOS2_and_tot_spec_regions}). The spectra for each of the regions are fitted with a 
single-temp \textbf{apec} model for a fixed Galactic absorption. The residual soft proton contamination and the instrumental Al line
 at 1.49 keV have been modeled by adding powerlaws and Gaussians respectively (separately for MOS1, MOS2 and PN) to the models. Best 
fit values for the temperature (kT), elemental abundance relative to the solar values, normalization of the \textbf{apec} model, 
X-ray luminosity (L$_{\rm x}$), and minimum reduced $\chi^{2}_{\nu}$ are given along with the degrees of freedom (DOF).}
\label{tab:XMM_A3395_total_regions_spectral_results}
\vskip 0.5cm
\centering
  \begin{threeparttable}
\begin{tabular}{c c c c c c}
 \hline
\hline
Region & kT & Abundance  &  \textbf{apec} norm. & L$_{\rm x}$ & $(\chi^{2}_{\nu})_{min}$ (DOF)\\
 & & & &(0.5-9.0 keV)& \\
 & (keV) & (relative to solar) & ($10^{-3}$ cm$^{-5}$) & ( $10^{43}$ ergs s$^{-1}$) & \\
\hline
NE       & $4.8\pm0.1$         & $0.42\pm0.05$       & $5.55\pm0.07$ & $4.15\pm0.05$ & 1.03 (1015) \\
SW       & $5.1\pm0.1$         & $0.35\pm0.05$       & $4.47\pm0.07$ & $3.35\pm0.04$ & 1.15 (922) \\
W        & $5.2\pm0.3$         & $0.3\pm0.1$         & $1.81\pm0.05$ & $1.33\pm0.03$ & 1.19 (421) \\
NW       & $4.8^{+0.3}_{-0.3}$ & $0.6^{+0.2}_{-0.1}$ & $3.5\pm0.1$   & $2.7\pm0.1$   & 1.10 (766) \\
Filament & $6.5^{+0.8}_{-0.6}$ & $0.3\pm0.2$         & $0.61\pm0.04$ & $0.48\pm0.02$ & 1.11 (227)\\[1ex]
\hline
\end{tabular}
     \begin{tablenotes}
\footnotesize
       \item[a] All errors are quoted at 90\% confidence level based on $\chi^{2}_{min}$+2.71.
     \end{tablenotes}
  \end{threeparttable}
\end{table}
\clearpage

\begin{table}
 \caption{Best fit parameters obtained from the spectral analysis of the full NE, SW, and W regions plus the filament connecting the 
NE and W regions using \textbf{Chandra} data (all regions as shown in 
Fig.~\ref{fig:A3395_unsharp_mask_comb_MOS1_MOS2_and_tot_spec_regions} and covered in Fig.~\ref{fig:Chandra_smoothedimage}). The 
spectra for each of the regions are fitted with the \textbf{wabs*apec} model for a fixed Galactic absorption. Best fit values for the 
temperature (kT), elemental abundance relative to the solar values, normalization of the \textbf{apec} model, X-ray luminosity 
(L$_{\rm x}$), and minimum reduced $\chi^{2}_{\nu}$ are given along with the degrees of freedom (DOF).}
\label{tab:Chandra_A3395_total_regions_spectral_results}
\vskip 0.5cm
\centering
  \begin{threeparttable}
\begin{tabular}{c c c c c c}
 \hline
\hline
Region & kT & Abundance  &  \textbf{apec} norm. & L$_{\rm x}$ & $(\chi^{2}_{\nu})_{min}$ (DOF)\\
 & & & &(0.5-9.0 keV)& \\
 & (keV) & (relative to solar) & ($10^{-3}$ cm$^{-5}$) & ( $10^{43}$ ergs s$^{-1}$) & \\
\hline
NE       & $5.8^{+0.6}_{-0.8}$ & $0.5\pm0.3$         & $5.9\pm0.4$   & $4.6\pm0.2$   & 0.97 (55) \\
SW       & $5.8^{+0.6}_{-0.5}$ & $0.3^{+0.3}_{-0.2}$ & $4.6\pm0.3$   & $3.4\pm0.1$   & 1.37 (55) \\
W        & $5.0^{+1.1}_{-0.9}$ & $0.0^{+0.2}_{-0.0}$ & $1.38\pm0.07$ & $0.84\pm0.07$ & 1.48 (49) \\
Filament & $6.3^{+2.2}_{-1.6}$ & $0.7^{+1.3}_{-0.7}$ & $0.5\pm0.1$   & $0.45\pm0.06$ & 1.20 (55)\\[1ex]
\hline
\end{tabular}
     \begin{tablenotes}
\footnotesize
       \item[a] All errors are quoted at 90\% confidence level based on $\chi^{2}_{min}$+2.71.
     \end{tablenotes}
  \end{threeparttable}
\end{table}
\clearpage

\begin{table}
 \caption{Best fit parameters obtained from the spectral analysis of elliptical annuli in the NE, SW, and W regions using 
\textbf{XMM-Newton} data. The spectra for all the annuli were fitted using the model \textbf{wabs*apec} for a fixed value of Galactic 
absorption and with elemental abundances set to the value of 0.42, 0.35, and 0.3 times the solar values for NE, SW, and W regions 
respectively. The residual soft proton contamination and the instrumental Al line at 1.49 keV have been modeled by adding powerlaws 
and Gaussians respectively (separately for MOS1, MOS2 and PN) to the models. Annulus number represents the position of the annulus 
from the innermost to the outermost annuli in increasing order. Values of temperature (kT), electron density (n$_{\rm e}$), pressure 
(P), and entropy (S) are listed.}
\label{tab:XMM_projected_annuli_spectral_results}
\vskip 0.5cm
\centering
  \begin{threeparttable}
\begin{tabular}{c c c c c c}
\hline
\hline
Region &Annulus Number& kT & n$_{\rm e}$ & P & S \\
 & & (keV) & ($10^{-4}$ cm$^{-3}$) & ($10^{-12}$ dyn cm$^{-2}$) & (keV cm$^{2}$) \\
\hline
NE & 1 & $5.1\pm0.3$         & $33.7\pm0.3$ & $27.6^{+2.5}_{-3.0}$ & $228^{+19}_{-24}$ \\
   & 2 & $4.4^{+0.5}_{-0.4}$ & $22.0\pm0.2$ & $15.6^{+2.0}_{-1.6}$ & $262^{+32}_{-26}$ \\
   & 3 & $4.7\pm0.4$         & $16.9\pm0.1$ & $12.8\pm1.2$         & $335\pm30$ \\
   & 4 & $4.4^{+0.4}_{-0.3}$ & $13.3\pm0.1$ &  $9.4^{+0.9}_{-0.7}$ & $363^{+35}_{-27}$ \\
   & 5 & $4.5\pm0.4$         & $10.5\pm0.1$ &  $7.6\pm0.7$         & $438\pm41$ \\
   & 6 & $4.3^{+0.4}_{-0.3}$ &  $8.8\pm0.1$ &  $6.0^{+0.6}_{-0.5}$ & $467^{+46}_{-36}$ \\
   & 7 & $4.8\pm0.4$         &  $6.7\pm0.1$ &  $5.2\pm0.5$         & $629\pm57$ \\
\hline
SW & 1 & $4.9^{+0.3}_{-0.5}$ &  $33.8\pm0.3$ & $26.4^{+1.9}_{-3.0}$ & $218^{+15}_{-24}$ \\
   & 2 & $5.1^{+0.3}_{-0.4}$ &  $20.3\pm0.2$ & $16.7^{+1.1}_{-1.5}$ & $321^{+21}_{-27}$ \\
   & 3 & $5.2^{+0.4}_{-0.5}$ &  $15.1\pm0.2$ & $12.5^{+1.1}_{-1.3}$ & $392^{+33}_{-41}$ \\
   & 4 & $6.0^{+0.4}_{-0.7}$ &  $11.7\pm0.1$ & $11.3^{+0.9}_{-1.4}$ & $541^{+40}_{-67}$ \\
   & 5 & $5.8^{+0.9}_{-0.8}$ &  $9.3\pm0.1$  & $8.7^{+1.5}_{-1.3}$  & $613^{+100}_{-90}$ \\
\hline
W & 1 & $5.2^{+0.9}_{-1.0}$ &  $30.0\pm0.6$ & $25.2^{+4.8}_{-5.3}$ &  $252^{+47}_{-51}$ \\
  & 2 & $5.2^{+1.0}_{-1.1}$ &  $19.2\pm0.4$ & $16.1^{+3.4}_{-3.7}$ &  $338^{+69}_{-76}$ \\
  & 3 & $5.4\pm0.9$         &  $14.4\pm0.2$ & $12.5\pm2.3$         &  $423\pm75$ \\
  & 4 & $6.1^{+1.4}_{-1.2}$ &  $11.8\pm0.3$ & $11.5^{+2.9}_{-2.5}$ &  $547^{+133}_{-116}$ \\[1ex]
\hline
\end{tabular}
    \begin{tablenotes}
\footnotesize
       \item[a] All errors are quoted at 90\% confidence level based on $\chi^{2}_{min}$+2.71.
     \end{tablenotes}
  \end{threeparttable}
\end{table}
\clearpage

\begin{table}
 \caption{Best fit parameters obtained from the spectral analysis of elliptical annuli in the NE, SW, and W regions using 
\textbf{Chandra} data. The spectra for all the annuli were fitted using the model \textbf{wabs*apec} for a fixed value of Galactic 
absorption and with elemental abundances set to the value of 0.42, 0.35, and 0.3 times the solar values for NE, SW, and W regions 
respectively. Annulus number represents the position of the annulus from the innermost to the outermost annuli in increasing order. 
Values of temperature (kT), electron density (n$_{\rm e}$), pressure (P), and entropy (S) are listed.}
\label{tab:Chandra_projected_annuli_spectral_results}
\vskip 0.5cm
\centering
  \begin{threeparttable}
\begin{tabular}{c c c c c c}
\hline
\hline
Region &Annulus Number& kT & n$_{\rm e}$ & P & S \\
 & & (keV) & ($10^{-4}$ cm$^{-3}$) & ($10^{-12}$ dyn cm$^{-2}$) & (keV cm$^{2}$) \\
\hline
NE & 1 & $4.2^{+0.9}_{-0.6}$ & $35.6\pm0.9$ & $24.0^{+5.5}_{-4.0}$ & $181^{+40}_{-28}$ \\
   & 2 & $5.0^{+1.2}_{-0.8}$ & $22.1\pm0.5$ & $17.5^{+4.5}_{-3.2}$ & $293^{+73}_{-52}$ \\
   & 3 & $5.7^{+1.1}_{-0.8}$ & $17.1\pm0.3$ & $15.7^{+3.2}_{-2.5}$ & $402^{+80}_{-60}$ \\
   & 4 & $4.9^{+0.8}_{-0.7}$ & $13.6\pm0.3$ & $10.7^{+2.0}_{-1.7}$ & $403^{+71}_{-60}$ \\
\hline
SW & 1 & $5.0^{+0.7}_{-0.6}$ &  $35.3\pm0.7$ & $28.1^{+4.5}_{-4.0}$ &  $215^{+33}_{-29}$ \\
   & 2 & $4.7^{+0.8}_{-0.6}$ &  $20.9\pm0.5$ & $15.8^{+3.0}_{-1.8}$ &  $289^{+53}_{-35}$ \\
   & 3 & $5.0^{+0.9}_{-0.7}$ &  $15.3\pm0.3$ & $12.2^{+2.5}_{-2.1}$ &  $377^{+74}_{-61}$ \\
   & 4 & $5.5^{+1.3}_{-0.9}$ &  $11.4\pm0.3$ & $10.0^{+2.6}_{-1.8}$ &  $500^{+124}_{-86}$ \\
   & 5 & $6.6^{+1.8}_{-1.3}$ &   $9.1\pm0.2$ &  $9.5^{+2.8}_{-2.2}$ &  $702^{+203}_{-155}$ \\
\hline
W & 1 & $3.2^{+1.0}_{-0.7}$ &  $27.6\pm1.6$ & $14.4^{+5.2}_{-4.0}$ &  $165^{+57}_{-43}$ \\
  & 2 & $5.1^{+1.8}_{-1.3}$ &  $18.3\pm0.7$ & $14.9^{+5.9}_{-4.2}$ &  $341^{+131}_{-92}$ \\[1ex]
\hline
\end{tabular}
    \begin{tablenotes}
\footnotesize
       \item[a] All errors are quoted at 90\% confidence level based on $\chi^{2}_{min}$+2.71.
     \end{tablenotes}
  \end{threeparttable}
\end{table}
\clearpage

\begin{table}
 \caption{Best fit parameters obtained from the deprojected spectral analysis of elliptical annuli in the NE, SW, and W regions using 
\textbf{XMM-Newton} data. The spectra for all annuli in each region were fitted together using the model \textbf{projct(wabs*apec)} 
for a fixed Galactic absorption and with elemental abundances set to the value of 0.42, 0.5, and 0.3 times the solar values for NE, 
SW, and W regions respectively.  The residual soft proton contamination and the instrumental Al line at 1.49 keV have been modeled by 
adding powerlaws and Gaussians respectively to the models. Annulus number represents the position of the annulus from the innermost 
to the outermost annuli in increasing order. Values of temperature (kT), electron density (n$_{\rm e}$), pressure (P), and entropy 
(S) are listed.}
\label{tab:XMM_deprojected_annuli_spectral_results}
\vskip 0.5cm
\centering
  \begin{threeparttable}
\begin{tabular}{c c c c c c}
\hline
\hline
Region &Annulus Number& kT & n$_{\rm e}$ & P & S \\
 & & (keV) & ($10^{-4}$ cm$^{-3}$) & ($10^{-12}$ dyn cm$^{-2}$) & (keV cm$^{2}$) \\
\hline
NE & 1 & $4.5^{+2.0}_{-1.2}$ & $23.0\pm1.3$ & $16.4^{+8.3}_{-5.5}$  & $255^{+124}_{-81}$ \\
   & 2 & $4.4^{+2.4}_{-1.4}$ & $20.6\pm1.2$ & $14.6^{+8.7}_{-5.4}$  & $272^{+157}_{-95}$ \\
   & 3 & $3.6^{+3.5}_{-1.0}$ & $16.6\pm0.8$ & $9.6^{+9.8}_{-3.0}$   & $256^{+258}_{-77}$ \\
   & 4 & $4.6^{+0.9}_{-1.3}$ & $16.2\pm0.5$ & $11.9^{+2.7}_{-3.9}$  & $334^{+72}_{-105}$ \\
   & 5 & $3.5^{+3.0}_{-0.7}$ & $12.7\pm0.5$ & $7.1^{+6.4}_{-1.7}$   & $297^{+263}_{-68}$ \\
   & 6 & $3.4^{+1.4}_{-0.7}$ & $11.8\pm0.5$ & $6.3^{+2.9}_{-1.6}$   & $300^{+135}_{-71}$ \\
   & 7 & $4.6\pm0.6$         & $14.8\pm0.3$ & $10.9\pm1.6$          & $355\pm50$ \\
\hline
SW & 1 & $4.3^{+1.7}_{-1.0}$ & $28.6\pm1.0$ & $19.7^{+8.2}_{-5.2}$ & $214^{+87}_{-54}$ \\
   & 2 & $5.5^{+1.6}_{-1.7}$ & $20.6\pm0.8$ & $18.1^{+6.0}_{-6.2}$ & $339^{+108}_{-112}$ \\
   & 3 & $4.2^{+1.7}_{-0.9}$ & $18.1\pm0.7$ & $12.0^{+5.3}_{-3.2}$ & $280^{+118}_{-70}$ \\
   & 4 & $4.8^{+1.9}_{-2.3}$ & $12.8\pm0.7$ & $9.9^{+4.4}_{-5.2}$  & $407^{+175}_{-206}$ \\
   & 5 & $5.0^{+1.7}_{-0.7}$ & $18.8\pm0.5$ & $15.0^{+5.5}_{-2.4}$ & $326^{+117}_{-51}$ \\
\hline
W & 1 & $3.8^{+3.9}_{-1.8}$ & $19.0\pm2.1$ & $11.5^{+13.2}_{-6.6}$  & $245^{+274}_{-132}$ \\
  & 2 & $6.7^{+6.1}_{-3.6}$ & $18.8\pm1.3$ & $20.0^{+19.6}_{-12.1}$ & $438^{+419}_{-254}$ \\
  & 3 & $3.2^{+2.7}_{-1.4}$ & $11.7\pm1.1$ & $6.1^{+5.6}_{-3.2}$    & $292^{+258}_{-146}$ \\
  & 4 & $5.4^{+2.6}_{-1.1}$ & $21.4\pm0.7$ & $18.4^{+9.4}_{-4.4}$   & $323^{+161}_{-74}$ \\[1ex]
\hline
\end{tabular}
    \begin{tablenotes}
\footnotesize
       \item[a] All errors are quoted at 90\% confidence level based on $\chi^{2}_{min}$+2.71.
     \end{tablenotes}
  \end{threeparttable}
\end{table}
\clearpage

\begin{table}
 \caption{Best fit parameters obtained from the deprojected spectral analysis of elliptical annuli in the NE, SW, and W regions using 
\textbf{Chandra} data. The spectra for all annuli in each region were fitted together using the model \textbf{projct(wabs*apec)} for 
a fixed Galactic absorption and with elemental abundances set to the value of 0.42, 0.35, and 0.3 times the solar values for NE, SW, 
and W regions respectively. Annulus number represents the position of the annulus from the innermost to the outermost annuli in 
increasing order. Values of temperature (kT), electron density (n$_{\rm e}$), pressure (P), and entropy (S) are listed.}
\label{tab:Chandra_deprojected_annuli_spectral_results}
\vskip 0.5cm
\centering
  \begin{threeparttable}
\begin{tabular}{c c c c c c}
\hline
\hline
Region &Annulus Number& kT & n$_{\rm e}$ & P & S \\
 & & (keV) & ($10^{-4}$ cm$^{-3}$) & ($10^{-12}$ dyn cm$^{-2}$) & (keV cm$^{2}$) \\
\hline
NE & 1 &   $3.6^{+3.4}_{-1.4}$ & $30.0\pm2.6$ & $17.3^{+1.8}_{-8.3}$   & $174^{+172}_{-78}$ \\
   & 2 &   $3.2^{+2.6}_{-1.0}$ & $23.3\pm1.9$ & $11.9^{+1.1}_{-4.8}$   & $182^{+157}_{-69}$ \\
   & 3 & $13.7^{+12.1}_{-7.8}$ & $16.6\pm2.5$ & $36.4^{+20.2}_{-26.2}$ & $981^{+972}_{-655}$ \\
   & 4 &   $5.0^{+0.7}_{-0.6}$ & $27.7\pm0.5$ & $22.0^{+3.3}_{-3.0}$   & $252^{+36}_{-33}$ \\
\hline
SW & 1 & $5.1\pm1.4$         &  $30.7\pm1.6$ & $25.0^{+8.0}_{-8.0}$ &  $241\pm73$ \\
   & 2 & $4.3\pm1.2$         &  $21.9\pm1.5$ & $15.2^{+5.4}_{-5.4}$ &  $256\pm85$ \\
   & 3 & $4.5\pm1.1$         &  $19.0\pm1.1$ & $13.6^{+4.0}_{-4.0}$ &  $291\pm81$ \\
   & 4 & $4.3^{+3.4}_{-1.2}$ &  $13.1\pm1.1$ &  $9.1^{+7.9}_{-3.2}$ &  $361^{+306}_{-118}$ \\
   & 5 & $6.4\pm1.2$         &  $18.6\pm0.5$ & $19.1^{+4.2}_{-4.1}$ &  $423^{+89}_{-88}$ \\
\hline
W & 1 & $2.4^{+0.8}_{-0.5}$ &  $19.5\pm1.5$ &  $7.5^{+3.1}_{-2.3}$ &  $155^{+60}_{-43}$ \\
  & 2 & $5.2^{+2.0}_{-1.2}$ &  $15.0\pm0.6$ & $12.4^{+5.4}_{-3.4}$ &  $395^{+166}_{-102}$ \\[1ex]
\hline
\end{tabular}
    \begin{tablenotes}
\footnotesize
       \item[a] All errors are quoted at 90\% confidence level based on $\chi^{2}_{min}$+2.71.
     \end{tablenotes}
  \end{threeparttable}
\end{table}
\clearpage

\begin{table}
\begin{center}
\caption{Flux densities of the WAT source.}
\vskip 0.5cm
\label{tab:radio_flux}
\begin{tabular}{l c c}
\tableline\tableline
Frequency & Flux density & Refs. \\
\tableline
(MHz) &  (mJy) & \\
\tableline
\tableline
408  &  $7860\pm680$ & a \\
843  &  $5000\pm500$ & b \\
1348 &  $3128\pm156$ & P \\
1410 &  $3200\pm160$ & c \\
2374 &  $1651\pm83$  & P \\ 
2650 &  $1800\pm90$  & c \\
2700 &  $1730\pm72$  & d \\
4850 &  $931\pm47$   & e \\
4850 &  $1021\pm54$  & f \\
5000 &  $870\pm44$   & d \\
8400 &  $460\pm23$   & g \\
\tableline
\end{tabular}

{a}: {\citep{1981MNRAS.194..693L}; flux density estimated from the integrated
      flux ratio in the Molonglo Transit Catalogue (\citep{2006AJ....131..100B})};
{b}: {integrated flux density at 843 MHz, measured with the Volume
      program from a MOST CUTS image as listed in Burgess \& Hunstead (2006); 
      Jones \& McAdam (1992) quote a somewhat higher value of 5610$\pm$505 mJy};
{c}: {\citep{1969AuJPh..22...79G}};
{d}: {\citep{1975AuJPA..34...55W}};
{e}: {\citep{1994ApJS...90..173G}};
{f}: {\citep{1994ApJS...91..111W}};
{g}: {PKS Catalogue 1990, as listed in NED (\citep{1990PKS...C......0W}), errors of $\sim$5\% have been assumed.};
{P}: {Present paper; ATCA observations. The 2374 MHz observation is missing flux and has not been used in the spectrum plot in 
Fig.~\ref{A3395_radio_spectrum}.}
\end{center}
\end{table}

\clearpage

\clearpage
\begin{table}
 \caption{Results of $\beta$-model fitting of the surface-brightness profiles for the components of A3395 and their 
bolometric X-ray luminosities.}
\label{tab:bol_xray_luminosity}
\vskip 0.5cm
\centering
  \begin{threeparttable}
\begin{tabular}{c c c c  c}
\hline
\hline
Region & $\beta$               & $r_{c}$ & $\rm F_{\rm x}(0)$ & $\rm L^{\rm b}_{\rm X}$ \\
 &  & ($10^{-2}$ Mpc) & ($10^{-13}$ erg cm$^{-2}$ s$^{-1}$) & ($10^{44}$ erg s$^{-1}$) \\
\hline
NE   & $0.71\pm0.02$  & $3.02\pm0.08$ & $7.2\pm0.2$ & $1.32\pm0.03$ \\
SW   & $0.43\pm0.01$  & $1.12\pm0.06$ & $8.7\pm0.5$ & $1.59\pm0.08$ \\
W    & $0.43\pm0.08$  & $1.4\pm0.4$   & $4.4\pm0.7$ & $0.88\pm0.14$ \\
NW   &       --       &       --      &      --     & $0.17\pm0.01$ \\[1ex]
\hline
\end{tabular}
     \begin{tablenotes}
\footnotesize
       \item[a] Errors are quoted at 68\% confidence level (1$\sigma$) based on $\chi^{2}_{min}$+1.00.
       \item[a] $\rm F_{\rm x}(0)$ $=$ X-ray flux at the centre, $\beta$ $=$ $-1/3$(slope-0.5) and $\rm r_{\rm c}$ $=$ core radius.
       \item[b] The bolometric X-ray luminosity for the NW region has been estimated by assuming a constant surface-brightness in an 
oblate ellipsoid with semi-major and semi-minor axes lengths equal to  $11.11^{\prime}$ and $4.45^{\prime}$ respectively.
     \end{tablenotes}
  \end{threeparttable}
\end{table}

\clearpage
\begin{table}
 \caption{Mass of hot gas for different regions of A3395.}
\label{tab:gas_mass_estimate}
\vskip 0.5cm
\centering
  \begin{threeparttable}
\begin{tabular}{c c c c c c}
\hline
\hline
Region & $\beta$               & $r_{c}$ & $\rho_{0}$ & r & $\rm M_{\rm gas } (r)$ \\
 &  & ($10^{-2}$ Mpc) & ($10^{13}$ $\rm M_{\odot}$ Mpc$^{-3}$) & (Mpc) & ($10^{13}$ $\rm M_{\odot}$) \\
\hline
NE   & $0.36\pm0.01$           & $6.0\pm0.6$  & $6.8\pm0.4$ & 0.5 & $0.54\pm0.07$ \\
     &                         &              &               & 1.0 & $2.1\pm0.3$ \\
SW   & $0.55\pm0.06$           & $11.4\pm2.2$ & $4.7\pm0.6$ & 0.5 & $0.4\pm0.1$ \\
     &                         &              &               & 1.0 & $1.1\pm0.5$ \\
W    & $0.36\pm0.07$           & $7.4\pm2.2$  & $3.9\pm0.4$   & 0.5 & $0.4\pm0.2$  \\
     &                         &              &               & 1.0 & $1.6\pm0.8$\\
NW   &           --            &     --       &      --       &  -- & $0.11\pm0.02$  \\[1ex]
\hline
\end{tabular}
     \begin{tablenotes}
\footnotesize
       \item[a] Errors are quoted at 68\% confidence level (1$\sigma$) based on $\chi^{2}_{min}$+1.00.
       \item[b] The gas mass for the NW region was calculated by assuming a constant density (derived from the \textbf{apec} 
normalization obtained from the spectral analysis of the NW region in \S\ref{sec:Total-Spec-Analy}) in an oblate ellipsoid with 
semi-major and semi-minor axes lengths equal to  $11.11^{\prime}$ and $4.45^{\prime}$ respectively.
     \end{tablenotes}
  \end{threeparttable}
\end{table}
\clearpage

\begin{table}
 \caption{Velocities and virial mass values for A3395 regions/subclusters}
\label{tab:A3395_gala_vel_distbn_results}
\vskip 0.5cm
\centering
  \begin{threeparttable}
\begin{tabular}{c c c c c}
\hline
\hline
Region/Subcluster & No. of galaxies & $\bar{\rm v}$ & $\sigma_{\rm v}$ & $\rm M_{\rm virial}$\\
 &  & (km s$^{-1}$) & (km s$^{-1}$) & ($10^{14}$ $\rm M_{\odot}$)\\
\hline
NE & 71 & $15170\pm64$ & $948\pm64$ & $8.1\pm1.1$ \\slide
SW & 32 & $15582\pm55$ & $405\pm55$ & $1.4\pm0.4$ \\
W & 15 & $15298\pm330$ & $842\pm330$ & $3.5\pm2.7$ \\
NW & 17 & $15110\pm130$ & -- & -- \\
 & 10 & $14540\pm70$ & -- & -- \\[1ex]

\hline
\end{tabular}
     \begin{tablenotes}
\footnotesize
       \item[a] Errors are quoted at 68\% confidence level (1$\sigma$) based on $\chi^{2}_{min}$+1.00.
       \item[b] The two values quoted for the mean velocity of the NW region are obtained, first one is from the velocity data 
given in Donnelly et al. (2001) while second one is from  \citep{1990ApJS...72..715T}. Results for all other regions
 are from Donnelly et al. (2001) alone.
     \end{tablenotes}
  \end{threeparttable}
\end{table}

\begin{table}
 \caption{Parameters from the bound system analysis of A3395 subcluster pairs}
\label{tab:A3395_bnd_system_analysis}
\vskip 0.5cm
\centering
  \begin{threeparttable}
\begin{tabular}{c c c}
\hline
\hline
pair  & $R_{p}$ &  $V_{r}$ \\
      & (kpc)   & (km s$^{-1}$) \\
\hline
NE-SW &  469  & $412\pm118$  \\
SW-W  &  386  & $283\pm384$ \\
NE-W  &  690  &  $129\pm393$ \\[1ex]
\hline
\end{tabular}
     \begin{tablenotes}
\footnotesize
       \item[a] Errors are quoted at 68\% confidence level (1$\sigma$) based on $\chi^{2}_{min}$+1.00.
     \end{tablenotes}
  \end{threeparttable}
\end{table}
\clearpage


\clearpage

\clearpage
 
\clearpage

\clearpage

\clearpage

 \clearpage

\clearpage

 \clearpage

\clearpage

\clearpage

\clearpage

 \clearpage

\clearpage

\end{document}